\documentclass[12pt]{article}

\usepackage{graphicx,psfrag,epsf,color}
\usepackage{amsmath,amssymb,amsfonts}
\usepackage{array}
\usepackage{cite}
\usepackage{multirow}

\newcommand{\rescm}{\widehat}

\begin{document}

\allowdisplaybreaks

\begin{titlepage}

\begin{flushright}
{\small
TUM-HEP-864/12\\
TTK-12-44\\
SFB/CPP-12-77\\
UWThPh-2012-31\\
IFIC/12-70\\[0.2cm]

27 October 2012}
\end{flushright}

\vskip1cm
\begin{center}
\Large\bf\boldmath
Non-relativistic pair annihilation of  
nearly mass degenerate neutralinos and charginos
I. General framework and S-wave annihilation
\end{center}

\vspace{0.8cm}
\begin{center}
{\sc M.~Beneke$^{a,b}$, C.~Hellmann$^{a,b}$} and 
{\sc P. Ruiz-Femen\'\i a$^{c,d}$}\\[5mm]
{\it ${}^a$Physik Department T31,\\
James-Franck-Stra\ss e, 
Technische Universit\"at M\"unchen,\\
D--85748 Garching, Germany\\
\vspace{0.3cm}
${}^b$Institut f\"ur Theoretische Teilchenphysik und 
Kosmologie,\\
RWTH Aachen University, D--52056 Aachen, Germany\\
\vspace{0.3cm}
${}^c$University of Vienna - Faculty of Physics \\
Boltzmanngasse 5, A-1090 Wien, Austria\\
\vspace{0.3cm}
${}^d$Instituto de F\'\i sica Corpuscular (IFIC), 
CSIC-Universitat de Val\`encia \\
Apdo. Correos 22085, E-46071 Valencia, Spain}\\[0.3cm]
\end{center}

\vspace{1cm}
\begin{abstract}
\vskip0.2cm\noindent
We compute analytically the tree-level annihilation rates of 
a collection of non-relativistic
neutralino and chargino two-particle states in 
the general MSSM, including the previously unknown off-diagonal 
rates. The results are prerequisites to the calculation of the 
Sommerfeld enhancement in the MSSM, which will be presented in 
subsequent work. They can also be used to obtain concise analytic 
expressions for MSSM dark matter pair annihilation in the present 
Universe for a large number of exclusive two-particle final states.
\end{abstract}
\end{titlepage}



\section{Introduction}
\label{sec:introduction}

The presence of dark matter (DM) in the Universe is one of the few empirical 
evidences that the current Standard Model of particle physics 
cannot be complete. The dark matter density $\Omega_{\rm DM} = 
0.111(6)/h^2$ ($h=0.710 (25)$) \cite{Beringer:1900zz} is now determined 
very precisely from various observations. While the particle nature and 
genesis of the dark matter relic density remains unknown, it is 
intriguing that it can be explained naturally through 
thermal production and freeze-out of a particle with electroweak interaction
strengths and mass of order of the TeV scale. In this simple scenario 
freeze-out occurs when the Universe cools below the mass of the 
particle and the DM particles become non-relativistic, with typical 
velocities of order $v\sim 0.2\, c$. The DM pair-annihilation cross 
section, which determines the relic density, can then be expanded  
in a Taylor series in $v$, and keeping only the first two terms is 
usually a very good approximation:
\begin{align}
\label{eq:sigmav}
 \sigma_\text{ann} \, v_\text{rel} \, \approx \,a + b\,v_\text{rel}^2 \,.
\end{align}
Here $v_\text{rel}=\vert\vec{v}_1-\vec{v}_2\vert$ 
denotes the relative velocity of the annihilating 
particles in their center-of-mass frame. Furthermore, when 
dark matter particles pair-annihilate in the present Universe, 
potentially revealing themselves in cosmic ray signatures,
the typical velocities are $v \sim 10^{-3} c$, and the 
annihilation occurs even deeper in the non-relativistic regime.

Among the many models that contain weakly interacting dark matter 
candidates at the TeV scale, the minimal supersymmetric standard 
model (MSSM) has been studied most extensively, and for quite 
some time~\cite{Jungman:1995df,Bertone:2004pz}. Several programs are available 
\cite{Gondolo:2004sc,Belanger:2010gh} which compute the 
annihilation cross section of the lightest neutralino, together 
with possible co-annihilation processes, in the MSSM numerically in 
the tree-level approximation. The observed relic density then 
provides a valuable constraint on the parameter space of the model, 
complementary to those from collider physics. Given the 
precision of $\Omega_{\rm DM}$, it seems desirable to compute 
the cross section parameters $a$, $b$ at the one-loop level. This, 
however, is a daunting task due to the complexity of 
complete one-loop calculations in the MSSM, and the number of 
individual annihilation processes that add up to the total cross 
section. Nevertheless, such calculations have been performed for 
certain scenarios where QCD corrections are the most important 
ones~\cite{Herrmann:2007ku,Herrmann:2009wk,Herrmann:2009mp}, 
or in certain approximations~\cite{Boudjema:2011ig,Chatterjee:2012db}. 
The calculation of the full electroweak corrections has been 
started  \cite{Baro:2007em,Baro:2009na}.

There exist situations when quantum corrections become exceedingly 
large and cannot be neglected. In non-relativistic scattering and 
annihilation of DM particles this happens when the Coulomb  
(Yukawa) force generated by massless (massive) particle exchange between 
the DM particles becomes strong at small relative velocities, 
a phenomenon also known as ``Sommerfeld effect''. In the MSSM 
this situation is naturally realized 
when the lightest neutralino (LSP) has mass above one TeV, 
in which case the neutralino is almost a pure gauge eigenstate, 
and degeneracies and co-annihilation effects in the neutralino-chargino
sector are generic. The 
Sommerfeld effect in the MSSM was first studied in the wino- and 
Higgsino-limit by Hisano et al. \cite{Hisano:2004ds,Hisano:2006nn}, 
and subsequently in ``minimal dark matter models'' \cite{Cirelli:2007xd} 
that resemble the MSSM in the above-mentioned limits. In these  
heavy dark matter scenarios the annihilation cross section can be 
enhanced by more than an order of magnitude, since the typical distance 
of DM particles at small velocity is within the long-range part of the 
Yukawa potential generated by exchange of the electroweak 
$W$ and $Z$ gauge bosons. The suggestion \cite{ArkaniHamed:2008qn}
that the Sommerfeld enhancement due to the exchange of a new, light particle 
may generate an excess in the cosmic ray positron spectrum has 
generated a resurge of interest in this effect.

The present work aims at improving the calculation of the dark matter 
annihilation cross section and relic abundance by including the Sommerfeld 
radiative corrections in the general MSSM, beyond the previously 
considered wino- and Higgsino-limit. The idea is that even when the 
Sommerfeld correction is not of order one, unlike in scenarios of TeV 
scale LSPs, it may still constitute the dominant radiative correction 
in a significantly larger portion of the MSSM parameter space. 
Related work has been undertaken recently in 
\cite{Drees:2009gt,Hryczuk:2010zi,Hryczuk:2011tq,Hryczuk:2011vi}. 
Our approach differs from or extends these works in several aspects.
\begin{itemize}
\item We use the non-relativistic effective theory approach to separate 
  the short-distance annihilation process from the long-distance 
  Sommerfeld effect, which is encoded in the matrix elements of local 
  four-fermion operators. The approach is very similar to the 
  NRQCD treatment of quarkonium annihilation \cite{Bodwin:1994jh}, except 
  that we deal with scattering states of several species of  
  particles interacting through the electroweak Yukawa force.
\item Since electroweak gauge boson exchange may change the 
  two-particle state (for instance, scatter a neutralino pair into 
  a pair of oppositely charged charginos), the annihilation process is 
  described by a matrix in the space of two-particle states, which 
  is not diagonal. The off-diagonal terms cannot be obtained from the 
  tree-level cross sections computed by numerical programs, and have not 
  been considered previously, except in the simplified situation of the 
  strict wino- and Higgsino-limit~\cite{Hisano:2004ds,Hisano:2006nn,Cirelli:2007xd,Hryczuk:2011vi}.
\item We compute the expansion of the short-distance annihilation 
  cross section analytically rather than numerically. The only 
  systematic previous analytic calculation \cite{Drees:1992am} 
  refers to the annihilation of two LSPs, but does not include 
  co-annihilation channels and the above-mentioned off-diagonal 
  annihilation matrix entries.
\item The non-relativistic annihilation cross section can be organized in 
  a partial wave expansion. The leading-order term $a$ in 
  (\ref{eq:sigmav}) contains the leading-order contributions from $S$-wave 
  annihilations, whereas the second term $b$ encodes both $S$- and $P$-wave 
  annihilation contributions. The Sommerfeld correction factor is 
  different for the $S$- and $P$-wave contribution. Its consistent 
  implementation therefore requires a separation of $b$ into its 
  two components, which has not been done before, but can be 
  implemented relatively easily within our analytic framework.
\end{itemize}
The present paper is devoted to the analytic calculation of the dominant 
$S$-wave annihilation coefficient $a$ and to outlining the general 
framework. The subleading term $b$ and the calculation of the Sommerfeld 
effect in the MSSM with almost degenerate neutralinos and charginos will be 
presented in \cite{Hellmann:2013jxa} and \cite{paperIII}, respectively.

The outline of this paper is as follows. In Sec.~\ref{sec:eft} 
we introduce the effective Lagrangian framework and the method of calculation. 
In the non-relativistic MSSM the short-distance annihilation 
process is encoded in the Wilson coefficient of a four-fermion operator. 
We introduce the required notation and discuss the expansion in 
the mass differences of the nearly degenerate neutralino and chargino 
species. Sec.~\ref{sec:results} discusses various examples of 
annihilation cross sections obtained from our analytic calculation. In this 
section we also perform checks by comparing some diagonal annihilation 
matrix entries with numerical cross sections and with 
\cite{Drees:1992am}. The complete analytic results are rather lengthy. 
We provide them in appendix~\ref{sec:appendix} together with the rules 
to construct the MSSM coupling factors of the various diagrams. 
In Sec.~\ref{sec:discussions} we explain why it is convenient to 
employ Feynman gauge despite the fact that this requires the computation 
of many unphysical final states. We also illustrate the importance 
of including the off-diagonal annihilation matrix entries in the 
computation of the Sommerfeld-corrected cross section on the 
example of a heavy wino-like MSSM parameter point.
We summarize in Sec.~\ref{sec:summary}.


\section{Effective Lagrangian and method of calculation}
\label{sec:eft}

\subsection{The Lagrangian in the effective theory}
\label{subsec:lagrangian}

We describe the kinematics and interactions of neutralinos and 
charginos moving 
at small velocities within a non-relativistic effective theory (EFT), the
non-relativistic MSSM (NRMSSM), that contains only nearly on-shell
non-relativistic chargino and neutralino modes, while the effects from higher
mass and virtual modes are encoded in the Wilson coefficients of
higher-dimensional operators. The neutralinos and charginos described
in the EFT approach are those whose masses are nearly degenerate with the
mass $m_{\text{LSP}}$ of the lightest neutralino.
The corresponding effective Lagrangian is given by
\begin{align}
 \mathcal L^\text{NRMSSM}
	=
   \mathcal L_\text{kin} + \mathcal L_\text{pot}
 + \delta\mathcal L_\text{ann} + \text{higher order terms}
 \ .
\end{align}
The kinetic part of the Lagrangian for $n_0\le 4$ non-relativistic
neutralino species and  $n_+\le 2$ non-relativistic chargino species is 
given by
\begin{eqnarray}
\mathcal L_\text{kin}&= &
 \sum\limits_{i=1}^{n_0}
  \xi^\dagger_i \left( i\partial_t - (m_i - m_{\text{LSP}}) + 
\frac{\vec\partial^{\,2}}{2 m_{\mathrm{LSP}} } \right) \xi_i
\nonumber\\
&& + \,
 \sum_{\psi=\eta,\zeta} \sum\limits_{j=1}^{n_+}
   \psi^\dagger_j \left( i\partial_t - (m_j - m_{\text{LSP}}) + 
\frac{\vec\partial^{\,2}}{2 m_{\mathrm{LSP}}} \right) \psi_j\,.
\label{eq:kin}
\end{eqnarray}
The fields $\xi_i$ and  $\psi_j=\eta_j, \zeta_j$ 
represent the non-relativistic two-component spinor fields of
non-relativistic neutralinos ($\chi^0_i$) and charginos ($\chi^-_j$ and
$\chi^+_j$), respectively.
This EFT setup with one reference mass scale, $m_{\mathrm{LSP}}$, is suited
for the description of (neutralino) dark matter annihilation processes in the
present Universe as well as for the computation of dark matter co-annihilation
reactions with further nearly mass-degenerate neutralinos and charginos in the
context of the relic abundance calculation. The EFT framework can easily 
be extended to the case where the non-relativistic particle species are
(nearly) mass-degenerate with respect to two distinct scales
$m_{\mathrm{ref}}^{\,(1,2)}$, with $m_{\mathrm{ref}}^{(1)} \ll m_{\mathrm{ref}}^{(2)}$.
In that case, the mass differences $(m_k - m_{\text{LSP}})$ in (\ref{eq:kin})
have to be replaced by $m_k - m_{\mathrm{ref,}\,k}$, where each
$m_{\mathrm{ref,}\,k}$ is given by one of the scales $m_{\mathrm{ref}}^{(1,2)}$. In
that way, an entirety of hydrogen-like two-particle states can be described,
within which a set of light, nearly mass-degenerate and another set of heavy,
nearly mass-degenerate particles exists. Our results for the absorptive part of
the Wilson coefficients, specified in Sec.~\ref{subsec:basis} and given in the
appendix, cover both the cases of a set of particles nearly mass-degenerate 
with the neutralino LSP and a set of non-relativistic hydrogen-like 
neutralino and chargino systems.

The term $\mathcal L_\text{pot}$ summarizes (instantaneous) Yukawa- and
Coulomb potential interactions that arise through the exchange
of SM gauge bosons and Higgs particles.
The generic form of $\mathcal L_\text{pot}$ reads  
\begin{align}
\mathcal L_\text{pot} = - \int d^3 \vec r ~ \Phi_{kl}^\dagger(x, \vec{r}\,)  
\, V_{ijkl}(r) \, \Phi_{ij}(x, \vec{r}\,)
\label{eq:pot}
\end{align}
where the fields $\Phi_{ij}$ describe a two-body state of the form
$\chi^0_i\chi^0_j$, $\chi^\mp_i\chi^\pm_j$, $\chi^0_i\chi^\pm_j$ or
$\chi^\pm_i\chi^\pm_j$ and a sum over repeated indices is implicit.
 $V_{ijkl}(r)$ thus represents the potential interactions among two-body states
$(i j)$ and $(k l)$ in configuration space, with $\vec{r}$ the spatial
3-vector denoting the relative distance in the two-body system.
The explicit form of the potentials between neutralino and chargino
species for a given MSSM point will be given elsewhere~\cite{paperIII}.

\subsection{Basis of the dimension-6 operators in $\delta \mathcal L_\text{ann}$}
\label{subsec:basis}
Within the NRMSSM, we aim to describe neutralino and chargino
pair-annihilation processes into two-particle final states of Standard Model
(SM) and (light) Higgs particles, which are not non-relativistic. The theory
will contain effects from virtual and higher-mass Higgs and SUSY particle modes 
as well, encoded in the EFT operator-coefficients and parameters.
The specific case of resonant $s$-channel pair-annihilation\ reactions can be
covered by adding a resonance width in the analytic results that we
give in the appendix.
Yet we exclude the case of accidental mass degeneracies of further
SUSY particles with the set of non-relativistic neutralinos and charginos.

The SM and light Higgs particle final states in the neutralino and chargino
pair-annihilation reactions are not described
within the non-relativistic effective theory, as they are characterized by
velocities outside the non-relativistic regime.
However, since the hard inclusive pair-annihilation processes take 
place within distances of order $1/m_{\text{LSP}}$, we can incorporate the 
short-distance annihilation
rates of non-relativistic neutralinos and charginos in the
effective theory through the absorptive part of Wilson coefficients of local
four-fermion operators in $\delta \mathcal L_\text{ann}$, 
following the approach of \cite{Bodwin:1994jh}.
The full annihilation rates in the non-relativistic effective theory are given
by the absorptive part of the matrix elements of these four-fermion 
operators. While the matrix elements of the operators themselves may encode
long-distance effects, giving rise to Sommerfeld enhancements,
the contribution to the hard annihilation reaction factors out in
the form of the Wilson coefficient.

\begin{figure}[t]
\begin{center}
\includegraphics[width=0.95\textwidth]{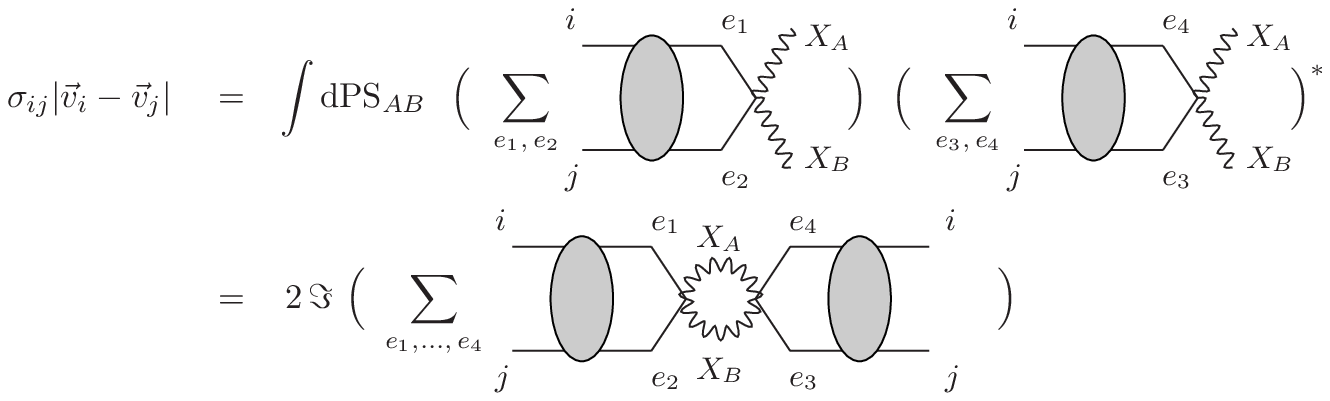}
\caption{ Diagrammatic picture for the relation among
          the annihilation amplitude and the absorptive
          part of the corresponding forward scattering amplitude
          in presence of long-range potential interactions.}
\label{fig:genericdiagram}
\end{center}
\end{figure}

In contrast to the application of this formalism to quarkonium annihilation in
QCD \cite{Bodwin:1994jh}, we are going to describe annihilations of
scattering states instead of bound states and allow for more than
one non-relativistic particle species.
The latter allows for the possibility, that (long-range) potential interactions
(indicated by the grey oval in Fig.~\ref{fig:genericdiagram})
lead to transitions from the  initially incoming two-particle state
$\chi_i \chi_j$ to another two nearly on-shell non-relativistic two-particle
state $\chi_{e_1} \chi_{e_2}$ prior to the annihilation reaction.
Unitarity relates the phase space integrated product of annihilation amplitudes
$\chi_i \chi_j \to X_A X_B$ in the first line of Fig.~\ref{fig:genericdiagram}
to the absorptive part of the forward scattering amplitude
$\chi_i \chi_j \rightarrow \chi_i \chi_j$ depicted in the second line, where
$X_A X_B$ generically denotes a pair of SM and light Higgs particles.
%
%
Note that due to the presence of the long-range potential interactions, the
hard annihilation reaction is determined by the absorptive part of the
$\chi_{e_1} \chi_{e_2} \to \chi_{e_4} \chi_{e_3}$ amplitude,
as can be seen in the second line of Fig.~\ref{fig:genericdiagram}.
It is worth to stress, that the $\chi_{e_1} \chi_{e_2}$ particle pair is
not necessarily equal to the $\chi_{e_4} \chi_{e_3}$ pair, such that apart
from true forward scattering reactions
$\chi_{e_1}\chi_{e_2}\to\chi_{e_1}\chi_{e_2}$, we encounter off-diagonal
$\chi_{e_1}\chi_{e_2}\to \chi_{e_4}\chi_{e_3}$ reactions as well.

In this paper we are concerned with the calculation of the absorptive part of
$\chi_{e_1} \chi_{e_2} \to X_A X_B \to \chi_{e_4} \chi_{e_3}$ 1-loop 
reactions, encoding the hard tree-level $\chi_{e_1} \chi_{e_2}$ and 
$\chi_{e_4} \chi_{e_3}$
annihilation processes. The full annihilation rates, including the long-range
effects shall be studied elsewhere \cite{paperIII}.
To determine the absorptive part of the Wilson coefficients, we evaluate
the absorptive part of the hard 
$\chi_{e_1} \chi_{e_2} \to \chi_{e_4} \chi_{e_3}$
1-loop scattering amplitude within the MSSM and match the result
with the tree-level matrix element of four-fermion operators contained in
$\delta \mathcal L_\text{ann}$ in the effective theory.
At tree-level the annihilation rates can be 
given separately for every final state $X_A X_B$, since the
tree-level processes are free from infrared divergences.  In 
higher-orders the formalism applies to the inclusive annihilation 
cross section \cite{Bodwin:1994jh}, or to suitably defined 
infrared-safe final states.

The leading-order contributions in $\delta \mathcal L_\text{ann}$
are given by dimension-6 four-fermion operators.
For instance, the specific dimension-6 four-fermion operator that encodes
scattering of a non-relativistic incoming neutralino 
pair  $ \chi^0_1  \chi^0_1$ in a $^1S_0$ partial-wave state 
into an outgoing $ \chi^0_1  \chi^0_1$ state
in the same $^1S_0$ partial-wave configuration is given by
\begin{align}
\delta \mathcal L_\text{ann}^{d = 6} \supset
\frac{1}{4} ~ f^{\chi^0 \chi^0 \rightarrow 
\chi^0 \chi^0}_{\lbrace 11 \rbrace \lbrace 11 \rbrace}(^1S_0) \ \
\xi^\dagger_1 \, \xi^c_1 \ \ \xi^{c \dagger}_1 \, \xi^{}_1 \,,
\end{align}
where the spinor $\xi^c$ is the charge conjugate of $\xi$,
$\xi^c=-i\sigma^2\,\xi^*$, and $\sigma^2$ specifies the second Pauli matrix.
Note that $\xi^{c \dagger}_1 \, \xi^{}_1$ represents the Lorentz invariant
bilinear built from the non-relativistic particle field $\xi^{}_1$, 
which destroys the incoming state of two identical $\chi_1^0$ particles.
The factor $1/4$ denotes a normalization factor which compensates the symmetry
factors arising from the number of identical contractions in the tree-level
$\chi^0_1 \chi^0_1 \to \chi^0_1 \chi^0_1$ matrix element. The symbol 
$f^{\chi^0 \chi^0 \rightarrow \chi^0 \chi^0}_{\lbrace 11 \rbrace 
\lbrace 11 \rbrace} (^1S_0)$
denotes the Wilson coefficient corresponding to the dimension-6 operator.
We can generalize the above expression to include all possible
spin-0 and spin-1 \emph{S}-wave four-fermion operators at leading order in
the non-relativistic expansion.
Written in a compact form, the contribution of dimension-6 operators
in $\delta \mathcal L_\text{ann}$ reads

\begin{align}
\delta \mathcal L_\text{ann}^{d = 6}  = 
\sum\limits_{ \chi \chi \rightarrow \chi \chi}
\sum\limits_{ s = 0,1 }~\frac{1}{4}~
f^{ \chi \chi \to \chi \chi }_{ \lbrace e_1 e_2\rbrace \lbrace e_4 e_3\rbrace }
       \left( {}^{2s+1}S_J \right) \
\mathcal O^{\chi \chi \to \chi \chi }_{ \lbrace e_4 e_3\rbrace 
\lbrace e_2 e_1\rbrace }
\left( {}^{2s+1}S_J \right) \,,
\label{eq:deltaL4fermion}
\end{align}
where $J=s$ for the case of $S$-wave operators considered here.
The first sum, taken over all non-relativistic $2 \to 2$ neutralino
and chargino scattering processes $\chi \chi \to \chi \chi$,
implies the consideration of neutral scattering reactions
as well as single-charged and double-charged processes.
The $\chi \chi \to \chi \chi$ reactions that we take into account
are summarized in Tab.~\ref{tab:scattering_reactions}.
\begin{table}[t]
\centering
\begin{tabular}{|c|c|c|}
\hline
   neutral reactions
 & single-charged reactions
 & double-charged reactions
\\
\hline\hline
   $\chi^0 \chi^0 \to \chi^0 \chi^0 $
 & $\chi^0 \chi^+ \to \chi^0 \chi^+$
 & $\chi^+ \chi^+ \to \chi^+ \chi^+$
\\
   $\chi^0 \chi^0 \to \chi^- \chi^+$
 & $\chi^- \chi^0 \to \chi^- \chi^0$
 & $\chi^- \chi^- \to \chi^- \chi^-$
\\
   $\chi^- \chi^+ \to \chi^0 \chi^0$
 &
 &
\\
   $\chi^- \chi^+ \to \chi^- \chi^+$
 &
 &
\\
\hline
\end{tabular}
\caption{Collection of all $\chi_{e_1} \chi_{e_2} \to \chi_{e_4} \chi_{e_3}$
         scattering reactions.
         The labels $e_i$ on the fields $\chi_{e_i}$ are
         suppressed in the above table. If $\chi_{e_i}$ represents a field
         $\chi^0_{e_i}$, the label $e_i$ can range over $e_i = 1, \ldots, n_0$,
         whereas $e_i = 1, \ldots, n_+$ for the case of a $\chi^\pm_{e_i}$ field.}
\label{tab:scattering_reactions}
\end{table}
The spin of the incoming and outgoing two-particle states can be either
$s=0$ or $s=1$, such that the terms in the above
Lagrangian $\delta \mathcal L^{d=6}_\text{ann}$ describe
${}^1S_0$ and ${}^3S_1$ partial-wave scattering reactions.
The $f_{ \lbrace e_1 e_2\rbrace \lbrace e_4 e_3\rbrace }^{ \chi \chi \to \chi \chi } \left( {}^{2s+1}S_J \right)$
denote the Wilson coefficients, that correspond to the four-fermion
operators
$\mathcal O^{\chi\chi \to \chi\chi }_{ \lbrace e_4 e_3\rbrace \lbrace e_2 e_1\rbrace } \left( {}^{2s+1}S_J \right)$.
The indices $e_1$ and $e_2$ ($e_3$ and $e_4$) refer to the neutralino or
chargino species of the incoming (outgoing) particles, and take the values 1 to
$n_0$ for neutralino species and 1 to $n_+$ for chargino species.
Note that the order of the labels $e_i$ on the Wilson coefficients and the
operators is not accidental in (\ref{eq:deltaL4fermion}). The labels on the
operators are given in the order, in which the field operators with 
label $e_i$ occur in the operator. In case of the corresponding 
Wilson coefficients, the indices refer to the actual scattering reaction
$\chi_{e_1} \chi_{e_2} \to \chi_{e_4} \chi_{e_3}$, that is described by 
the operators. For the basis of the operators see Tab.~\ref{tab:fields}.
\begin{table}[t]
\centering
%
\begin{tabular}{|c|c|c|}
\hline
   $\chi_{e_1} \chi_{e_2} \to \chi_{e_4} \chi_{e_3}$
 & $\mathcal O^{ \chi \chi \to \chi \chi}_{ \lbrace e_4 e_3\rbrace \lbrace e_2 e_1\rbrace } \left( {}^1 S_0\right)$
 & $\mathcal O^{ \chi \chi \to \chi \chi}_{ \lbrace e_4 e_3\rbrace \lbrace e_2 e_1\rbrace } \left( {}^3 S_1\right)$
\\[0.8ex]
\hline\hline
   $ \chi^0 \ \chi^0 \to \chi^0 \ \chi^0 $
 & $ \xi^\dagger_{e_4} \, \xi^c_{e_3} \ \ \xi^{c \dagger}_{e_2} \, \xi^{}_{e_1}$
 & $ \xi^\dagger_{e_4} \vec\sigma \xi^c_{e_3} \ \ \xi^{c \dagger}_{e_2} \vec\sigma \, \xi^{}_{e_1}$
\\
   $ \chi^0 \ \chi^0 \to \chi^- \chi^+ $
 & $ \eta^\dagger_{e_4} \, \zeta^{c}_{e_3} \ \ \xi^{c \dagger}_{e_2} \, \xi^{}_{e_1}$
 & $ \eta^\dagger_{e_4} \vec\sigma \zeta^{c}_{e_3} \ \ \xi^{c \dagger}_{e_2} \vec\sigma \, \xi^{}_{e_1}$
\\
$ \chi^- \chi^+ \rightarrow \chi^0 \ \chi^0 $
 &  $ \xi^\dagger_{e_4} \, \xi^c_{e_3} \ \ \zeta^{c \dagger}_{e_2} \, \eta^{}_{e_1}$
 & $ \xi^\dagger_{e_4} \vec\sigma \xi^c_{e_3} \ \ \zeta^{c \dagger}_{e_2} \vec\sigma \, \eta^{}_{e_1}$
\\ 
   $ \chi^- \chi^+ \to \chi^- \chi^+ $
 & $ \eta^\dagger_{e_4} \, \zeta^{c}_{e_3} \ \ \zeta^{c \dagger}_{e_2} \, \eta^{}_{e_1}$
 & $ \eta^\dagger_{e_4} \vec\sigma \zeta^{c}_{e_3} \ \ \zeta^{c \dagger}_{e_2} \vec\sigma \, \eta^{}_{e_1}$
\\
 &
 &
\\ 
   $ \chi^0 \chi^+ \to \chi^0 \chi^+ $
 & $ \xi^\dagger_{e_4} \, \zeta^{c}_{e_3} \ \ \zeta^{c \dagger}_{e_2} \, \xi^{}_{e_1}$
 & $ \xi^\dagger_{e_4} \vec\sigma \zeta^{c}_{e_3} \ \ \zeta^{c \dagger}_{e_2} \vec\sigma \, \xi^{}_{e_1}$
\\
   $ \chi^- \chi^0 \to \chi^- \chi^0 $
 & $ \eta^\dagger_{e_4} \, \xi^c_{e_3} \ \ \xi^{c \dagger}_{e_2} \, \eta^{}_{e_1}$
 & $ \eta^\dagger_{e_4} \vec\sigma \xi^c_{e_3} \ \ \xi^{c \dagger}_{e_2} \vec\sigma \, \eta^{}_{e_1}$
\\
 &
 &
\\ 
   $ \chi^+ \chi^+ \to \chi^+ \chi^+ $
 & $ \zeta^\dagger_{e_4} \, \zeta^c_{e_3} \ \ \zeta^{c \dagger}_{e_2} \, \zeta^{}_{e_1}$
 & $ \zeta^\dagger_{e_4} \vec\sigma \zeta^c_{e_3} \ \ \zeta^{c \dagger}_{e_2} \vec\sigma \, \zeta^{}_{e_1}$
\\
   $ \chi^- \chi^- \to \chi^- \chi^- $
 & $ \eta^\dagger_{e_4} \, \eta^c_{e_3} \ \ \eta^{c \dagger}_{e_2} \, \eta^{}_{e_1}$
 & $ \eta^\dagger_{e_4} \vec\sigma \eta^c_{e_3} \ \ \eta^{c \dagger}_{e_2} \vec\sigma \, \eta^{}_{e_1}$
\\[0.8ex]
\hline
\end{tabular}
\caption{Four-fermion operators for leading-order $S$-wave
         $\chi_{e_1}\chi_{e_2}\to\chi_{e_4}\chi_{e_3}$ transitions.
         The indices $e_i, i = 1, \ldots, 4$ on the $\chi$-fields
         are suppressed in the first column.
         In addition to the specified operators there are redundant ones,
         which are obtained by interchanging the field-operator symbols 
         $\xi, \eta$ or $\zeta$ (but not the labels)
         at the first and second and/or the third and
         fourth position in the operator $\mathcal O^{\chi\chi\to\chi\chi}$.
         For example, for ${}^1S_0$ 
         $\chi^0 \chi^+ \to \chi^0 \chi^+$ operators one of the three classes of
         field-interchanged operators is given by the ${}^1S_0$
         $\chi^+ \chi^0 \to \chi^+ \chi^0$ operators
         $ \zeta^\dagger_{e_4}\,\xi^{c}_{e_3}\, \xi^{c \dagger}_{e_2} \, \zeta^{}_{e_1}$.}
\label{tab:fields}
\end{table}
%
The $\chi$ in the labels $\chi\chi\to\chi\chi$ of the operators and Wilson
coefficients in (\ref{eq:deltaL4fermion}) should indicate the particular
particle species $\chi^0$ and $\chi^\pm$, whose $\chi_{e_1} \chi_{e_2} \to \chi_{e_4} \chi_{e_3}$ scattering reaction is described,
see Tab.~\ref{tab:scattering_reactions}.
A summation over the indices $e_i$ is implicit in (\ref{eq:deltaL4fermion}).
The normalization factor $1/4$ in (\ref{eq:deltaL4fermion}) ensures that the
tree-level transition matrix element for $^1S_0$-wave scattering is given by
\begin{align}
  \langle \chi_{l} \chi_{k}
  & \vert
        \int d^4 x
      \sum\limits_{\chi\chi \to \chi\chi}
      \frac{1}{4} ~
      f^{\chi\chi \to \chi\chi}_{ \lbrace e_1 e_2\rbrace \lbrace e_4 e_3\rbrace }
        (^1S_0) \ \
      \mathcal O^{\chi\chi \to \chi\chi}_{ \lbrace e_4 e_3\rbrace \lbrace e_2 e_1\rbrace }(^1S_0) (x) \ 
    \vert
 \chi_{i} \chi_{j} \rangle
     \vert_\text{tree}
\nonumber\\
  & = 
      (2 \pi)^4 \delta^{(4)}(p_\text{in} - p_\text{out}) \ 
      2 \ f^{\chi \chi \to \chi \chi}_{ \lbrace i j\rbrace \lbrace l k\rbrace }
        (^1S_0)
\label{eq:normalization}
\end{align}
for all $\chi_i\chi_j \to \chi_l\chi_k$ reactions at leading order 
in the non-relativistic effective theory.
In (\ref{eq:normalization}) we have assumed that the incoming and outgoing
two-particle states $\chi_i\chi_j$ and $\chi_l\chi_k$ both reside in
a $^1S_0$-wave configuration with normalised spin state
$\frac{1}{\sqrt{2}}\left(\vert\uparrow\downarrow\,\rangle - \vert\downarrow\uparrow\,\rangle \right)$.
A similar relation for the tree-level
transition matrix element of $^3S_1$-wave scattering in the effective theory
holds for all $\chi_i\chi_j \to \chi_l\chi_k$ reactions. Note that in order
to derive (\ref{eq:normalization}) one has to take into 
account relations among Wilson coefficients of different operators, 
which will be deduced in the next paragraph.

There are redundancies in $\delta \mathcal L^{d=6}_{\text{ann}}$,
(\ref{eq:deltaL4fermion}), as several operators can describe one specific
scattering reaction with a $\chi_{e_1}$ and a
$\chi_{e_2}$ ($\chi_{e_4}$ and $\chi_{e_3}$) particle in the initial (final) state.
This redundancy is associated with operators that arise from interchanging
the single-particle field operators at the first and second and/or third and
fourth position in a given $\mathcal O^{\chi\chi \to \chi\chi}$.
The corresponding Wilson coefficients are related to each other, as
they encode the same information on a given specific scattering reaction.
Consequently, the redundancy manifests itself in symmetry relations among the
Wilson coefficients under exchange of the labels $e_{1} \leftrightarrow e_{2}$
and/or $e_{4} \leftrightarrow e_{3}$. These relations read
\begin{align}
\nonumber
   f_{ \lbrace e_2 e_1\rbrace \lbrace e_4 e_3\rbrace }^{ \chi_{e_2} \chi_{e_1} \to \chi_{e_4} \chi_{e_3} }
                                              \left( {}^{2s+1}S_J \right)
	=
 \eta_s \  f_{ \lbrace e_1  e_2\rbrace \lbrace e_4 e_3\rbrace }^{ \chi_{e_1} \chi_{e_2} \to \chi_{e_4} \chi_{e_3} }
                                               \left( {}^{2s+1}S_J \right)
\ ,
\\
   f_{ \lbrace e_1 e_2\rbrace \lbrace e_3 e_4\rbrace }^{ \chi_{e_1} \chi_{e_2} \to \chi_{e_3} \chi_{e_4} }
                                              \left( {}^{2s+1}S_J \right)
	=
 \eta_s \  f_{ \lbrace e_1  e_2\rbrace \lbrace e_4 e_3\rbrace }^{ \chi_{e_1} \chi_{e_2} \to \chi_{e_4} \chi_{e_3} }
                                               \left( {}^{2s+1}S_J \right)
\ ,
\label{eq:WilsonCoeffSymmetry}
\end{align}
with
\begin{equation}
 \eta_s \ = \
	\begin{cases}
		 \phantom{-} 1 & \text{for } s = 0 \\
 			   - 1 & \text{for } s = 1
	\end{cases}
\ \ .
\label{eq:etasdefinition}
\end{equation}
To exemplify the origin of the first relation in (\ref{eq:WilsonCoeffSymmetry}),
let us consider the terms in $\delta \mathcal L^\text{d=6}_\text{ann}$ that account
for \emph{S}-wave $\chi^0 \chi^0 \to \chi \chi$ reactions at leading order in
the non-relativistic velocity expansion:
\begin{align}
\label{eq:symmetries4fermionoperators}
  \sum\limits_{e_1,\dots, e_4} \
        \frac{1}{4} ~
        f_{ \lbrace e_1 e_2\rbrace \lbrace e_4 e_3\rbrace }^{ \chi^0 \chi^0 \to \chi \chi } \left( {}^{2s+1}S_J \right) ~ ~
        \mathcal O^{(s)\,\chi\chi}_{ \lbrace e_4 e_3\rbrace }
        \ \xi^{c \dagger}_{e_2} ~ \Gamma^{(s)} \, \xi_{e_1}\,,
\end{align}
where the operator $\mathcal O^{(s)\,\chi\chi}_{ \lbrace e_4 e_3\rbrace }$ stands for the 
two-field operator that creates the outgoing state, and $\Gamma^{(s)}$ is given
by $\Gamma^{(s=0)}= 1_{2\times2}$ and $\Gamma^{(s=1)}= \vec\sigma$ in case of
$^1S_0$ and $^3S_1$ operators, respectively.
Using the definition of the spinor $\xi^c$ we can write
\begin{align}
 \xi^{c \dagger}_{e_2} ~ \Gamma^{(s)} \, \xi_{e_1} 
	=
 \xi^{c \dagger}_{e_1} ~ \sigma^2 \Gamma^{(s)\top } \sigma^2 \, \xi_{e_2} 
	= 
 \eta_s ~ \xi^{c \dagger}_{e_1} ~ \Gamma^{(s)} \, \xi_{e_2} 
\ ,
\end{align}
with $\eta_s$ defined as in (\ref{eq:etasdefinition}).
After renaming the labels $e_1$ and $e_2$, the terms in
(\ref{eq:symmetries4fermionoperators}) can be written as
\begin{align}
\nonumber
	\sum\limits_{e_1,\dots, e_4} \
        \frac{1}{4} ~
	\eta_s \, f_{ \lbrace e_2 e_1\rbrace \lbrace e_4 e_3\rbrace }^{ \chi^0 \chi^0 \to \chi \chi } \left( {}^{2s+1}S_J \right) ~ ~
        \mathcal O^{(s)\,\chi\chi}_{ \lbrace e_4 e_3\rbrace }
        ~ \xi^{c \dagger}_{e_2} ~ \Gamma^{(s)} \, \xi_{e_1}
\ .
\end{align}
Comparing to the original expression in (\ref{eq:symmetries4fermionoperators}),
we arrive at the relation
\begin{align}
   f_{ \lbrace e_2 e_1\rbrace \lbrace e_4 e_3\rbrace }^{ \chi^0 \chi^0 \to \chi \chi } \left( {}^{2s+1}S_J \right)
	=
 \eta_s \  f_{ \lbrace e_1  e_2\rbrace \lbrace e_4 e_3\rbrace }^{ \chi^0 \chi^0 \to \chi \chi } \left( {}^{2s+1}S_J \right)
\ .
\end{align}
This equation as well as the more comprehensive relations in
(\ref{eq:WilsonCoeffSymmetry}) imply that the Wilson coefficients of $^1S_0$
operators have to be symmetric under the exchange
$e_1\leftrightarrow e_2$, whereas Wilson coefficients of $^3S_1$ operators are
antisymmetric under $e_1\leftrightarrow e_2$.
The same statement applies to the exchange $e_3\leftrightarrow e_4$ in case of
outgoing states.
In processes with identical incoming or outgoing particles, the above
relations in (\ref{eq:WilsonCoeffSymmetry}) imply the vanishing of the
$^3S_1$ Wilson coefficients. This rephrases the well-known fact that a pair of 
identical spin-1/2 particles cannot build a ${}^3S_1$ state.

Finally, a further property of the Wilson coefficients under the exchange of
the particle labels is directly inherited from the hermiticity of the
non-relativistic Lagrangian:
\begin{align}
   f_{ \lbrace e_1 e_2\rbrace \lbrace e_4 e_3\rbrace }^{ \chi \chi  \to \chi \chi } \left( {}^{2s+1}S_J \right)
	=
  \  \left[ f_{ \lbrace e_4  e_3\rbrace \lbrace e_1 e_2\rbrace }^{ \chi \chi \to \chi \chi } \left( {}^{2s+1}S_J \right) \right]^*
\ .
\label{eq:hermiticity}
\end{align}

\subsection{Matching condition} 
\label{sec:match}
The Wilson coefficients of the four-fermion operators in
$\delta \mathcal L_\text{ann}$ are determined by the matching condition
\begin{align}
\nonumber
 \mathcal A(\chi_i\chi_j \to \chi_l \chi_k) \ 
  &\vert_\text{MSSM, perturbative}
\ = \
 \ \sum \ 
      \frac{1}{4}~ 
      f^{\chi\chi \to \chi \chi}_{\lbrace e_1 e_2 \rbrace \lbrace e_4 e_3 \rbrace}(^{2s+1}L_J) \
\\
&
\times
   \langle \chi_l \chi_k | \
      \mathcal O^{ \chi\chi \to \chi\chi}_{\lbrace e_4 e_3 \rbrace \lbrace e_2 e_1 \rbrace}
        (^{2s+1}L_J) \ 
   | \chi_i \chi_j \rangle \
  \vert_\text{NRMSSM, perturbative}
\ .
\label{eq:matchingcond}
\end{align}
For this equation to hold, we have to use the same (non-relativistic)
normalization of the incoming and outgoing states in both the full theory
and the NRMSSM. 
Here we will determine the contributions to the Wilson coefficients that
describe the tree-level annihilation reactions into exclusive 
SM and light Higgs two-body final states $X_A X_B$, which we shall denote as
$\hat{f}^{\chi\chi \to X_A X_B \to \chi \chi}(^{2s+1}L_J)$.
The unitarity of the S-matrix at the diagrammatic level establishes a relation
among the tree-level annihilation rate for 
$\chi_i \chi_j \to X_A X_B$ and the imaginary part
of the 1-loop forward-scattering reaction 
$\chi_i \chi_j \to X_A X_B \to \chi_i \chi_j$:
\begin{eqnarray}
\label{eq:defabsorptivepartforward}
&&  \int[\text{dPS}_{AB}] \,
      | \mathcal A (\chi_i\chi_j \to X_A X_B)|^2
= 
\,2 \ \Im \left[ \mathcal A (\chi_i\chi_j \to X_A X_B\to \chi_i\chi_j) \right]
\\
\nonumber
&& \hspace*{1cm} 
= 2 \ \sum \
           \frac{1}{4}~
       \Im \left[
           f^{\chi\chi \to X_A X_B \to \chi\chi}_{\lbrace e_1 e_2 \rbrace \lbrace e_4 e_3 \rbrace}(^{2s+1}L_J) 
           \right]   
      \langle \chi_i\chi_j|
          \mathcal O^{\chi\chi \to \chi\chi}_{\lbrace e_4 e_3 \rbrace \lbrace e_2 e_1 \rbrace}
             (^{2s+1}L_J)
      | \chi_i\chi_j \rangle \,.\quad
\end{eqnarray}
We generalize this and define the absorptive part of amplitude
$\mathcal A (\chi_i\chi_j \to X_A X_B \to \chi_l\chi_k)$ as well as the
absorptive part of the Wilson coefficients in the following way:
\begin{eqnarray}
&&  \int[\text{dPS}_{AB}] \
      \mathcal A (\chi_i\chi_j \to X_A X_B)
    \times
      \mathcal A (\chi_l\chi_k \to X_A X_B)^*
\nonumber
\\
&& \hspace*{1cm} = 
  2 \ \left[ \mathcal A (\chi_i\chi_j \to X_A X_B \to \chi_l\chi_k) \right]
         |_\text{absorptive}
\nonumber
\\[0.2cm]
&& \hspace*{1cm}= 
  2 \ \sum \
       \frac{1}{4}~
       \hat{f}^{\,\chi\chi \to X_A X_B  \to \chi\chi}_{\lbrace e_1 e_2 \rbrace \lbrace e_4 e_3 \rbrace}(^{2s+1}L_J) \
    \langle \chi_l\chi_k|
      \mathcal O^{ \chi\chi \to \chi\chi}_{\lbrace e_4 e_3 \rbrace \lbrace e_2 e_1 \rbrace}
         (^{2s+1}L_J)
    | \chi_i\chi_j \rangle \, ,\qquad
\label{eq:defabsorptivepart}
\end{eqnarray}
where we have introduced the notation
\begin{align}
\hat{f}^{\chi\chi \to X_A X_B \to \chi\chi}_{\lbrace ij \rbrace \lbrace lk \rbrace}(^{2s+1}L_J) \
= \
f^{\,\chi\chi \to X_A X_B \to \chi\chi}_{\lbrace ij \rbrace \lbrace lk \rbrace}(^{2s+1}L_J) \ |_\text{absorptive} \ .
\end{align}
With this definition, the absorptive part of a Wilson coefficient that encodes
a $\chi_i\chi_j \to \chi_i\chi_j$ forward-scattering reaction coincides with its
imaginary part.

We make use of the defining relations to determine the absorptive
part of the Wilson coefficients $\hat{f}^{\chi\chi \to X_A X_B \to \chi \chi}$
from the product of the full-theory tree-level
annihilation amplitudes integrated over the final state particles' phase-space,
as given in the first line of (\ref{eq:defabsorptivepart}).
Technically this is achieved by considering all 1-loop scattering amplitudes
$\chi_i \chi_j \to X_A X_B \to \chi_l \chi_k$ with a specific SM
or Higgs particle-pair $X_A X_B$ in the intermediate state and by applying the
Cutkosky rules to the $X_A$ and $X_B$ propagators.
The resulting expression coincides with the first line of
(\ref{eq:defabsorptivepart}).
To determine the absorptive part of the Wilson coefficients,
the expression has to be expanded in the non-relativistic momenta
of the external particles as well as in their mass differences
and an appropriate spin-projection has to be performed.

\subsection{Expansion in mass differences in $\delta \mathcal L_\text{ann}$}
\label{subsec:massdiffexp}
To simplify the notation, we shall replace the indices $(e_1,e_2,e_3,e_4)$ by
(1,2,3,4) throughout this section. Further, we shall adopt the convention that
particles $1$ and $4$ in the reaction $\chi_1 \chi_2 \to \chi_4 \chi_3$ share the
same reference mass scale $m$, while particles $2$ and $3$ have masses closer to
the reference scale $\overline{m}$.
Introducing two distinct mass scales for the particle species allows us
to consider pair annihilations of two particles with similar mass
($m \sim \overline m$), but also pair annihilation of a hydrogen-like
two-particle system where one of the particles is much lighter
(though still heavy enough to be considered as non-relativistic). 
According to these assignments, we define
\begin{align}
\nonumber
 m_1 \ =& \ m -\delta m \, , \hspace{7ex}
 m_2 \ = \ \overline m -\delta \overline m \, ,
\\
 m_4 \ =& \ m +\delta m \, , \hspace{7ex}
 m_3 \ = \ \overline m +\delta \overline m \, ,
\label{eq:mei}
\end{align}
with
\begin{align}
 m \ =& \ \frac{m_1 + m_4}{2} \, ,  \hspace{7ex}
 \overline m \ = \ \frac{m_2 + m_3}{2} \ ,
\label{eq:mmbar}
\end{align}
such that the mass differences read
\begin{align}
 \delta m \ =& \ \frac{m_4 - m_1}{2} \, , \hspace{7ex}
 \delta \overline m \ = \ \frac{m_3 - m_2}{2} \, .
\label{eq:deltam}
\end{align}
The results for the Wilson coefficients presented in the appendix adopt
the definitions (\ref{eq:mei}--\ref{eq:deltam}). If for a given process
$\chi_i \chi_j \to \chi_l \chi_k$ it turns out that the reverse condition,
$m_i \sim m_k \sim m$ and $m_j \sim m_l \sim \overline{m}$, 
is more meaningful given the actual values of the masses, one can make use of
the symmetry properties (\ref{eq:WilsonCoeffSymmetry}) to relate the Wilson
coefficients for $\chi_i \chi_j \to \chi_l \chi_k$ to those of
$\chi_i \chi_j \to \chi_k \chi_l$, which would then conform to the prescription
above, {\it i.e.} $m$ would be equal to the average 
of the mass of the particle associated with field 1 and the mass of 
the particle associated with field 4,  $m= ( m_i +m_k )/2$.
Note that the mass differences $\delta m$ and $\delta \overline m$ in
(\ref{eq:deltam}) obviously vanish in case of diagonal scattering reactions
$\chi_1 \chi_2 \to \chi_1 \chi_2$, such that $m = m_1$ and
$\overline m = m_2$ in that case.

The absorptive parts of the Wilson coefficients are obtained by matching
amplitudes for the  process $\chi_1\chi_2 \to \chi_4\chi_3$ with
on-shell external states. This implies that the energy-conservation relation
in the center-of-mass system,
\begin{align}
\sqrt{s} = E_1(\vec{p}^{\,2}) +  E_2(\vec{p}^{\,2})
         = E_4(\vec{p}^{\;\prime 2}) +  E_3(\vec{p}^{\;\prime 2})
\ ,
\label{eq:Econserv}
\end{align}
with $E_i(\vec{p}^{\,2})= \sqrt{m_i^2 + \vec{p}^{\,2}}$ and $\vec{p}$
($\vec{p}^{\;\prime}$) the incoming (outgoing) particles' momentum in the
center-of-mass system, is fulfilled.
Using (\ref{eq:mei}--\ref{eq:deltam}) and $M\equiv m+\overline{m}$, the
expansion of the energy-conservation relation~(\ref{eq:Econserv}) for
non-relativistic momenta $\vec{p}^{\,2}$ and $\vec{p}^{\,\prime 2}$ reads
\begin{align}
\sqrt{s} = M  - \delta m - \delta \overline m + \frac{\vec{p}^{\;2}}{2\mu}
              + \ldots
         = M  + \delta m + \delta \overline m + \frac{\vec{p}^{\;\prime 2}}{2\mu}
              + \ldots \ ,
\label{eq:EconservNR}
\end{align}
where $\mu= m\,\overline{m}/M$ and terms of order $\vec{p}^{\,4}/\mu^3$ 
and $(\delta m/M \times \vec{p}^{\,2}/\mu )$ have been dropped.
This can be rewritten as
\begin{align}
 \frac{\vec{p}^{\,\prime 2}}{2\mu} \ = \
 \frac{\vec{p}^{\,2}}{2\mu}- 2 \delta m - 2 \delta \overline m+ \ldots \ .
\label{eq:ppofp}
\end{align}
From (\ref{eq:ppofp}) we see that a consistent expansion which treats both
$\vec{p}^{\,2}$ and $\vec{p}^{\,\prime 2}$ as small quantities of the same order
requires that the mass differences $\delta m, \delta \overline m$ are also
formally considered of order $\vec{p}^{\,2}/\mu$ in the expansion 
of the amplitudes.
Note that an expansion in mass differences is only required for the 
off-diagonal scattering reactions where the  incoming and outgoing
$\chi\chi$ states are different, as $\delta m = \delta \overline m = 0$ for
$\chi_1 \chi_2 \to \chi_1 \chi_2$ reactions.

The amplitude for a generic process
$\chi_{1}\chi_{2} \to X_A X_B \to \chi_{4}\chi_{3}$
then depends on the hard scales $(m,\,\overline m)$ and on the small scales
$(\vec{p}^{\;2}/\mu,\,\vec{p}^{\;\prime\,2}/\mu,\,\vec{p}\cdot\vec{p}^{\,\prime}/\mu,\,\delta m,\,\delta \overline m) \sim {\cal O} (\mu v^2)$, 
where $v$ stands for the relative velocity in the two-particle system.
In the following we enumerate the steps to obtain the absorptive part 
of the Wilson coefficients from the process 
$\chi_{1}\chi_{2} \to X_A X_B \to \chi_{4}\chi_{3}$, 
including the subleading ${\cal O}(v^2)$ terms, which will be presented in 
\cite{Hellmann:2013jxa}.
\begin{enumerate}
 \item 
The absorptive part of the 1-loop scattering amplitude $\chi_1 \chi_2 \to X_A X_B \to \chi_4 \chi_3$ with a SM
or Higgs final state $X_A X_B$  is computed by applying the Cutkosky rules
to the $X_A$ and $X_B$ propagators. The result is written in terms of
the mass scales introduced above, and expanded in the small scales retaining terms up to ${\cal O}(v^2)$.
 \item To ${\cal O}(v^2)$ the result contains scalar products 
with at most two powers of $\vec{p}$ and  $\vec{p}^{\; \prime}$. For the 
spin-1 configuration, the scalar products also involve the
spin-polarization vectors $\vec{n}$ and  $\vec{n}^{\,\prime}$ of the incoming ($\chi_{1}\chi_{2}$) 
and outgoing ($\chi_{4}\chi_{3}$) states, respectively. The generic form of the result for spin-1 incoming and outgoing states reads
\begin{align}
\nonumber
& \left\{ \, c_0({}^3S_1) + c_1({}^3S_1) \, \delta m + c_2({}^3S_1) \, 
\delta \overline m + c_3({}^3S_1) \,\vec{p}^{\;2} 
+ c_4({}^3S_1) \, \vec{p}^{\;\prime\,2}  \, \right\}
 \, \vec{n} \cdot \vec{n}^{\,\prime} 
\nonumber
\\& 
+ c_5({}^3P_0) \, (\vec{p} \cdot  \vec{n})\; (\vec{p}^{\,\prime} \cdot \vec{n}^{\,\prime} )
+ c_6({}^3P_1) \, [p,n]^{k} \,[p^{\prime},n^{\prime}]^k
+ c_7({}^3P_2) \, p^{\{i} \, n^{j \}} \, p^{\,\prime \{ i} \, n^{\prime j \} }
\nonumber
\\& 
+ c_8({}^3S_1,{}^3P_1) \, n^{k} \, [ p^{\prime},n^{\prime} ]^k
 + c_9({}^3P_1,{}^3S_1) \, [p,n]^k  n^{\prime \, k}
\nonumber
\\& 
+ c_{10}({}^3S_1,{}^3D_1) \, p^{\,\prime \{ i} \, p^{\prime j\} } \, 
n^i\,n^{\prime j} + c_{11}({}^3D_1,{}^3S_1) \, p^{ \{ i} \, p^{ j\} } \, 
n^i\,n^{\prime j} \, ,
\label{eq:absA}
\end{align}
where we have introduced the notation 
$[a,b]^k \equiv  \varepsilon^{ijk} a^i b^j$
and $a^{\{i} \, b^{j\}} \equiv a^i b^j + a^j b^i - 2 \,
\vec{a}\cdot\vec{b} \, \delta^{ij}/3$, corresponding
to $J=1$ and $J=2$ Cartesian tensors, respectively. The spin-polarization 
vector $\vec{n}$ is introduced by replacing the spinor matrix 
$[\xi{\xi^c}^\dagger]_{ij}$ of an incoming  
two-neutralino state by $\frac{1}{\sqrt{2}}\, 
\vec{n}\cdot \vec{\sigma}_{ij}$. Similar replacements apply to 
outgoing two-particle states and states involving charginos. 
The coefficients $c_i$ are functions of $m$ and $\overline{m}$. The first term, $c_0$, gives the leading-order contribution,
where all the others count as ${\cal O}(v^2)$. We have further specified the quantum numbers ${}^3 L_J$
of each term, which matches the angular-momentum configuration of the incoming state, equal to that
of the outgoing state except for the $c_{8-11}$ terms (the first quantum number between parentheses refers then
to the incoming state, the second to the outgoing one). 
For spin-0 incoming and outgoing states, the result simplifies to
\begin{align}
&  c_0({}^1S_0) + c_1({}^1S_0) \, \delta m + c_2({}^1S_0) \, \delta \overline m + c_3({}^1S_0) \,\vec{p}^{\;2} 
+ c_4({}^1S_0) \, \vec{p}^{\;\prime\,2}  
+ c_5({}^1P_1) \, \vec{p} \cdot \vec{p}^{\,\prime} \, .
\quad
\label{eq:absAspin0}
\end{align}
We have not considered the possibility of spin-0 to spin-1 transitions 
between incoming and outgoing states in the hard annihilation
process, though the transitions ${}^3 S_1 \to {}^1 P_1$ and 
${}^3 P_{0,1} \to {}^1 S_0,\,{}^1 P_1$ are also allowed at ${\cal O}(v)$
by angular-momentum conservation. Such spin-changing transitions in the 
hard annihilation part of the full forward scattering amplitude 
(see Fig.~\ref{fig:genericdiagram}) will also require spin-changing 
potential interactions in the long-range part, in order to bring
the spin of the two-particle state after annihilation back to the spin 
of the (left-most) incoming state. Since the non-relativistic 
spin-changing potentials carry an additional $v$-suppression, such 
transitions are only relevant for the calculation of the 
annihilation rates at ${\cal O}(g^2 v^2)$. At present we ignore 
${\cal O}(v^2)$ effects that arise from subleading non-Coulomb (non-Yukawa) 
potentials and consider only those from the short-distance annihilation.
Likewise, the terms $c_{8-11}$ included in (\ref{eq:absA}) imply a change of the orbital angular momentum which 
must be compensated by a potential interaction which is also $v$-suppressed in the non-relativistic limit, and can be ignored
for our purposes. 
\item By virtue of the energy-conservation relation (\ref{eq:EconservNR}), 
we rewrite powers of $\vec{p}^{\;2}$ and $\vec{p}^{\;\prime\,2}$ as
\begin{align}
\nonumber
\vec{p}^{\;2} \ =& \ \frac{1}{2} \, ( \, \vec{p}^{\;2} + \vec{p}^{\;\prime\,2} \,) + \frac{2m \overline m }{M}
 ( \,  \delta m +  \delta \overline m \,) + \dots \ ,
\\
\vec{p}^{\;\prime\,2} \ =& \ \frac{1}{2} \, ( \, \vec{p}^{\;2} + \vec{p}^{\;\prime\,2} \,) - \frac{2 m \overline m }{M}
 ( \,  \delta m +  \delta \overline m \,) + \dots\ ,
\end{align}
such that the coefficients multiplying $\vec{p}^{\;2}$ and 
$\vec{p}^{\;\prime\,2}$ become equal. This convention is adopted in order 
that the Wilson coefficients of the dimension-8 operators with 
derivatives also have the symmetry property (\ref{eq:hermiticity}) 
under the exchange of the incoming and outgoing states.
\item Finally, the Wilson coefficients are identified by comparing the expanded expression for
the absorptive part of the amplitude ${\cal A}(\chi_{1}\chi_{2} \to X_A X_B \to \chi_{4}\chi_{3})$
with the amplitude for the same process computed with the dimension 6 and dimension 8 EFT operators 
in $\delta\mathcal L_\text{ann}$.
\end{enumerate}
The explicit expressions for the leading-order $S$-wave coefficients are 
provided in the appendix. While the Wilson coefficients refer to the 
inclusive annihilation rates, summed over all accessible final states, 
the calculation is performed for individual final states, which are 
therefore also given separately. Such final-state separated results should 
be of interest to the calculation of primary decay spectra of dark matter 
annihilation in the present Universe.

\section{Results}
\label{sec:results}
We have performed a number of dedicated numeric and analytic checks
of our results for the absorptive parts of the Wilson coefficients.
As these expressions also encode the absorptive part of
$\chi_{e_1} \chi_{e_2} \to X_A X_B \to \chi_{e_1} \chi_{e_2}$ 
forward scattering
reactions, which are related to the tree-level annihilation cross section
for $\chi_{e_1} \chi_{e_2} \to X_A X_B$ processes (see
(\ref{eq:defabsorptivepartforward})),
a comparison of the analytic non-relativistic
approximation to the tree-level annihilation cross section with
results from a numeric code can be carried out for the diagonal 
entries of the annihilation coefficients. We discuss this in
Sec.~\ref{subsec:numericchecks}.

In addition, we can relate our analytic expressions for partial-wave separated
neutralino LSP pair-annihilation cross sections to existing analytic results
available in the literature\cite{Hisano:2004ds,Hisano:2006nn,Drees:1992am}. 
This will be briefly discussed in Sec.~\ref{subsec:analyticchecks}. 
No checks are available in the general case for the 
off-diagonal entries of the annihilation coefficient matrix.

\subsection{Numerical comparison with {\sc MadGraph} }
\label{subsec:numericchecks}
The expansion of the exclusive, spin-averaged center-of-mass frame
$\chi_{e_1} \chi_{e_2} \to X_A X_B$ tree-level pair-annihilation cross section
in the non-relativistic momentum $\vec p$ of the $\chi_{e_i}$ 
particles is given by
\begin{eqnarray}
\label{eq:app_xsection}
\sigma^{\chi_{e_1} \chi_{e_2} \to X_A X_B} ~ v_{\rm rel}
& =& 
 \hat f(^1S_0) + 3~\hat f(^3S_1)
\\
&& 
\nonumber
\hspace*{-3cm} +\, \frac{\ \vec p^{~2}}{M^2} ~ \Bigl(
        \hat f(^1P_1) + \frac{1}{3}~\hat f(^3P_0) + \hat f(^3P_1)
      + \frac{5}{3}~\hat f(^3P_2)
       + \hat g(^1S_0) + 3~\hat g(^3S_1)
                   \Bigr)
   + \mathcal O(\vec p^{~4}) \ . \quad
\end{eqnarray}
Here $v_{\rm rel} = \vert \vec v_{e_1} - \vec v_{e_2}\vert$ is 
the relative velocity of the
$\chi_{e_1}\chi_{e_2}$ pair and $\vec v_{e_i}$ denotes the velocity of
particle $\chi_{e_i}$ in the center-of-mass frame of the 
annihilation reaction. We have suppressed the superscripts
${\chi_{e_1}\chi_{e_2} \to X_A X_B \to \chi_{e_1} \chi_{e_2}}$ on the
Wilson coefficients $\hat f$ in (\ref{eq:app_xsection}), 
where these expressions
explicitly refer to the exclusive (tree-level) annihilation rates.
Further note that (\ref{eq:app_xsection}) contains
not only the leading-order $S$-wave Wilson coefficients 
$\hat f(^{2s+1}S_J)$ with spin configuration $s = 0,1$ of the incoming 
two-body system, but also includes $P$-wave and next-to-next-to-leading 
order $S$-wave coefficients (denoted with $\hat g$).
For analytic results on those coefficients we refer the reader to
\cite{Hellmann:2013jxa}.

In the non-relativistic limit the relation between the relative velocity 
$v_{\rm rel}$ and the particle momentum
$\vec p$ in the center-of-mass frame of the $\chi_{e_1}\chi_{e_2}$ 
annihilation reaction is approximated by
\begin{equation}
 v_{\rm rel} =  \vert \vec v_{e_1} - \vec v_{e_2} \vert
= \vert \vec p \,\vert ~ \left(
               \frac{m_{e_1} + m_{e_2}}{m_{e_1} m_{e_2}} + 
\mathcal O(\vec p^{\,2}) \right) \ .
\label{eq:v_rel}
\end{equation}
Together with (\ref{eq:app_xsection}), this relation allows us to express the
first two coefficients, $a$ and $b$, in the Taylor expansion of the
$\chi_{e_1} \chi_{e_2} \to X_A X_B$ center-of-mass frame annihilation cross
section with respect to the relative velocity,
\begin{align}
\label{eq:app_aplusbv2}
 \sigma^{\chi_{e_1} \chi_{e_2} \to X_A X_B} ~ v_{\rm rel}
    \ = \
  a + b~v_{\rm rel}^2 \ + \ \mathcal O(v_{\rm rel}^4) \ ,
\end{align}
in terms of the partial-wave separated Wilson coefficients
$\hat f^{\chi_{e_1}\chi_{e_2} \to X_A X_B \to \chi_{e_1}\chi_{e_2}}(^{2s+1}L_J)$ and
$\hat g^{\chi_{e_1}\chi_{e_2} \to X_A X_B \to \chi_{e_1}\chi_{e_2}}(^{2s+1}L_J)$.
The coefficient $a$ is given by (leading order) $S$-wave Wilson coefficients
only,
\begin{align}
\label{eq:app_aparam}
 a \ = \ \hat f(^1S_0) + 3~\hat f(^3S_1) \,.
\end{align}
The coefficient $b$ receives both $P$-wave and next-to-next-to-leading
order $S$-wave Wilson coefficient contributions,
\begin{align}
\nonumber
 b \ =& \ 
     \frac{m_{e_1}^2 m_{e_2}^2}{ M^2 \left( m_{e_1} + m_{e_2} \right)^2 }\,
      \Bigl(\,
        \hat f(^1P_1) + \frac{1}{3}~\hat f(^3P_0) + \hat f(^3P_1)
      + \frac{5}{3}~\hat f(^3P_2)
\\
&\phantom{\frac{ M^2 \left( m_{e_1} + m_{e_2} \right)^2 }{m_{e_1}^2 m_{e_2}^2}\Bigl(~}
       + \hat g(^1S_0) + 3~\hat g(^3S_1)
       \Bigr)  \ .
\label{eq:app_bparam}
\end{align}
The parameters $a$ and $b$ in (\ref{eq:app_aplusbv2}) can also 
be extracted numerically from computer codes that
determine the center-of-mass frame annihilation 
cross sections. This is done by considering the cross section's
behaviour for small relative velocities of the annihilating particle pair 
and performing a parabola fit to 
$\sigma^{\chi_{e_1}\chi_{e_2} \to X_A X_B}~v_{\rm rel}$, which provides
the corresponding coefficients $a$ and $b$. Note, however, that a 
separation of the coefficient $b$ into its constituent $P$-wave and 
next-to-next-to-leading order $S$-wave contributions, as given in
(\ref{eq:app_bparam}), cannot be achieved with the sole knowledge of the cross
section. Likewise, the separation of the $S$-wave contributions for the spin
singlet and triplet configurations, as performed in (\ref{eq:app_aparam})
and (\ref{eq:app_bparam}), requires intervention at the amplitude level, which
is not straightforward for the publicly available computer codes.

In the absence of threshold effects, resonances or enhanced radiative 
corrections, the knowledge of the coefficients
$a$ and $b$ in $\chi_{e_1} \chi_{e_2}$ annihilation processes allows 
for a rather accurate calculation of the present-day relic abundance. 
Yet the separation of $b$ into $P$- and $S$-wave contributions 
is required for a consistent treatment of the Sommerfeld enhancement 
at  $\mathcal O(v^2)$ because the long-range interactions responsible for
the Sommerfeld effect depend on the quantum numbers of the incoming state.
Our analytic approach allows us to perform this separation by 
construction.

We perform a numeric check of our results for the
$\chi_{e_1} \chi_{e_2} \to X_A X_B$ tree-level annihilation cross sections as
given in (\ref{eq:app_xsection}) for all initial state two-particle pairs in
Tab.~\ref{tab:scattering_reactions} into all accessible SM and Higgs
two-particle final states. We consider several MSSM spectra, 
which we compute using the spectrum calculator
{\tt SuSpect}~\cite{Djouadi:2002ze} and its implementation of the 
{\it phenomenological MSSM}, a model with 27 free parameters.
For each spectrum, we obtain the coefficients $a$ and $b$ in
(\ref{eq:app_aparam}) and (\ref{eq:app_bparam}) from our analytic 
calculation, and compare them with the corresponding coefficients extracted 
purely numerically using {\sc MadGraph}~\cite{Alwall:2011uj} to 
calculate the cross sections.
Our results for the coefficient $a$ agree with the corresponding
numeric expression extracted from {\sc MadGraph} data at permille level. 
Similarly, we find agreement of the coefficients $b$ derived 
with (\ref{eq:app_bparam}) and extracted from {\sc MadGraph} data 
at $1\%$ up to permille level, where the
level of agreement slightly varies depending on the initial- and final-state
particles. In addition, the level of agreement on the parameter $b$ depends on
the interval of the $v_{\rm rel}$ variable used for the parabola fit to the
{\sc MadGraph}  data, which for the numbers quoted above is taken as
$v_{\rm rel}/c = [0, 0.4]$. We find that the non-relativistic approximation 
is reliable for single-particle velocities up to $v_{e_i}/c\sim 0.3$.  
For such velocities the absolute error of the non-relativistic approximation to
$\sigma^{\chi_{e_1} \chi_{e_2} \to X_A X_B}~v_{\rm rel}$ with respect to 
the unexpanded $\sigma^{\chi_{e_1} \chi_{e_2} \to X_A X_B}~v_{\rm rel}$ 
expression lies within the level of a few percent.
Therefore the non-relativistic approximation has an acceptable
accuracy for calculations in the early Universe during the time of
$\chi_{e_i}$-decoupling, as the mean velocity of the $\chi_{e_i}$ 
in that period was around $v_{e_i}/c \sim 0.2$.

\begin{figure}[t]
\includegraphics[width=0.475\textwidth]{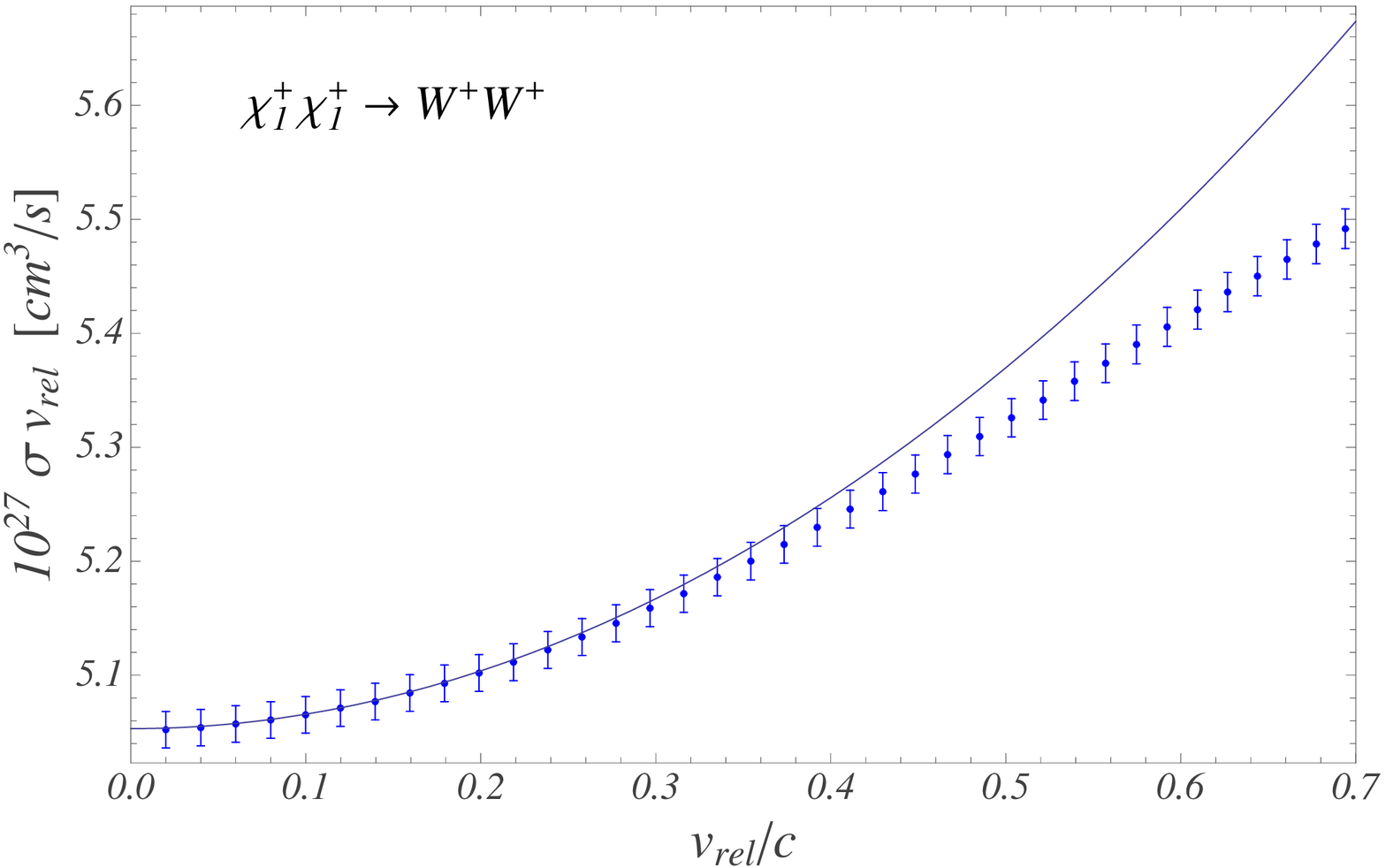} \ \ 
\includegraphics[width=0.475\textwidth]{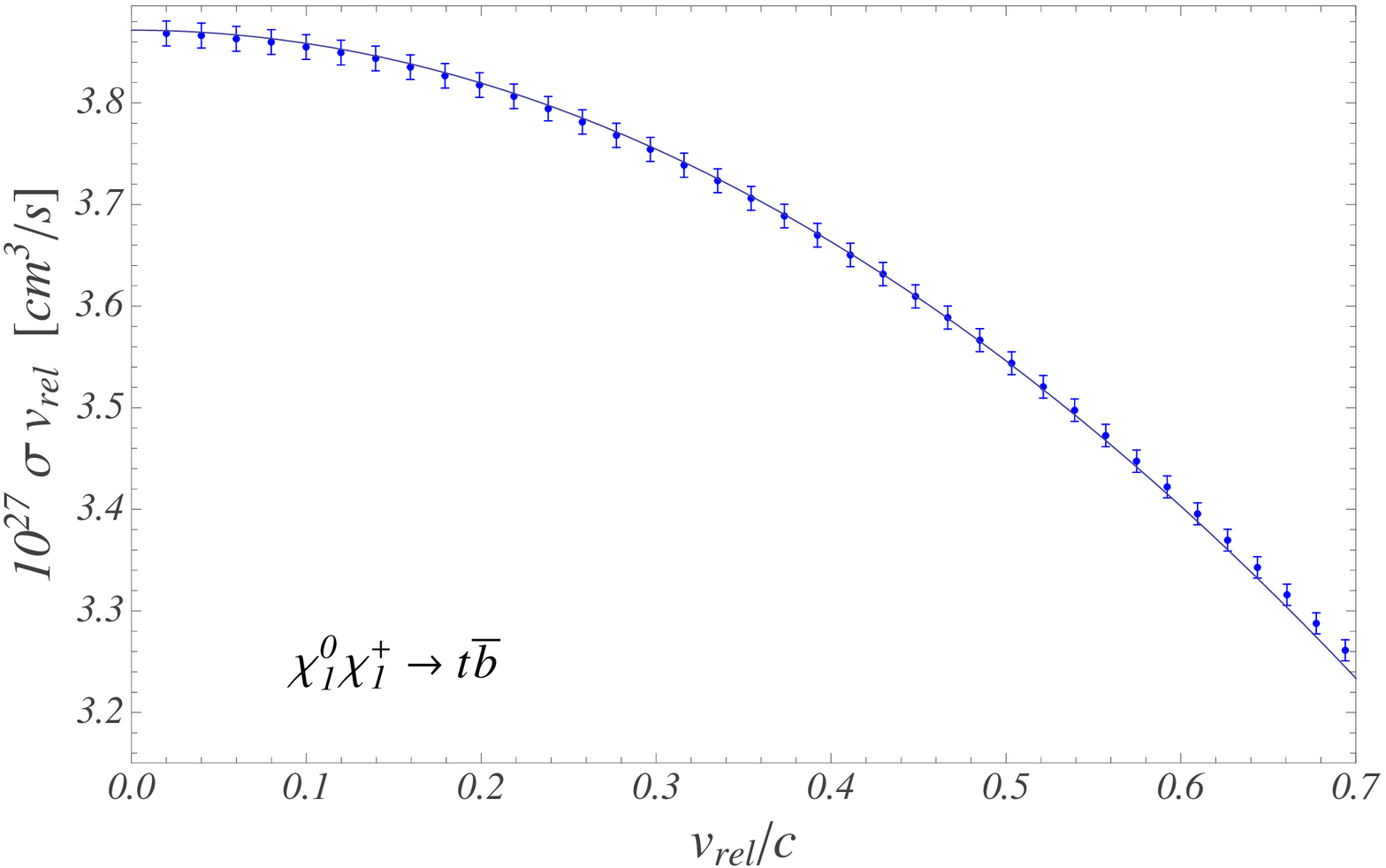}
\caption{Numeric comparison of the non-relativistic approximation 
(solid lines) to the tree-level annihilation cross-section times relative
velocity, $\sigma\, v_{\rm rel}$, for $\chi^+_1 \chi^+_1 \to W^+ W^+$ (left) and
$\chi^0_1 \chi^+_1 \to t \overline b$ (right) reactions with the
corresponding unexpanded annihilation cross section produced with
{\sc MadGraph}.
The numeric errors on the latter are taken to
be $\sigma\, v_{\rm rel}/\sqrt{N}$, where $N = 10^5$ gives the number of
events used in the {\sc MadGraph} calculation of each cross section value.
$v_{\rm rel}$ is given by $v_{\rm rel} = \vert \vec v_{e_1} - \vec v_{e_2} \vert$ for
the $\chi_{e_1} \chi_{e_2} \to X_A X_B$ process.
The underlying MSSM spectrum is a wino-like neutralino LSP scenario,
generated with the spectrum calculator {\tt SuSpect}.
The masses of the $\chi^0_1$ and $\chi^+_1$ are given by
$m_{\chi^0_1} = 2748.92\,\mbox{GeV}$ and $m_{\chi^+_1} = 2749.13\,\mbox{GeV}$.}
\label{fig:xsectionsMG_1}
\end{figure}

\begin{figure}[t]
\includegraphics[width=0.475\textwidth]{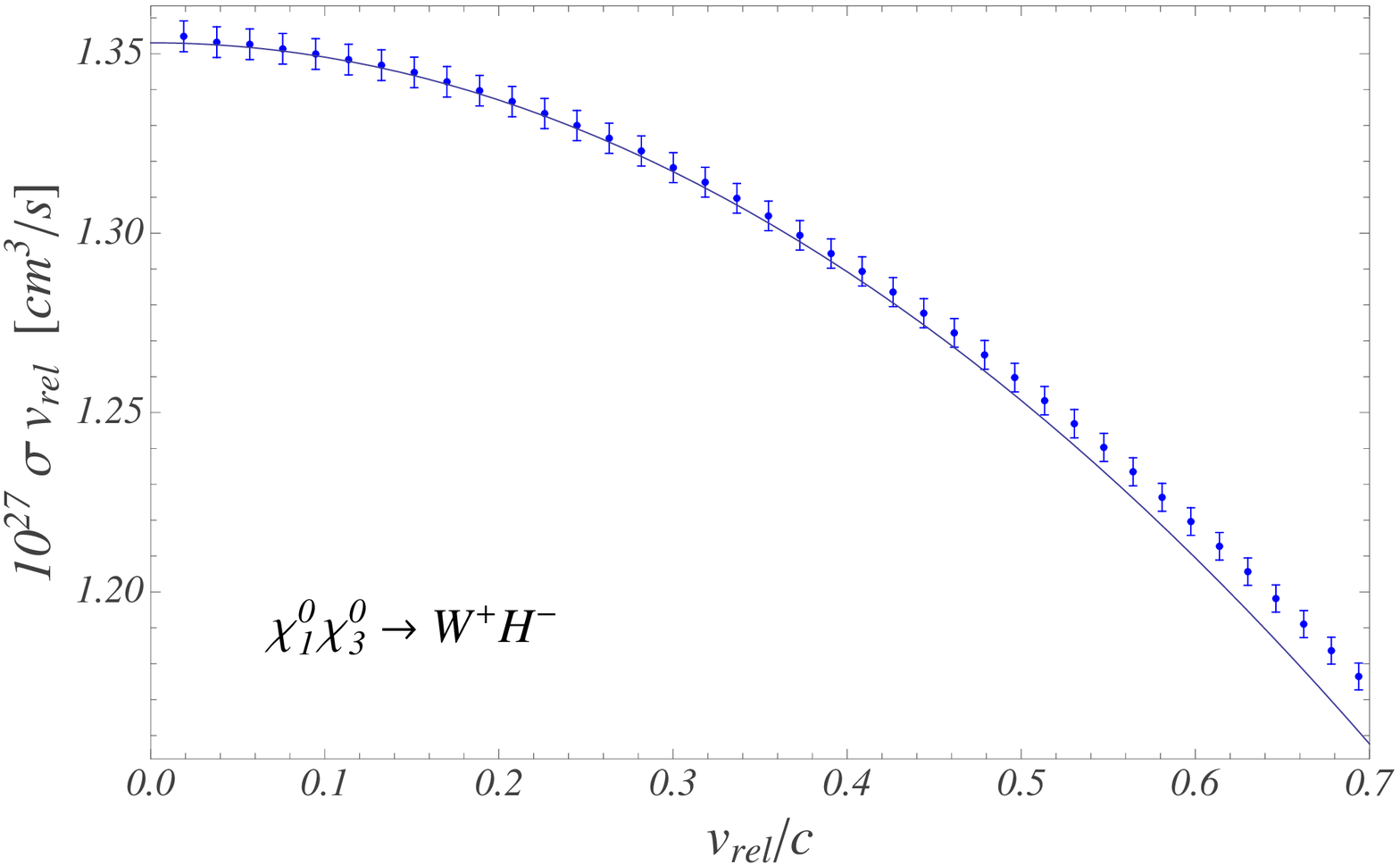} \ \ \includegraphics[width=0.475\textwidth]{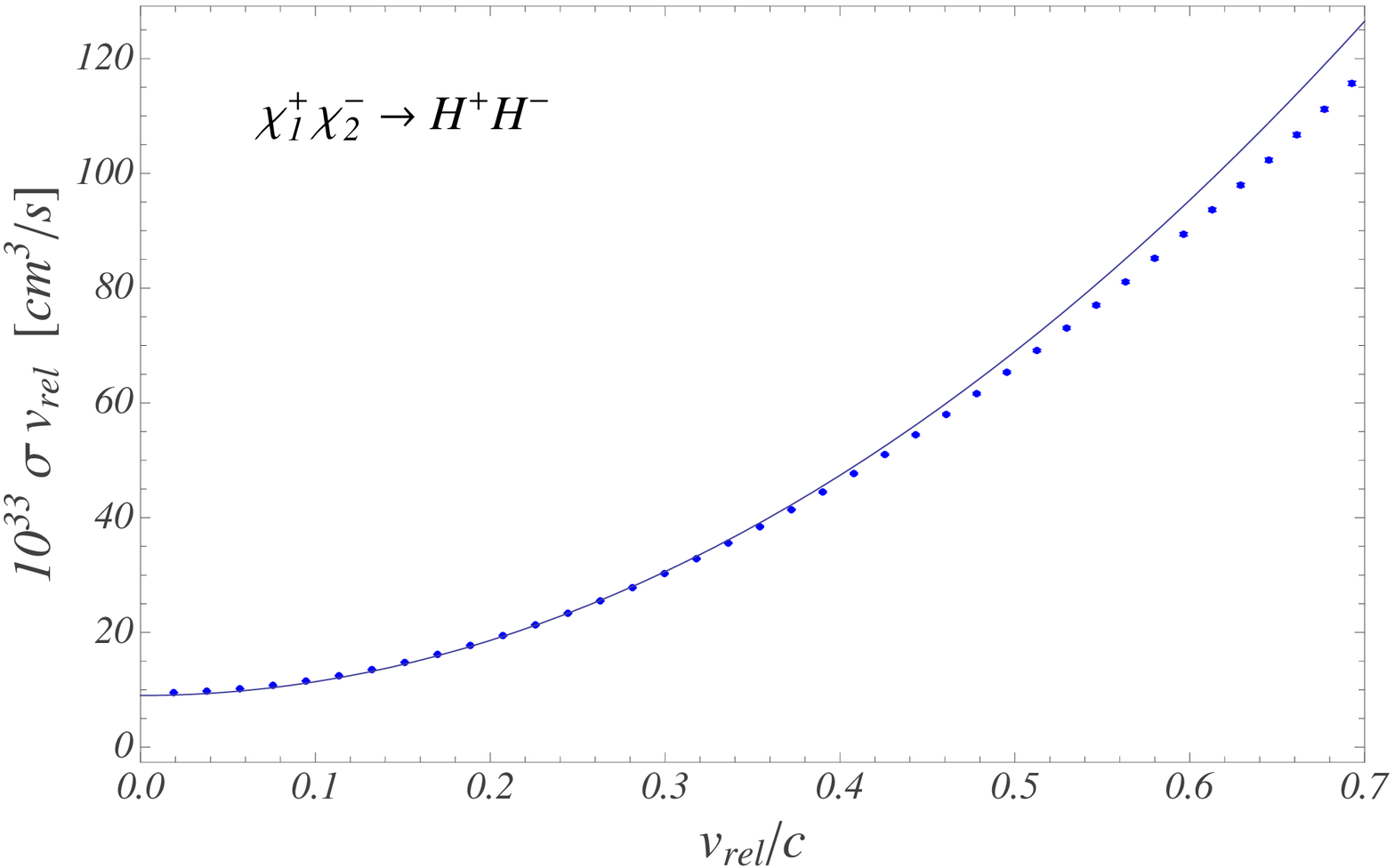}
\caption{Numeric comparison of the non-relativistic approximation (solid line)
to $\sigma\,v_{rel}$ for the two neutral hydrogen-like two-body states
$\chi^0_1\chi^0_3 \to W^+ H^-$(left) and
$\chi^+_1\chi^-_2 \to H^+ H^-$(right) to data produced with {\sc MadGraph}.
Again, we take the errors on the {\sc MadGraph} data to
be $\sigma\, v_{\rm rel}/\sqrt{N}$, where $N = 10^5$ gives the number of
events used in the {\sc MadGraph} calculation of each cross section value.
The process on the right-hand side is dominated by
$P$-wave annihilations.
The underlying MSSM spectrum is the same as in the plots in
Fig.~\ref{fig:xsectionsMG_1}, where the masses of the $\chi^0_{3}$ and
$\chi^-_2$  are given by $m_{\chi^0_3} = 3061.99\,\mbox{GeV}$ and
$m_{\chi^-_2} =  3073.31\,\mbox{GeV}$. The mass of the Higgs particles
$H^\pm$ takes the value $m_{H^\pm} = 167.29\,\mbox{GeV}$.}
\label{fig:xsectionsMG_2}
\end{figure}

Selected results of our numeric check with {\sc MadGraph} are presented in
Fig.~\ref{fig:xsectionsMG_1} and Fig.~\ref{fig:xsectionsMG_2}, where the
underlying SUSY spectrum contains a wino-like neutralino LSP
with mass $m_{\chi^0_1} = 2748.92\,\mbox{GeV}$ and an almost mass-degenerate
wino-like chargino with $m_{\chi^+_1} = 2749.13\,\mbox{GeV}$.
Fig.~\ref{fig:xsectionsMG_1} shows tree-level annihilation cross sections
that are relevant in the calculation of the neutralino LSP relic abundance
including co-annihilations. The plot on the left-hand side displays the
annihilation cross section times the relative velocity for the double-charged
annihilation reaction $\chi^+_1 \chi^+_1 \to W^+ W^+$.
For $v_{\rm rel}/c \lesssim 0.4$ our analytic, non-relativistic
approximation nicely reproduces the numeric, unexpanded cross section 
$\sigma^{\chi^+_1  \chi^+_1 \to W^+ W^+}~v_{\rm rel}$. Furthermore, 
as the absolute curvature in this $S$-wave dominated reaction is rather small
compared to the coefficient $a$, even the absolute error that one would 
make in using the non-relativistic approximation instead of the full 
cross section is only of the order of $2\%$ for $v_{\rm rel}/c \sim 0.6$.
The coefficient $b$ for this reaction, calculated using (\ref{eq:app_bparam}),
is given by $b c^2 = 1.27\cdot10^{-27}\,$cm$^3$\,s$^{-1}$. Its $P$- and $S$-wave 
contributions are of the same order and read
$b_P c^2 = 2.95\cdot10^{-27}\,$cm$^3$\,s$^{-1}$ 
and $b_S c^2 = -1.68\cdot10^{-27}\,$cm$^3$\,s$^{-1}$.
The plot on the right-hand side in Fig.~\ref{fig:xsectionsMG_1} depicts
the single-charged annihilation reaction $\chi^0_1 \chi^+_1 \to t \overline b$
with (massive) fermionic final states. As it receives 
significant leading order
$S$-wave contributions, this annihilation process is also relevant in the
neutralino LSP relic abundance calculation including co-annihilation 
processes. Here it turns out that the $b$ coefficient is $S$-wave
dominated, as the contributions from $P$-waves are suppressed by 
five orders of magnitude.
Let us stress that our analytic results for the Wilson coefficients
include the full mass-dependence of the final state particles and can be
applied to MSSM scenarios with flavour off-diagonal sfermion generation mixing
as well.

The plots in Fig.~\ref{fig:xsectionsMG_2} show that our results can not only
be used to describe pair annihilations of nearly mass-degenerate incoming
particles $\chi_{e_1} \chi_{e_2} \to X_A X_B$, but also apply to annihilations
of a non-relativistic ``hydrogen-like'' $\chi_{e_1} \chi_{e_2}$ 
two-particle system of non-degenerate-in-mass constituents.
The plot on the left hand side in Fig.~\ref{fig:xsectionsMG_2} corresponds to
the pair annihilation of a hydrogen-like $\chi^0_1 \chi^0_3$  state into a
$W^+ H^-$ final state, with $m_{\chi^0_3} = 3061.99\,\mbox{GeV}$, which is
again dominated by leading-order $S$-wave contributions. The curvature is
driven negative by the next-to-next-to-leading order $S$-wave contributions to
the coefficient $b$, given by $b_S c^2 = -5.29\cdot10^{-28}\,$cm$^3$\,s$^{-1}$.
The $P$-wave contributions are, however, of the same order 
and read $b_P c^2 = 1.30\cdot10^{-28}\,$cm$^3$\,s$^{-1}$.
The right plot in Fig.~\ref{fig:xsectionsMG_2} again refers to a hydrogen-like
incoming two-body system, $\chi^+_1 \chi^-_2$, where in this case the
annihilation $\chi^+_1 \chi^-_2 \to H^+ H^-$ is $P$-wave dominated:
the $P$-wave contribution to the coefficient $b$ is given by
$b_P c^2 = 2.48\cdot 10^{-31}\,$cm$^3$\,s$^{-1}$. Both the leading and
next-to-next-to-leading order $S$-wave contributions are strongly suppressed and
of the order $\mathcal O(10^{-33}\,$cm$^3$\,s$^{-1}{})$, respectively.
The mass of the second chargino is given by $m_{\chi^-_2} =  3073.31\,\mbox{GeV}$.

Generically, if the coefficient $a$ in the expansion (\ref{eq:app_aplusbv2}) is
suppressed with respect to the coefficient $b$, the curvature and hence the
corresponding non-relativistic annihilation process is $P$-wave dominated. This
property derives from the fact, that the leading-order coefficient $a$ is
related to the product of the leading order $S$-wave contributions to the
tree-level annihilation amplitude with its complex conjugate.
As the next-to-next-to-leading order $S$-wave contributions to the coefficient
$b$ result from the product of leading order with next-to-next-to-leading order
$S$-wave contributions in the annihilation amplitudes, a suppressed 
coefficient  $a$ indicates a small next-to-next-to-leading order $S$-wave 
contribution to the coefficient $b$ as well.

\subsection{Analytic checks}
\label{subsec:analyticchecks}
In \cite{Drees:1992am}, the authors performed a calculation of the
neutralino relic abundance in minimal supergravity models. In the
appendix, they give a complete summary of all partial-wave
separated tree-level helicity amplitudes in 
$\chi^0_1 \chi^0_1 \to X_A X_B$ pair
annihilations. These comprehensive results for tree-level neutralino LSP
pair-annihilations are also referenced and (partly) quoted in the (SUSY)
particle dark matter reviews \cite{Jungman:1995df} and \cite{Bertone:2004pz},
and easily extend to $\chi^0_{e_1} \chi^0_{e_1} \to X_A X_B$
annihilations. Hence, these results allow for an explicit analytic check of
our expressions for the different partial-wave contributions to a
neutralino $\chi^0_{e_1} \chi^0_{e_1} \to X_A X_B$ annihilation cross section.
The partial-wave coefficients that can be cross-checked in that way
correspond to $^1S_0$-, $^3P_0$-, $^3P_1$- and $^3P_2$-wave
$\chi^0_{e_1}\chi^0_{e_1} \to X_A X_B$ annihilation reactions, and the
leading order and next-to-next-to-leading order $^1S_0$-wave
contributions can be compared separately.
As already inferred from (\ref{eq:WilsonCoeffSymmetry}) and noted at the end of
Sec.~\ref{subsec:basis}, there are no $^3S_1$ and $^1P_1$ partial-wave
contributions for annihilation reactions of identical incoming particles, which
is the case covered by \cite{Drees:1992am}. Our expressions for the 
partial-wave separated $\chi^0_{e_1} \chi^0_{e_1} \to X_A X_B$ 
annihilation cross sections into all possible SM and Higgs final states, 
obtained from (\ref{eq:app_xsection}), agree with the corresponding terms
derived from the helicity amplitudes in
\cite{Drees:1992am}\footnote{
The only minor discrepancies that we find are related to $P$-wave contributions:
our results for $^3P_1$-wave $\chi^0_{e_1}\chi^0_{e_1} \to H^+ H^-$
annihilations correspond to a factor 2 instead of a factor 4 in the second term 
of Eq.~(A27b) in \cite{Drees:1992am}.
In the case of $^3P_0$-wave $\chi^0_{e_1}\chi^0_{e_1} \to f\overline{f}$ reactions,
our results correspond to a factor $\sqrt{2/3}$ instead of a factor $\sqrt{6}$
in the second term in the first line of Eq.~(A29b) in \cite{Drees:1992am}.
}.
We note that our results for annihilations into a pair of fermions
include the case of flavour-off-diagonal sfermion mixing as well, which is
covered in \cite{Jungman:1995df} and \cite{Bertone:2004pz}, but was not yet
included in \cite{Drees:1992am}, wherein only flavour-diagonal right-left
sfermion mixing was taken into account, although it is straightforward to
extend these results to the general flavour-off-diagonal case.

The comparison with analytic results for inclusive leading-order $^1S_0$- and
$^3S_1$-wave pair-annihilation reactions of a pure wino-like neutralino
$\chi^0_1$ and its mass-degenerate chargino partners $\chi^\pm_1$ into all
possible SM and Higgs final states considered in Ref.~\cite{Hisano:2006nn} provides 
another useful check of our results for the absorptive part of the Wilson
coefficients. The results in \cite{Hisano:2006nn} comprise all possible 
neutral, as well as single and double charged inclusive pair-annihilation 
reactions. The masses of the SM and Higgs particle final states are
set to zero, such that the corresponding results can be understood as the
leading-order term in an expansion in $m_{\text{SM}}/m_{\chi^0_1}$ and
$m_{\text{Higgs}}/m_{\chi^0_1}$. Furthermore, all supersymmetric 
particle states
heavier than $\chi^0_1$ and $\chi^\pm_1$ are treated as completely decoupled.
We agree with all results for the inclusive annihilation reactions given in
\cite{Hisano:2006nn}. In particular we agree with the results in
\cite{Hisano:2006nn} that refer to leading-order $^1S_0$-wave
$\chi^0_1\chi^0_1 \to \chi^-_1\chi^+_1$ as well as
$\chi^-_1\chi^+_1 \to \chi^0_1\chi^0_1$ reactions, which can be related to the
Wilson coefficients $\hat f^{\chi^0_1\chi^0_1 \to \chi^-_1\chi^+_1}(^1S_0)$ and
$\hat f^{\chi^-_1\chi^+_1 \to \chi^0_1\chi^0_1}(^1S_0)$, 
therewith permitting an explicit check of some of our Wilson coefficients
 encoding off-diagonal scattering reactions.\footnote{
The authors of Ref.~\cite{Hisano:2006nn} also provided analytic results for
exclusive leading-order $^1S_0$-wave annihilation reactions for both the cases
of a wino-like and a Higgsino-like neutralino LSP scenario in a previous 
work~\cite{Hisano:2004ds}. We agree with the results for all diagonal
$\chi_{e_1} \chi_{e_2} \to \chi_{e_1} \chi_{e_2}$ reactions.
A typo in the off-diagonal terms in Eq.~(28) of \cite{Hisano:2004ds} was fixed
in \cite{Hisano:2006nn}, and the latter agrees with our findings.
In the Higgsino-like scenario, we get differing expressions for off-diagonal
$\chi^0_i \chi^0_i \to \chi^-_1 \chi^+_1$ and
$\chi^-_1 \chi^+_1 \to \chi^0_i \chi^0_i$ reactions in the
case of $W^+ W^-$ and $Z Z$ final states for both $i=1,2$:
our results are a factor $4$ and a factor $2$ larger, respectively, than the
corresponding expressions presented in \cite{Hisano:2004ds}.
}

\section{Discussion}
\label{sec:discussions}

\subsection{Unitary vs Feynman gauge}
\label{subsec:gauge}

The computation of the absorptive parts of the Wilson coefficients for 
forward-scattering reactions, 
$\chi_{e_1} \chi_{e_2} \to X_A X_B \to \chi_{e_1}\chi_{e_2}$, 
has been performed using both the unitary and Feynman gauge. The 
results agree numerically, which provides a further check of our calculation.
For the off-diagonal reactions, where the incoming and 
outgoing states are different, the use of unitary gauge for final 
states with two massive vector bosons in the final state introduces 
enhanced $1/M_V^4$ and $1/M_V^2$ terms proportional to the mass 
differences between the incoming and outgoing particle species,
which must cancel in the final result. Similarly, a cancellation of $1/M_V^2$
enhanced terms in off-diagonal rates with one massive vector boson in the
final state has to take place. However, for these cancellations to occur, one
has to also expand the SUSY mixing matrices systematically in the gauge boson
masses $M_V$. In the same way, the
mass differences between the incoming and outgoing particles have to be
expanded in $M_V$ and in the differences of soft SUSY breaking
parameters $M_1$, $M_2$, $\mu$, if these differences are small.
The latter expansions must be done differently depending on how many
neutralinos and charginos are (nearly) mass-degenerate. The presentation of the
results computed with unitary gauge then has to distinguish among many cases 
and also consider diagonal and off-diagonal terms separately, since for the
diagonal terms it is desirable to keep the full mass dependence as well as
unexpanded mixing matrices. We thus find it more convenient to 
use Feynman gauge
for the calculation of the off-diagonal reactions, which allows to keep the
coupling matrices unexpanded and a more concise presentation of the results. 
The price for this is that one must compute a large number of unphysical 
final states containing pseudo-Goldstone Higgs and ghost particles, 
see Tab.~\ref{tab:XAXBstates}.

\subsection{Off-diagonal terms}
\label{subsec:offdiag}

Our framework aims to describe the annihilation of a pair of non-relativistic
charginos or neutralinos ($\chi_i\chi_j$) into SM and light Higgs particles
pairs ($X_A X_B$) including potential interactions between all nearly
mass-degenerate $\chi\chi$ states, that can produce a Sommerfeld enhancement
of the rates. A contribution to these enhanced annihilation rates is given by
the imaginary part of the amplitude for a process of the type,
\begin{align}
\chi_i \chi_j \to \ldots \to \chi_{e_1} \chi_{e_2} \to X_A X_B \to \chi_{e_4} \chi_{e_3} \to \ldots  \to \chi_i  \chi_j 
\ ,
\label{eq:off-diag}
\end{align}
where the intermediate states involved in the short-distance annihilation,
$\chi_{e_1} \chi_{e_2}$ and $\chi_{e_4} \chi_{e_3}$, can be different 
(off-diagonal annihilation terms), compare to Fig.~\ref{fig:genericdiagram} 
for a figurative illustration. In a recent work~\cite{Hryczuk:2010zi}, 
a general formalism which also aims to compute the Sommerfeld-enhanced 
annihilation rates 
for a coupled system of neutralino and chargino pairs, has been presented
which, however, does not implement the possibility of off-diagonal transitions
in the hard part of the annihilation process. We show in this section 
that the off-diagonal terms can indeed be relevant, and should be accounted 
for in the calculation of the Sommerfeld enhanced rates.

Naively, if the final state $X_A X_B$ is allowed for 
both $\chi_{e_1} \chi_{e_2}$
and $\chi_{e_4} \chi_{e_3}$, given one particular partial-wave configuration of
the two-body systems, the off-diagonal absorptive amplitude can be of the same
size as the diagonal absorptive amplitude, {\it i.e.}
\begin{align}
  \int[\text{dPS}_{AB}] \ &
      \mathcal A (\chi_{e_1}\chi_{e_2} \to X_A X_B)
    \times
      \mathcal A (\chi_{e_4}\chi_{e_3} \to X_A X_B)^*
\nonumber
\\
\sim \
&
  \int[\text{dPS}_{AB}] \
      | \mathcal A (\chi_{e_1}\chi_{e_2} \to X_A X_B)|^2 \ ,
\label{eq:offdiagabs}
\end{align}
since the phase-space integration involves very similar kinematics. An example
of the latter is given by the annihilation rates of wino-like neutralino dark
matter, where the wino-like neutralino $(\chi^0_1)$ is highly degenerate with
its charged $SU(2)_L$ partners $(\chi^\pm_1)$. In such scenario the spin-0
$\chi^0_1\chi^0_1$ system mixes with the $\chi^-_1 \chi^+_1$ state through
$W$-boson exchange. The inclusive annihilation rates that have to be fed into
the calculation of the enhanced rates for the spin-0 $\chi^0_1\chi^0_1$ channel
in the wino limit read
\begin{align}
[\mathcal A (\chi^0_1\chi^0_1 \to \chi^0_1\chi^0_1 )\,(^1S_0)]|_{\rm abs} 
\ =& \
2 ~ f^{00 \to 00}_{\lbrace 1 1 \rbrace \lbrace 1 1 \rbrace}~(^{1}S_0) 
\ = \ 
\frac{4\pi\alpha_2^2}{m_{\chi^0}^2} \ ,
\label{eq:diagwino1}
\\
 [\mathcal A (\chi^-_1\chi^+_1 \to \chi^-_1\chi^+_1)\,(^1S_0)]|_{\rm abs}
\ =& \
2 ~ f^{-+ \to -+}_{\lbrace 1 1 \rbrace \lbrace 1 1 \rbrace}~(^{1}S_0) 
\ = \ 
\frac{3\pi\alpha_2^2}{m_{\chi^0}^2} \ ,
\label{eq:diagwino2}
\\
[\mathcal A (\chi^0_1\chi^0_1 \to \chi^-_1\chi^+_1 )\,(^1S_0)]|_{\rm abs} 
\ =& \
2 ~ f^{00 \to -+}_{\lbrace 1 1 \rbrace \lbrace 1 1 \rbrace}~(^{1}S_0) 
\ = \ 
 \frac{2\pi\alpha_2^2}{m_{\chi^0}^2} \ ,
\label{eq:offdiagwino}
\end{align}
where $\alpha_2 = g_2^2/4\pi$, $g_2$ denotes the $SU(2)_L$ gauge coupling,
and all gauge boson and Higgs-particle masses are treated as massless.
We see explicitly that the off-diagonal term (\ref{eq:offdiagwino}) is
of the same order as the diagonal reactions (\ref{eq:diagwino1}--\ref{eq:diagwino2}).

In order to stress the importance of the off-diagonal annihilation terms, we have computed the 
thermally averaged effective annihilation cross section, $\langle \sigma_{\rm eff} v\rangle$,
which enters the Boltzmann equation
for the calculation of the dark matter yield, for the same wino-like scenario used for the checks
with {\sc{MadGraph}} presented in Sec.~\ref{subsec:numericchecks}, and compared to the results obtained when the off-diagonal terms 
are switched off by hand. For the necessary formulas to compute
$\langle \sigma_{\rm eff} v\rangle$, including co-annihilation effects, we
refer the reader  to~\cite{Gondolo:1990dk,Griest:1990kh}. The
annihilation rates have been calculated using~(\ref{eq:app_xsection}) with
Wilson coefficients multiplied by the appropriate Sommerfeld factors computed
solving the coupled-channel Schr\"odinger equation for each partial-wave. The
details about the calculation of the Sommerfeld enhancement factors from the
long-range interactions will be given in a future publication~\cite{paperIII}. 

\begin{figure}[t]
\begin{center}
\includegraphics[width=0.75\textwidth]{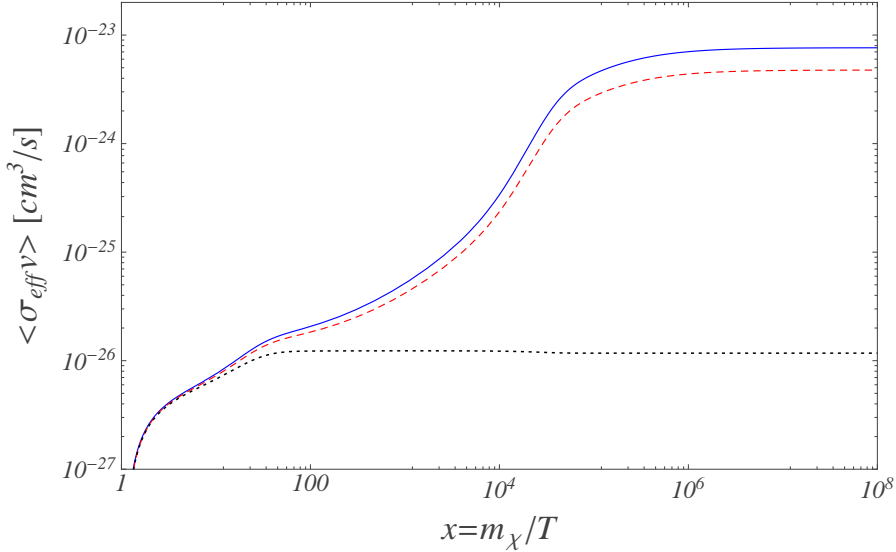} 
\caption{Thermally averaged effective annihilation cross section as a function
of $x=m_{\chi}/T$ with fixed $m_{\chi}=2748.92$~GeV, including the Sommerfeld
effect (solid blue line). The same quantity computed with the off-diagonal 
perturbative annihilation rates set to zero is depicted by the dashed 
red line. The perturbative result is also shown as a dotted line for 
comparison. The decrease of $\langle \sigma_{\rm eff} v\rangle$ towards 
$x\sim 1$ is due to  large negative ${\cal O}(\vec{p}^{\,2})$ terms in 
the $\chi\chi$ annihilation rates, which are unphysical because the 
non-relativistic expansion (\ref{eq:app_xsection}) for the annihilation 
rate becomes unreliable for large temperatures. }
\label{fig:sigmaeff}
\end{center}
\end{figure}

We observe from Fig.~\ref{fig:sigmaeff} that removing the off-diagonal annihilation terms decreases
the thermally averaged cross section by a factor larger than 1.5 at small temperatures. The corresponding
thermal relic abundance of the dark matter in the present Universe, $\Omega_{\rm DM} h^2$,
obtained by numerical integration of the Boltzmann equation, gets then increased by approximately $20$\%,
if the off-diagonal reactions are neglected. The latter represents thus a sizeable effect
which has to be accounted for in such a scenario. 

%
%

\section{Summary}
\label{sec:summary}

The calculation of the thermal relic abundance of the lightest neutralino
as a promising dark matter candidate within the MSSM places 
strong bounds on the MSSM parameter space, assuming that the observed 
cosmic dark matter has
particle nature and is composed solely of the neutralino LSP.
Given the expected future experimental accuracy of the measurement of 
the cosmic dark matter abundance observed today, 
radiative corrections to the pure
tree-level annihilation cross section, entering the relic abundance calculation
as a central ingredient, should eventually be taken into account. 
The inclusion of
1-loop corrections to the annihilation cross section as well as the systematic
treatment of Sommerfeld enhancements, has recently been a field of
elaborate studies in the literature.
Similarly, in the context of dark matter annihilation processes in the present
Universe relevant in indirect detection, the above types of radiative
corrections to the neutralino pair-annihilation cross section 
have been studied extensively.

In this paper we take advantage of the non-relativistic nature of the
annihilating neutralinos in the present Universe as well as during thermal
dark matter decoupling in the early Universe, which introduces a clear
separation of energy scales in all annihilation processes of interest. The
latter property allows us to set up an effective field theory (the NRMSSM) of
non-relativistic neutralinos and charginos, that provides an appropriate setup
for a systematic investigation of radiative corrections to neutralino LSP pair
annihilation processes both in the present and the early Universe, taking
co-annihilations with nearly mass-degenerate neutralinos and charginos into
account.
As a first step in the explicit construction of the NRMSSM we have 
derived fully
analytic formulas for the absorptive part of the Wilson coefficients of 
four-fermion operators in the effective theory pertaining to $S$-wave 
annihilation, that encode the hard annihilation rates of
$\chi_{e_1} \chi_{e_2} \to X_A X_B$ processes 
(see (\ref{eq:deltaL4fermion}) and Tab.~\ref{tab:fields}).
Our results separately include leading-order $^1S_0$- and $^3S_1$-wave as well
as all $P$-wave and next-to-next-to-leading order $S$-wave 
Wilson coefficients and
apply to general neutralino and chargino states in the MSSM.
Flavour off-diagonal sfermion generation mixing can be covered, and we keep 
the full mass dependence of all SM and Higgs particles.
Analytic results for the absorptive part of leading order $S$-wave Wilson
coefficients are presented in the appendix.
Results for $P$-wave and next-to-next-to-leading order $S$-wave 
coefficients will be given in a future publication \cite{Hellmann:2013jxa}.
By taking into account charge-neutral annihilation processes of a 
chargino pair as well as singly charged and doubly charged 
annihilation reactions of non-relativistic
neutralinos and charginos, we extend the analytic results for partial wave
decomposed neutralino LSP pair-annihilation cross sections given in the
literature \cite{Drees:1992am}.

We have shown that the non-relativistic expansion to ${\cal O}(v^2)$ 
produces accurate results up to $v_{\rm rel}\sim 0.6$, which is 
sufficient for relic density computations, and certainly for dark 
matter annihilation in the present Universe. Our analytic results 
may therefore substitute for time-consuming numerical computations.

Our aim is to apply the effective field theory formalism to the calculation of
Sommer\-feld-enhanced (co-)annihilation cross sections in the neutralino relic
abundance calculation. As scattering prior to the annihilation process 
can lead to transitions from an incoming particle pair to another 
nearly mass-degenerate neutralino or chargino two-particle state, 
a proper treatment of the Sommerfeld effect requires 
the knowledge of the absorptive part of off-diagonal annihilation rates,
$\chi_{e_1} \chi_{e_2} \to X_A X_B \to \chi_{e_4} \chi_{e_3}$, for all possible
SM and Higgs two-particle states $X_A X_B$
(see Fig.~\ref{fig:genericdiagram}). To the best of our knowledge 
we present for the first time analytic results that allow for a 
systematic treatment of all these off-diagonal rates 
in Sommerfeld-enhanced (co-)annihilation
reactions for general masses and composition of the $\chi_{e_i}$ particles. 
The implications of these results 
for MSSM relic density calculations will be studied in a forth-coming 
publication~\cite{paperIII}.

\subsubsection*{Note added}
The present arXiv version replaces an incorrect version of Figure 4 and fixes
some typos which are also present in the journal publication. For an explicit
list of errata see the JHEP erratum \cite{Beneke:2012tg}.

\subsubsection*{Acknowledgements}
We would like to thank N. Baro for discussions and some cross checks
of the tree-level annihilation cross sections with the private code Sloops 
\cite{Baro:2007em,Baro:2008em,Baro:2009na}.
The work of M.B. was supported in part by the DFG
Sonder\-for\-schungs\-bereich/Trans\-regio~9 ``Computergest\"utzte
Theoreti\-sche Teilchenphysik''. 
C.H. greatly acknowledges the support by the ``Deutsche Telekom Stiftung''.
The work of P.~R. is partially supported by MEC (Spain) under grants FPA2007-60323 and FPA2011-23778 and by the
Spanish Consolider-Ingenio 2010 Programme CPAN (CSD2007-00042).
Feynman diagrams have been drawn with the packages 
{\sc Axodraw}~\cite{Vermaseren:1994je} and 
{\sc Jaxo\-draw}~\cite{Binosi:2008ig}.

\appendix

\section{Absorptive parts of   
Wilson coefficients of 
dim\-ension-6 operators in $\delta \mathcal L_\text{ann}$}
\label{sec:appendix}

We present the leading order contributions to the  absorptive part,
$\hat{f}^{\, \chi\chi \to \chi\chi}(^{2s+1}L_J)$,
of the Wilson coefficients that correspond to the local four-fermion operators
given in Tab.~\ref{tab:fields}.
The $\hat{f}^{\, \chi_{e_1}\chi_{e_2} \to \chi_{e_4}\chi_{e_3}}(^{2s+1}L_J)$ encode the
absorptive part of hard $2\to 2$ scattering reactions of an incoming particle
pair  $\chi_{e_1} \chi_{e_2}$ of non-relativistic charginos or neutralinos in a
given $^{2s+1}L_J$ partial-wave state into an outgoing non-relativistic
$\chi_{e_4} \chi_{e_3}$-pair in the same partial-wave configuration.
They allow to reproduce the inclusive tree-level center-of-mass frame
annihilation cross sections of a non-relativistic
$\chi_{i} \chi_{j}$-pair\footnote{The covered $\chi_i \chi_j$-states have been collected in Tab.~\ref{tab:scattering_reactions}.}
into SM and light Higgs two-body final states $X_A X_B$,
expanded in the relative velocity of the annihilating particle pair.
The general case includes off-diagonal processes
$\chi_{e_1} \chi_{e_2} \to X_A X_B \to \chi_{e_4} \chi_{e_3}$ with
$\chi_{e_1} \chi_{e_2} \neq \chi_{e_4} \chi_{e_3}$, for all pairs 
of non-relativistic neutralinos and charginos.
Since the $\hat f^{\chi\chi \to \chi\chi}(^{2s+1}L_J)$ are
infrared-safe at leading order, we are able to give analytic results 
for the individual contributions 
$\hat f^{\chi\chi \to X_A X_B \to \chi\chi}(^{2s+1}L_J)$ 
pertaining to an exclusive final state $X_A X_B$.

\subsection{Notation and definitions}
\begin{figure}[t]
\begin{center}
\includegraphics[width=0.27\textwidth]{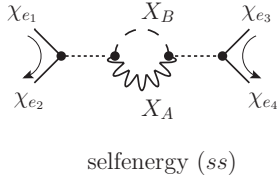}
\caption{ Generic selfenergy-diagram in $\chi\chi \to X_A X_B \to \chi\chi$
          reactions. Particles $X_A$ and $X_B$ represent any two-body 
          final state of SM and Higgs particles, which can be produced 
          on-shell in $\chi\chi \to X_A X_B$ annihilations.}
\label{fig:genericselfenergy}
\end{center}
\end{figure}

\begin{figure}[b]
\begin{center}
\includegraphics[width=0.95\textwidth]{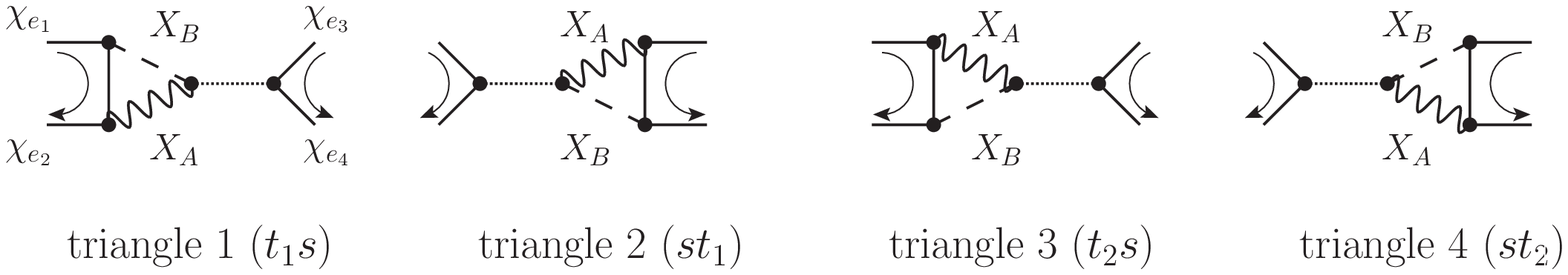}
\caption{ Generic triangle-diagrams in $\chi\chi \to X_A X_B \to \chi\chi$
          reactions.}
\label{fig:generictriangles}
\end{center}
\end{figure}

\begin{figure}[t]
\begin{center}
\includegraphics[width=0.95\textwidth]{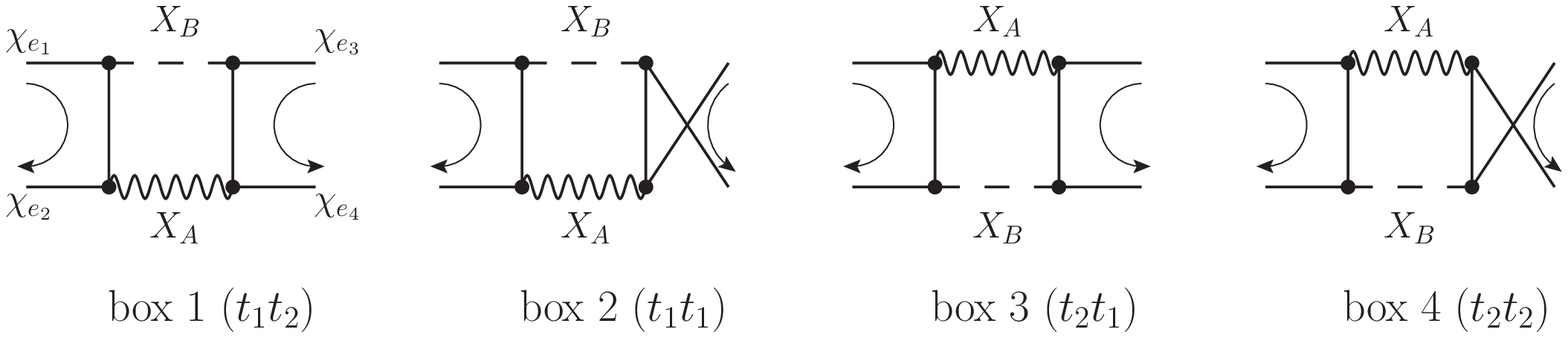}
\caption{ Generic box-diagrams in $\chi\chi \to X_A X_B \to \chi\chi$
          reactions.}
\label{fig:genericboxes}
\end{center}
\end{figure}

\begin{figure}[b]
\begin{center}
\includegraphics[width=0.95\textwidth]{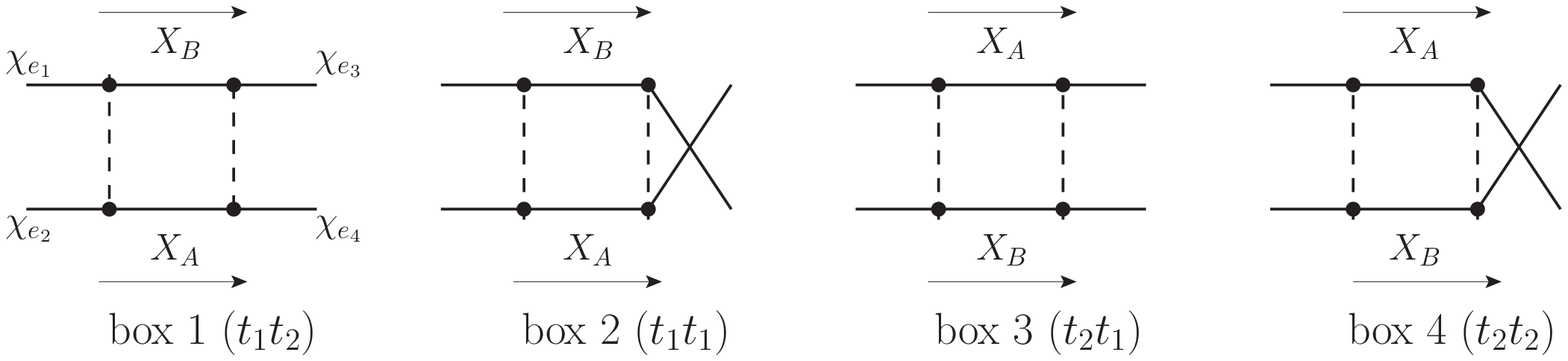}
\caption{ Generic box-diagrams in $\chi\chi \to X_A X_B \to \chi\chi$
          reactions, with $X_A X_B$ a pair of SM fermions.}
\label{fig:genericboxes_fermions}
\end{center}
\end{figure}

Recall that the calculation is performed in Feynman gauge. 
Hence the two-particle final states $X_A X_B$ that we account for
can be classified to be of
vector-vector ($VV$), vector-scalar ($VS$), scalar-scalar ($SS$),
fermion-antifermion ($ff$) or
ghost-anti-ghost ($\eta \bar \eta$) type. They are listed in
Tab.~\ref{tab:XAXBstates}.
\begin{table}[t]
\begin{tabular}{c | c c c c c}
 \hline
 $\chi\chi \to \chi\chi$
&
$V V$
&
$V S$
&
$S S$
&
$f f$
&
$\eta \bar{\eta}$
  \\
 \hline\hline
$
\begin{array}{c}
\chi^0\chi^0 \to \chi^0\chi^0\\
\chi^-\chi^+ \to \chi^-\chi^+\\
\chi^0\chi^0 \to \chi^-\chi^+\\ 
\chi^-\chi^+ \to \chi^0\chi^0\\
\end{array} 
$
&
$
\begin{array}{c}
W^+ W^-, \\
Z Z, \\
\gamma \gamma, Z \gamma
\end{array}
$
&
$
\begin{array}{c}
Z h^0, Z H^0, \\
\gamma h^0, \gamma H^0, \\
Z G^0, Z A^0, \\
\gamma G^0, \gamma A^0, \\
W^+ G^-, W^+ H^-, \\
 W^- G^+, W^- H^+ \\
\end{array}
$
&
$
\begin{array}{c}
\!h^0 h^0, h^0 H^0, H^0 H^0, \\
G^0 h^0, A^0 h^0 \\
G^0 H^0, A^0 H^0,\\
G^0 G^0, G^0 A^0, A^0 A^0 \\
G^+ G^-, G^+ H^-, \\
H^+ G^-, H^+ H^-, \\
\end{array}
$
&
$
\begin{array}{c}
u^J \bar u^I, \\
d^J \bar d^I, \\
e^J \bar e^I, \\
\nu^J \bar\nu^I \\
\end{array}
$
&
$
\begin{array}{c}
\eta^+ \bar\eta^+, \\
\eta^- \bar\eta^-, \\
\eta^Z \bar\eta^Z \\
\end{array}
$
  \\
 \hline
$
\begin{array}{c}
\chi^0 \chi^+ \to \chi^0 \chi^+
\end{array}
$
&
$
\begin{array}{c}
W^+ Z, \\
W^+ \gamma
\end{array}
$
&
$
\begin{array}{c}
Z G^+, \gamma G^+, \\
Z H^+, \gamma H^+, \\
W^+ h^0, W^+ H^0, \\
W^+ G^0, W^+ A^0 \\
\end{array}
$
&
$
\begin{array}{c}
G^+ h^0, G^+ H^0, \\
H^+ h^0, H^+ H^0, \\
G^+ G^0, G^+ A^0, \\
H^+ G^0, H^+ A^0 \\
\end{array}
$
&
$
\begin{array}{c}
u^J \bar d^I, \\
\nu^J \bar e^I \\
\end{array}
$
&
$
\begin{array}{c}
\eta^+ \bar\eta^Z, \\
\eta^Z \bar\eta^-, \\
\eta^+ \bar\eta^F, \\
\eta^F \bar\eta^- \\
\end{array}
$
  \\
 \hline
$
\begin{array}{c}
\chi^+ \chi^+ \to \chi^+ \chi^+
\end{array}
$
&
$
\begin{array}{c}
W^+ W^+
\end{array}
$
&
$
\begin{array}{c}
W^+ G^+, \\
W^+ H^+
\end{array}
$
&
$
\begin{array}{c}
G^+ G^+, \\
G^+ H^+, \\
H^+ H^+
\end{array}
$
&
&
 \\
  \hline
\end{tabular}
\caption{Particle pairs $X_A X_B$ in
$\chi\chi \to X_A X_B \to \chi\chi$ scattering reactions
(abbreviated as $\chi\chi \to \chi\chi$), that we account for
in the calculation of the absorptive part of the Wilson coefficients,
classified according to their type: $VV, VS, SS,
ff$ and $\eta \bar\eta$.
Negatively charged processes, corresponding to the charge-conjugates of the 
singly or doubly
positively charged reactions above are not explicitly written.}
\label{tab:XAXBstates}
\end{table}
The determination of the absorptive part of the Wilson coefficients
for the processes $\chi_{e_1}\chi_{e_2} \to X_A X_B \to \chi_{e_4}\chi_{e_3}$
requires the calculation of a large number of Feynman diagrams.
To be able to present the results in an efficient manner
it is convenient to make use of the classification in
$VV$-, $VS$-, $SS$-, $ff$- and $\eta \bar\eta$- type $X_A X_B$ particle
states and to further subdivide the contributing diagrams according to their
topology.
In each of the classes under consideration there arise
generic 1-loop amplitudes with selfenergy-, triangle- and box-topology
shown in
Figs.~\ref{fig:genericselfenergy}--\ref{fig:genericboxes_fermions}.
%
%
%
The generic self\-energy-diagram as well as the four generic
triangle- and box-diagrams cover all possible kinematic
configurations\footnote{The case of four different triangle-
and four different box-diagrams in Fig.~\ref{fig:generictriangles}
and Fig.~\ref{fig:genericboxes} applies to non-identical
particles $X_A\neq X_B$. For identical particles $X_A = X_B$,
triangle (box) 1 and 3 as well as triangle (box) 2 and 4 coincide.
In this case only one of the identical
diagrams must be taken into account to compute the corresponding
$\hat f^{\chi\chi \to X_A X_A \to \chi\chi}$ coefficients.
This rule incorporates the symmetry factor of 1/2 in the cross-section
for identical final-state particles, that one would take into account in the
conventional calculation of
the tree-level $\chi_{e_1} \chi_{e_2} \to X_A X_A$ annihilation rate.
\label{fn:XAXA}}
that can arise in a $\chi_{e_1}\chi_{e_2} \to X_A X_B \to \chi_{e_4}\chi_{e_3}$
1-loop amplitude.
Note that we have assigned specific directions for the fermion flow in
each diagram in Figs.~\ref{fig:genericselfenergy}--\ref{fig:genericboxes_fermions}, indicated by the arrows,
as it is convenient in the context of calculations involving both Dirac and
Majorana fermions, following the Feynman rules for fermion-number violating
interactions set out in \cite{Denner:1992vza}.
The depicted fermion flows establish our convention to arrange
the external fermion states $\chi_{e_i}, i = 1,\ldots,4$ in descending 
order, see Tab.~\ref{tab:fields}.

We calculate analytically the absorptive part of any of the contributing
selfenergy-, triangle- and box-amplitudes, subject to our convention for the
fermion flows.
Thereby we consider generic external Majorana fermions, generic t- and
u-channel exchanged Majorana fermions or sfermions, generic 
$X_A X_B$ states of type $VV, VS, SS, ff$ and $\eta\overline\eta$, 
and hence use generic `place-holder' coupling factors at each vertex.
This allows us to determine the generic form of those terms in the contributions
to the $\hat f^{\chi\chi \to X_A X_B \to \chi\chi}(^{2s+1}L_J)$, that are associated with 
the kinematics of the $\chi\chi \to X_A X_B \to \chi\chi$ reaction, where
each of these kinematic terms multiplies a certain combination of the
place-holder coupling factors.
In particular, these kinematic contributions are generic in the sense that 
they apply to both the cases of external and internal Majorana and Dirac
fermions.

A specific diagram's contribution to the absorptive part of a particular
$\chi_{e_1}\chi_{e_2}\to X_A X_B \to\chi_{e_4}\chi_{e_3}$ 
MSSM 1-loop process is obtained by replacing the generic place-holder 
coupling factors with their actual expressions in the above described 
generic Majorana fermion $2\to 2$ scattering reactions.
Note that by choosing these coupling factors properly, all 
$\chi_{e_1}\chi_{e_2} \to X_A X_B \to \chi_{e_4}\chi_{e_3}$ processes with
external and internal Majorana or Dirac fermions can be covered, although
the kinematic contributions are calculated referring to the generic Majorana
fermion $2\to2$ scattering reaction. Hence, the absorptive part of the 
Wilson coefficient, which encodes the absorptive part 
of a $\chi_{e_1}\chi_{e_2} \to X_A X_B \to \chi_{e_4}\chi_{e_3}$
scattering reaction, with the incoming and outgoing two-particle states in a
$^{2s+1}L_J$ partial-wave configuration, can be written as
\begin{eqnarray}
\nonumber
&&
\hspace*{-2cm} 
  \hat{f}^{\, \chi_{e_1}\chi_{e_2} \to X_A  X_B\to \chi_{e_4}\chi_{e_3}}_{\lbrace e_1 e_2 \rbrace \lbrace e_4 e_3 \rbrace}(^{2s+1}L_J)
\\[0.2cm]\nonumber
 \ = \ 
 \frac{\pi \alpha_2^2}{M^2} ~
 \Biggl(
& &
 \sum\limits_{n} \sum\limits_{i_1, i_2}
       b^{\, \chi_{e_1}\chi_{e_2} \to X_A X_B\to \chi_{e_4}\chi_{e_3} }_{n, \, i_1 i_2} \
       B^{\, X_A X_B }_{n, \, i_1 i_2} (^{2s + 1}L_J)
\\\nonumber
& &+ 
 \sum\limits_{\alpha = 1}^{4} \sum\limits_{n} \sum\limits_{i_1, i_2}
       c^{(\alpha) \, \chi_{e_1}\chi_{e_2} \to X_A X_B\to \chi_{e_4}\chi_{e_3} }_{n, \, i_1 i_2} \
       C^{(\alpha) \,  X_A X_B }_{n, \, i_1 i_2} (^{2s + 1}L_J)
\\
& &+
 \sum\limits_{\alpha = 1}^{4} \sum\limits_{n} \sum\limits_{i_1, i_2}
       d^{(\alpha) \, \chi_{e_1}\chi_{e_2} \to X_A X_B\to \chi_{e_4}\chi_{e_3} }_{n, \, i_1 i_2} \
       D^{(\alpha) \, X_A X_B }_{n, \, i_1 i_2} (^{2s + 1}L_J)
 \Biggr)
 \ .
\label{eq:genericstructureWilson}
\end{eqnarray}
Here $\alpha_2 = g_2^2/4\pi$, where $g_2$ denotes the $SU(2)_L$ gauge
coupling. The sums in the first line on the right-hand side of
(\ref{eq:genericstructureWilson}) collect all contributions 
from selfenergy-amplitudes. Similarly, the
second (third) line gives the triangle- (box-) amplitudes' contributions.
We use the index $\alpha$ to enumerate expressions related to the four
different triangle- and box-amplitudes,\footnote{For identical particles 
$X_A = X_B$ the index $\alpha$ has to be taken from $1$ to $2$ only, 
see footnote~\ref{fn:XAXA}.} according to the labelling of the 
diagrams in Figs.~\ref{fig:generictriangles}--\ref{fig:genericboxes_fermions}.
Further, we indicate the kinematic factors of  the generic $2 \to 2$ 
Majorana fermion scattering
amplitudes within a given class and topology with capital letters
($ B^{}_{n, \, i_1 i_2}, C^{(\alpha)}_{n, \, i_1 i_2}, 
D^{(\alpha)}_{n, \, i_1 i_2}$).
These are the quantities that include the kinematics of the process
and hence encode the $^{2s+1}L_J$ partial-wave specific information.
The process-specific coupling factors that multiply the kinematic
factors are denoted with lowercase letters
$( b^{}_{n, \, i_1 i_2}, c^{(\alpha)}_{n, \, i_1 i_2}, 
d^{(\alpha)}_{n, \, i_1 i_2} )$. Depending on the type of the particles 
$X_A$ and $ X_B$ as well as the topology,
there is a fixed number of different coupling-factor expressions that 
can occur, together with the corresponding kinematic factors.
The different contributions are enumerated with the index $n$
in (\ref{eq:genericstructureWilson}) above.
Finally, in each of the processes there is a certain set of
particle species that can be exchanged in the $s$- or the $t$-channels of
the contributing amplitudes. These are labelled with the indices $i_1$ 
and $i_2$.

The generic structure of the Wilson coefficients in
(\ref{eq:genericstructureWilson}) suggests to give the coupling
factors and the kinematic factors separately. A recipe for the
construction of the coupling factors
$ b^{}_{n, \, i_1 i_2}, c^{(\alpha)}_{n, \, i_1 i_2}, d^{(\alpha)}_{n, \, i_1 i_2}$ in any
of the covered reactions is given in Sec.~\ref{sec:app_couplingfactors}.
Analytic results for the kinematic factors
$  B^{}_{n, \, i_1 i_2}, C^{(\alpha)}_{n, \, i_1 i_2}, 
D^{(\alpha)}_{n, \, i_1 i_2}$ for the leading-order $^1S_0$ and $^3S_1$ 
partial-wave configurations can be found 
in Sec.~\ref{sec:app_kinematicfactors}. These expressions depend on the 
masses of the external and internal  particles in a particular
$\chi_{e_1}\chi_{e_2} \to X_A X_B \to \chi_{e_4}\chi_{e_3}$ process.
However, the kinematic factors are generic in the sense that their form is the
same for all possible external two-body states $\chi_{e_1}\chi_{e_2}$ and
$\chi_{e_3}\chi_{e_4}$ of neutralinos or charginos and all $X_A X_B$ particles
within one of the classes $VV, VS, SS, ff$ or $\eta\overline\eta$.

The coupling and kinematic factors will depend on the supersymmetric particles'
mixing matrices and masses, respectively. We adopt the same notation as in
\cite{Rosiek:1989rs} and hence introduce the chargino and neutralino mixing
matrices $Z_\pm$ and $Z_N$ defined via
\begin{align}
 Z_-^T ~ M_{\chi^\pm} ~ Z_+ \ &= \
  \left(\begin{array}{cc}
   m_{\chi^+_1} & \\
   & m_{\chi^+_2}
  \end{array}\right) \ ,
\\
 Z_N^T ~ M_{\chi^0} ~Z_N \ &= \
  \left(\begin{array}{cccc}
   m_{\chi^0_1} & & & \\
   & m_{\chi^0_2} & & \\
   & & m_{\chi^0_3} & \\
   & & & m_{\chi^0_4} \\
  \end{array}\right) \ ,
\end{align}
where $M_{\chi^\pm}$ and $M_{\chi^0}$ denote the chargino and neutralino mass
matrices, respectively (for details regarding the mass matrix expressions refer
to \cite{Rosiek:1989rs}). $m_{\chi^+_j}, j = 1,2$,  and
$m_{\chi^0_i}, i = 1,\ldots 4$ indicate the masses in the mass eigenstate basis of
charginos and neutralinos.

In order to properly apply the formulas for coupling and kinematic factors
in Sec.~\ref{sec:app_couplingfactors} and Sec.~\ref{sec:app_kinematicfactors}
given a specific MSSM spectrum, it is important to note that the NRMSSM and
hence the analytic expressions for the Wilson coefficients explicitly rely on
the positivity of all mass parameters. This derives from the fact that 
the NRMSSM Lagrangian is obtained by extracting the high-energy fluctuations
(of the order of the particle mass) from the relativistic fields, which
yields the non-relativistic kinetic term $\mathcal L_\text{kin}$ shown in
(\ref{eq:kin}). For species other than the LSP, the procedure leads to the
mass-difference terms $(m_i-m_{\text{LSP}})$ in (\ref{eq:kin}). If any of
the $m_i$ in $\mathcal L_\text{kin.}$ is negative, then the corresponding
mass difference counts as ${\cal O}(m_{\text{LSP}})$, an indication
that the parametrization used to relate the relativistic and
non-relativistic fields for that particle species is not the appropriate
one. The simplest way to obtain the NRMSSM Lagrangian in case that the mass
$m_{\chi_{e_i}}$ of one or several of the external $\chi_{e_i}$ particles 
happens to be negative for a given MSSM spectrum, is to perform a field
redefinition of the corresponding MSSM fields that yields mass terms with
positive mass parameters. Such a field 
redefinition affects the chargino and neutralino
mixing matrices, which are mapped in the following way:
\begin{align}
 Z_\pm \ \ &\rightarrow \ \ \widetilde Z_\pm
 \ = \ 
 Z_\pm ~ \cdot ~
          \left(\begin{array}{cccc}
             \sqrt{\mbox{sgn}(m_{\chi^+_1})} & \\
             & \sqrt{\mbox{sgn}(m_{\chi^+_2})}
          \end{array}\right) \ ,
\\
 Z_N \ \ &\rightarrow \ \ \widetilde Z_N
 \ = \
 Z_N ~ \cdot ~
         \left(\begin{array}{cccc}
            \sqrt{\mbox{sgn}(m_{\chi^0_1})} & & & \\
            & \sqrt{\mbox{sgn}(m_{\chi^0_2})} & & \\
            & & \sqrt{\mbox{sgn}(m_{\chi^0_2})} & \\
            & & & \sqrt{\mbox{sgn}(m_{\chi^0_2})}
 \end{array}\right) \ .
\end{align}
(We define $\sqrt{-1}=i$.) The redefined mixing matrices  $\widetilde Z_\pm$ 
and $\widetilde Z_N$ as well as the corresponding positive mass parameters 
for all MSSM neutralino and chargino fields should be used within the 
expressions given in Sec.~\ref{sec:app_couplingfactors} and
Sec.~\ref{sec:app_kinematicfactors}.

\subsection{Coupling factors}
\label{sec:app_couplingfactors}
\begin{figure}[t]
\begin{center}
\includegraphics[width=0.95\textwidth]{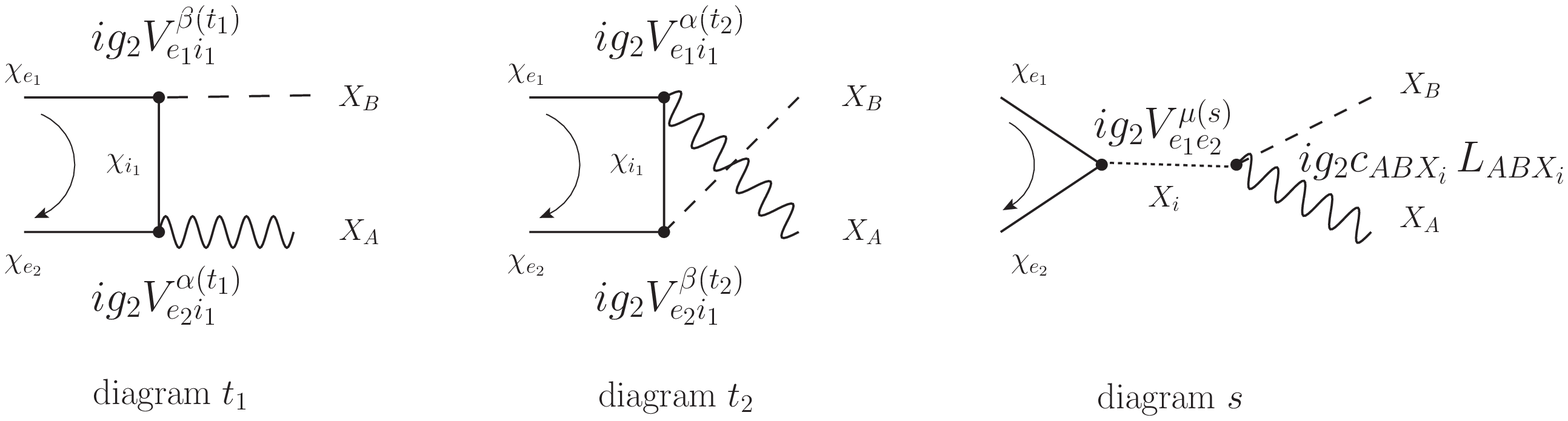}
\caption{ Generic tree-level amplitudes
          in $\chi\chi \to X_A X_B$ annihilations,
          referring to $VV, VS$ and $SS$-type final state particles $X_A X_B$.
          The generic form of $s$-channel exchange
          diagrams for $X_A X_B = \eta \bar\eta$ final states
          agrees with the $s$-channel diagram above.
          The vertex-factors $V^{\rho(d)}_{ei}$ are defined as
    $V^{\rho(d)}_{ei} = \gamma^\rho (r^{(d)}_{ei} + q^{(d)}_{ei}\gamma_5 )$,
          if attached to a three-point vertex with a gauge-boson
          (with Lorentz-index $\rho$), and
    $V^{\rho(d)}_{ei} = (r^{(d)}_{ei} + q^{(d)}_{ei}\gamma_5 )$,
          if associated with a vertex that involves a scalar particle
          $X_A$, $X_B$ or $X_i$.
          Here the expression $r^{(d)}_{ei}$($q^{(d)}_{ei}$) either
          denotes a vector or scalar (an axial-vector or pseudo-scalar)
          type of coupling factor.
          For the definition of $c_{ABX_i}$ and the Lorentz structures
          $L_{ABX_i}$ we refer to Tab.~\ref{tab:app_Labi_structures} below.}
\label{fig:genericamplitudes}
\end{center}
\end{figure}

\begin{figure}[t]
\begin{center}
\includegraphics[width=0.95\textwidth]{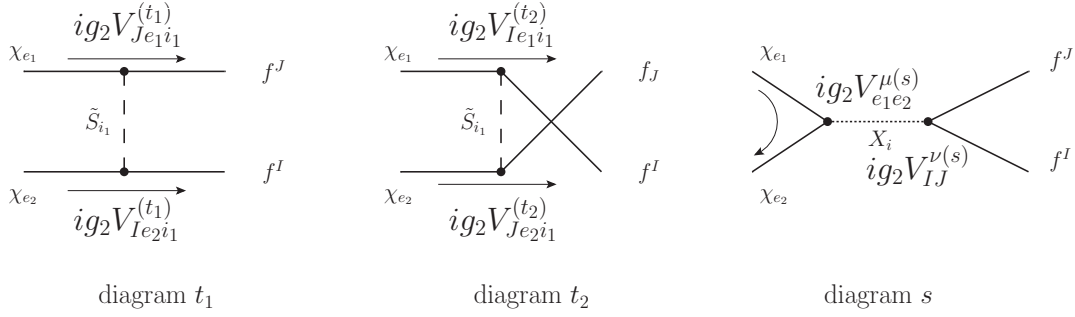}
\caption{ Generic tree-level amplitudes
          in $\chi\chi\to X_A X_B$ annihilations, with
          $X_A X_B = f^I f^J$.
          For the definition of $V^{\rho(d)}_{ei}$ see
          Fig.~\ref{fig:genericamplitudes}.
          The generic vertex factor $V^{(d)}_{Kei}$ is defined as
          $V^{(d)}_{Kei} = r^{(d)}_{Kei} + q^{(d)}_{Kei} \gamma_5$,
          such that the $r^{(d)}_{Kei}$ ($q^{(d)}_{Kei}$) denote 
          coupling factors of scalar (pseudo-scalar) type.
        }
\label{fig:genericamplitudes_fermions}
\end{center}
\end{figure}

By construction, the absorptive part
$\hat f^{\, \chi_{e_1}\chi_{e_2}\to X_AX_B \to \chi_{e_4}\chi_{e_3}}$
of an individual Wilson coefficient is associated with
the product
$\mathcal A^{(0)}_{\chi_{e_1}\chi_{e_2}\to X_A X_B} \times (\mathcal A^{(0)}_{\chi_{e_4}\chi_{e_3}\to X_A X_B})^*$
of Born-level annihilation amplitudes $\mathcal A^{(0)}$ related to
$\chi_{e_i}\chi_{e_j} \to X_A X_B$ reactions, integrated over the $X_A X_B$
two-particle phase space, see~(\ref{eq:defabsorptivepart}).
Each of the tree-amplitudes $\mathcal A^{(0)}_{\chi\chi\to X_A X_B}$ receives
contributions from diagrams with $t$-channel neutralino or chargino
exchange as well as from diagrams with $s$-channel Higgs-particle or
gauge-boson exchange, such as the generic diagrams shown in
Fig.~\ref{fig:genericamplitudes}.
%
%
In case of fermionic final states $X_A X_B$, instead of neutralino or chargino
$t$-channel exchange, $t$-channel sfermion-exchange occurs, as depicted in
Fig.~\ref{fig:genericamplitudes_fermions}.
%
%
Note, that in Fig.~\ref{fig:genericamplitudes} and
Fig.~\ref{fig:genericamplitudes_fermions} we again have established a specific
fermion flow, which in particular coincides with the convention for the
fermion flow associated with the incoming two particles in the 1-loop
amplitudes in Figs.~\ref{fig:genericselfenergy}--\ref{fig:genericboxes_fermions}.

A contribution to the amplitude 
$\mathcal A^{(0)}_{\chi\chi \to X_A X_B}$ involves a
product of two coupling factors, coming from the two vertices in the
tree-level diagrams. The generic form of these vertices is indicated
in Fig.~\ref{fig:genericamplitudes} and
Fig.~\ref{fig:genericamplitudes_fermions}.
It is especially convenient to write all vertex factors 
in any of the amplitudes contributing to the non-relativistic
$\chi \chi \to X_A X_B \to \chi \chi$ scattering-processes
as a combination of (axial-) vector or (pseudo-) scalar coupling factors,
instead of using left- and right-handed couplings, as it is common in
calculations related to the MSSM.
The reason for that is, that in the non-relativistic limit, either
the contributions to the annihilation amplitudes involving the axial-vector
(pseudo-scalar) coupling will be suppressed with respect to the corresponding
contributions related to the vector (scalar) coupling, or vice versa,
such that the use of (axial-) vector and (pseudo-) scalar couplings  allows
for a clearer understanding of leading and suppressed contributions in
the non-relativistic scattering regime that we aim to study.

Each of the coupling factors $b^{}_n, c^{(\alpha)}_n$ and $d^{(\alpha)}_n$ that
occur in (\ref{eq:genericstructureWilson}) is given by a product of 
two coupling factors, $r$ or $q$, arising in an individual diagram
in $\mathcal A^{(0)}_{\chi_{e_1}\chi_{e_2}\to X_A X_B}$, and the complex 
conjugate of another such two-coupling factor product originating from
$\mathcal A^{(0)}_{\chi_{e_4}\chi_{e_3}\to X_A X_B}$.
In the following, we give a recipe how to construct the coupling factors
in (\ref{eq:genericstructureWilson}) for a specific
$\chi_{e_1}\chi_{e_2} \to X_A X_B \to \chi_{e_4}\chi_{e_3}$ reaction, such
that taken together with the kinematic factors in
Sec.~\ref{sec:app_kinematicfactors}, they allow to determine the absorptive
part of the Wilson coefficients $\hat f$:
\begin{enumerate}
 \item Draw all tree-level diagrams that contribute to
 $\chi_{e_1}\chi_{e_2} \to X_A X_B$ and $\chi_{e_4}\chi_{e_3} \to X_A X_B$
 annihilation amplitudes, analogous to the generic diagrams sketched in
 Fig.~\ref{fig:genericamplitudes} or Fig.~\ref{fig:genericamplitudes_fermions}.
 In particular, assign the same fermion flow as indicated for the generic
 diagrams.
 \item Determine the process-specific (axial-) vector and/or (pseudo-) scalar
 coupling factors, that arise instead of the generic $q^{(d)}_{ei}$ or 
 $r^{(d)}_{ei}$
 place-holder expressions at the generic amplitudes' vertex factors.
 As the $\chi\chi \to X_A X_B$ processes may involve Majorana as well as Dirac
 fermions, and the latter involve a conserved fermion-number flow, note the
 following rules:
 \begin{itemize}
  \item[$a)$] If the direction of the fermion-number flow related to a Dirac
   particle coincides with the direction of the fermion flow 
   (fixed as in Fig.~\ref{fig:genericamplitudes} and
   Fig.~\ref{fig:genericamplitudes_fermions}), the $\chi\chi\to X_A X_B$
   process specific coupling factors at the vertices are directly
   deduced from the corresponding 
   interaction terms in the underlying Lagrangian.
   These coupling factors are given later in (\ref{eq:app_couplings_ncX}--\ref{eq:chifScouplings}).
  \item[$b)$] Otherwise, if the fermion-number flow is antiparallel to the
   indicated fermion flow, vector coupling factors at vertices attached to a
   Dirac fermion line, are given by a factor $-1$ times the expression for
   the vector coupling given in (\ref{eq:app_couplings_ncX}--\ref{eq:app_couplings_nnX}).
   Axial-vector, scalar and pseudo-scalar coupling factors are unchanged with
   respect to case $a)$ above.
 \end{itemize}
 \item Build all possible two-coupling factor products, including possible signs
 related to vector couplings, as far as the case in 2b) above applies,
 that can arise in each single diagram.
 \item Multiply each of the two-coupling factor products, that arise in the
 $\mathcal A^{(0)}_{\chi_{e_1} \chi_{e_2}\to X_A X_B}$ amplitude, with the complex
 conjugate of each of the two-coupling factor products, arising in
 $\mathcal A^{(0)}_{\chi_{e_4} \chi_{e_3}\to X_A X_B}$.
 As a result, all possible coupling factor combinations that can occur in
 $\hat f^{\, \chi_{e_1}\chi_{e_2} \to X_A X_B \to \chi_{e_4} \chi_{e_3}}$ are obtained.
\end{enumerate}
Rule $2b)$ arises in the following way for the case of diagram $s$ in Figs.~\ref{fig:genericamplitudes}-\ref{fig:genericamplitudes_fermions}:
according to our convention for the
fermion flow in Fig.~\ref{fig:genericamplitudes}, we obtain an expression
$-\overline{v}(p_1) \Gamma u(p_2)$ for the incoming particles' spinor chain
if the case under $2b)$ applies, where $\Gamma$ denotes the
involved Dirac-matrix structure. The minus sign accounts for our convention
for the order of the external fermion states. This expression can
be rewritten as
\begin{align}
 - \overline v(p_1) \Gamma u(p_2)
\ = \
 \overline v(p_2) \,C ~ \Gamma^T C^{-1} u(p_1) \ ,
\label{eq:app_rule_vectorcoupling}
\end{align}
wherein $C$ denotes the charge conjugation matrix. Using
\begin{align}
 C ~ \Gamma^T C^{-1}
\ = \
 \left\lbrace\begin{array}{l l}
{        -    \Gamma} & {\text{  for } \Gamma \ = \ \gamma_\mu \ ,} \\
{\phantom{-}  \Gamma} & {\text{  for } \Gamma \ = \ 1, \gamma_5, \gamma_\mu\gamma_5 \ ,}
             \end{array}\right.
\end{align}
the origin of the minus sign rules for vector couplings under $2b)$ above
becomes obvious. For diagrams with $t$-channel exchange, a similar derivation also confirms rule $2b)$.

Let us introduce the shorthand $a \tilde a$ to indicate the diagrams
$a$ and $\tilde a$ in the $\chi_{e_1}\chi_{e_2}\to X_A X_B$ and 
$\chi_{e_4}\chi_{e_3}\to X_A X_B$ processes, respectively, to which the coupling factors
in a specific coupling factor combination are related. Both $a$ and $\tilde a$
can be given by $s, t_1$ or $t_2$, see Figs.~\ref{fig:genericamplitudes}--\ref{fig:genericamplitudes_fermions}.
Coupling factor combinations originating from $s s$ lead to the
$b$ factors, that correspond to the generic selfenergy-amplitude in
Fig.~\ref{fig:genericselfenergy}.\footnote{Note, that in case of identical
particles $X_A = X_B$, all coupling factor expressions $b$ have to be multiplied
with a symmetry factor 1/2, which incorporates the symmetry factor associated
with the selfenergy amplitudes in case of identical particles $X_A = X_B$ in the
loop.}
We label coupling factor combinations, that originate from $t_1 s$, $s t_1$,
$t_2 s$ and $s t_2$ with the superscript $\alpha = 1,\ldots, 4$, respectively.
These coupling factor combinations, related to one $t$-channel and one $s$-channel
exchange diagram give rise to the $c^{(\alpha)}$ expressions in
(\ref{eq:genericstructureWilson}). The $\alpha = 1,\ldots ,4$ label-convention
for the specific coupling factor combinations allows to correctly allocate the
$c^{(\alpha)}$ to their corresponding generic triangle-amplitude
`triangle $\alpha$' in Fig.~\ref{fig:generictriangles}.
Coupling factor combinations originating from $t_1 t_2$, $t_1 t_1$, $t_2 t_1$ and
$t_2 t_2$ are labelled with superscript $\alpha = 1, \ldots, 4$, and give rise to
the $d^{(\alpha)}$ expressions. As in case of the $c^{(\alpha)}$, this convention
correctly assigns $d^{(\alpha)}$ expressions to their corresponding
`box $\alpha$' amplitude in Fig.~\ref{fig:genericboxes} or
Fig.~\ref{fig:genericboxes_fermions}.

We introduce the index $n$ in order to label the different coupling factor
combinations for a given fixed $a \tilde a$. Each individual $n$ is given
by a character-string, where  the $i$th character gives the type ($r$ or $q$)
of the coupling factor which is related to the $i$th
vertex in the particular diagram $a \tilde a$ in
Figs.~\ref{fig:genericselfenergy}--\ref{fig:genericboxes_fermions}.
The vertices of box-amplitudes are enumerated according to the respective
attached external particles $\chi_{e_i}, i = 1,\ldots,4$. In case of selfenergy-
and triangle-diagrams with inner vertices our vertex-enumeration convention is
from top to bottom and left to right.
Vertex factors of type $c_{ABX_i}$ are not specified in the corresponding string
$n$, because the nature of the particles $X_A$, $X_B$ and $X_i$ involved
in the diagram completely characterize that coupling. For triangles with
$X_A X_B = VV, VS$ or $SS$, for example, the index $n$ will range over strings 
$rrr, qqr, ...$, where the characters $r$ or $q$ indicate the type of coupling
of the external particles to the $X_A X_B$ pair and to the single $s$-channel
exchanged particle species, see Fig.~\ref{fig:generictriangles}.

To further specify the coupling factor combinations for given $a \tilde a$ and
$n$, we use the labels $i_1$ and $i_2$ to indicate the particle species that
are internally exchanged in diagrams $a$ and $\tilde a$.
Therewith, the coupling factor combinations $b_{n\,i_1 i_2}$,
$c^{(\alpha)}_{n\,i_1 i_2}$ and $d^{(\alpha)}_{n\,i_1 i_2}$ that should enter in
(\ref{eq:genericstructureWilson}), together with the generic kinematic factor
expressions given in App.~\ref{sec:app_kinematicfactors}, can be unambiguously
determined.

In order to completely fix our conventions, we summarize in the following
the expressions for the (axial-) vector and (pseudo-) scalar coupling factors,
that arise in the three-point interactions of charginos and neutralinos with
SM and Higgs particles. The definitions of the coupling factors assume that
we take $\chi^+_i$ to be the particle and  $\chi^-_i$ its
anti-particle, such that the Dirac fermion number flow, indicated by the arrow
on the Dirac fermion line for a chargino, will always refer to the
direction of $\chi^+_i$ flow. The latter convention agrees with that of Rosiek
\cite{Rosiek:1989rs}.
%
\begin{figure}[t]
\begin{center}
\includegraphics[width=0.95\textwidth]{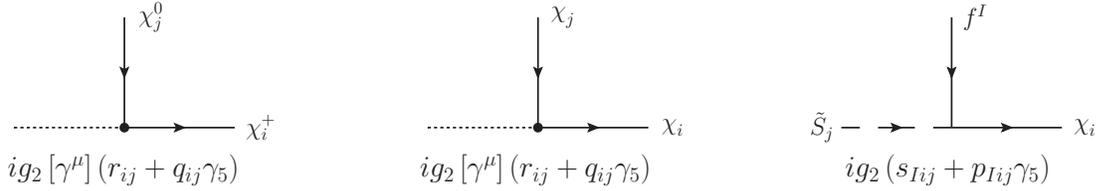}
\caption{ Generic form of the vertex factors in three-point interactions
          of neutralinos and charginos with SM and Higgs particles,
          upon which our definition of the (axial-) vector and (pseudo-) scalar
          coupling factors given in the text is based.}
\label{fig:generic_vertices}
\end{center}
\end{figure}
%

The generic form for the vertex factor, that describes the 3-point interaction
of an incoming neutralino $\chi^0_j$, an outgoing chargino $\chi^+_i$ and
either an incoming charged Higgs particle $G^+$ or $H^+$ or an incoming
$W^+$-boson is given in the left-most diagram 
in Fig.~\ref{fig:generic_vertices}.
Note that the gamma matrix $\gamma^\mu$ in the vertex factor has to be omitted
in case of interactions with the charged Higgs particles. The specific scalar
and pseudo-scalar or vector and axial-vector coupling factors, that have to be
replaced for the generic $r_{ij}$ and $q_{ij}$ couplings therein read
\begin{align}
\nonumber
 s^{H_m^+}_{ij} \, ( p^{H^+_m}_{ij}   ) \ = \
&
           - \frac{1}{2} \Bigl[
              \phantom{-} Z_H^{2m} \left(
                     \widetilde Z_N^{4j *} \widetilde Z_+^{1i *}
                   + \frac{1}{\sqrt 2} \widetilde Z_+^{2i *}
                     ( \widetilde Z_N^{2j *} + \tan \theta_W \widetilde Z_N^{1j *})
                                 \right)
\\\nonumber
&
         \phantom{- \frac{1}{2} \ \ \Bigl[}
                       \pm Z_H^{1m} \left(
                     \widetilde Z_N^{3j} \widetilde Z_-^{1i}
                   - \frac{1}{\sqrt 2} \widetilde Z_-^{2i}
                     ( \widetilde Z_N^{2j} + \tan \theta_W \widetilde Z_N^{1j})
                                 \right)
                        \ \Bigr] \ ,
\\\nonumber
 v^W_{ij} \ = \
&
 \phantom{-} \frac{1}{2} \left(
                      \widetilde Z_N^{2j *} \widetilde Z_-^{1i}
                   + \widetilde Z_N^{2j} \widetilde Z_+^{1i *}
                   + \frac{1}{\sqrt 2} \widetilde Z_N^{3j *} \widetilde Z_-^{2i}
                   - \frac{1}{\sqrt 2} \widetilde Z_N^{4j} \widetilde Z_+^{2i *}
                        \right) \ ,
\\
 a^W_{ij} \ = \
&
 \phantom{-} \frac{1}{2} \left(
                      \widetilde Z_N^{2j *} \widetilde Z_-^{1i}
                    - \widetilde Z_N^{2j} \widetilde Z_+^{1i *}
                    + \frac{1}{\sqrt 2} \widetilde Z_N^{3j *} \widetilde Z_-^{2i}
                    + \frac{1}{\sqrt 2} \widetilde Z_N^{4j} \widetilde Z_+^{2i *}
                        \right) \ ,
\label{eq:app_couplings_ncX}
\end{align}
where $H_1^+\equiv H^+$ and $H_2^+\equiv G^+$, and the mixing matrices are
defined as in Ref.~\cite{Rosiek:1989rs}.
The generic form of the three point interaction of either two neutralinos
or two charginos with an electrically neutral gauge boson or Higgs particle
is depicted in the second diagram in Fig.~\ref{fig:generic_vertices}.
Again, the gamma-matrix $\gamma^\mu$ has to be omitted in the vertex factor
if the corresponding reaction refers to interactions with the neutral
Higgs particles.
In the case of an incoming $\chi^+_j$ and an outgoing $\chi^+_i$, the
(axial-) vector and \mbox{(pseudo-)} scalar couplings, that encode interactions 
with the neutral scalar and pseudo-scalar Higgs particles ($h^0, H^0, G^0, A^0$),
the $Z$-boson or the photon are given by the following expressions:
\begin{align}
\nonumber
 s^{H^0_m}_{ij} \, ( p^{H^0_m}_{ij} ) \ = \
&
            - \frac{1}{2 \sqrt 2} \Bigl[
                   \, Z_R^{1m} \Bigl( \widetilde Z_-^{2j *} \widetilde Z_+^{1i *} \pm \widetilde Z_-^{2i} \widetilde Z_+^{1j} \Bigr)
                    + Z_R^{2m} \Bigl( \widetilde Z_-^{1j *} \widetilde Z_+^{2i *} \pm \widetilde Z_-^{1i} \widetilde Z_+^{2j} \Bigr)
                               \, \Bigr] \ ,
\\\nonumber
 s^{A^0_m}_{ij} \, ( p^{A^0_m}_{ij} ) \ = \
&
            - \frac{i}{2 \sqrt 2} \Bigl[
                   \, Z_H^{1m} \Bigl( \widetilde Z_-^{2j *} \widetilde Z_+^{1i *} \mp  \widetilde Z_-^{2i} \widetilde Z_+^{1j} \Bigr)
                    + Z_H^{2m} \Bigl( \widetilde Z_-^{1j *} \widetilde Z_+^{2i *} \mp  \widetilde Z_-^{1i} \widetilde Z_+^{2j} \Bigr)
                              \,   \Bigr] \ ,
\\\nonumber
 v^Z_{ij} \ = \
&
            - \frac{1}{4 c_W} \Bigl(
                    \widetilde Z_-^{1i} \widetilde Z_-^{1j *}
                  + \widetilde Z_+^{1i *} \widetilde Z_+^{1j} 
                  + 2  ( c_W^2 - s_W^2 ) \delta_{ij}
                        \Bigr) \ ,
\\\nonumber
 a^Z_{ij} \ = \
&
            \phantom{-} \frac{1}{4 c_W} \Bigl(
                    \widetilde Z_+^{1i *} \widetilde Z_+^{1j} 
                  - \widetilde Z_-^{1i} \widetilde Z_-^{1j *}
                        \Bigr) \ ,
\\\nonumber
 v^\gamma_{ij} \ = \
&
           - s_W \delta_{ij} \ ,
\\
 a^\gamma_{ij} \ = \
&
           \phantom{-} 0 \ ,
\label{eq:app_couplings_ccX}
\end{align}
where $H_1^0 \equiv H^0$, $H_2^0 \equiv h^0$ and $A_1^0\equiv A^0$, $A_2^0\equiv G^0$.
Finally, three-point interactions of an incoming $\chi^0_j$ and an
outgoing $\chi^0_i$ with a (pseudo-) scalar Higgs particle or the $Z$-boson
involve the following (axial-) vector or (pseudo-) scalar coupling factors
\begin{align}
\nonumber
 s^{(0) , H^{0}_m}_{ij} \, ( p^{(0), H^{0}_m}_{ij}   )  = \
&
            \frac{1}{4} \,\Bigl[ \,
                         \left(
                    Z_R^{2m} \, \widetilde Z_N^{4i *}
                   - Z_R^{1m} \, \widetilde Z_N^{3i *}
                         \right)
                         \left(
                  \widetilde Z_N^{2j *} - \tan\theta_W \, \widetilde Z_N^{1j *}
                         \right)
                       + \left( i \leftrightarrow j \right)\,
                    \Bigr]
                    \pm c.c.\,,
\\\nonumber
 s^{(0), A^{0}_m}_{ij}  (  p^{(0), A^{0}_m}_{ij}  )  = \
&
            \frac{i}{4} \,\Bigl[\,
                         \left(
                  \, Z_H^{2m} \, \widetilde Z_N^{4i *}
                   - Z_H^{1m} \, \widetilde Z_N^{3i *}
                         \right)
                         \left(
                  \widetilde Z_N^{2j *} - \tan\theta_W \, \widetilde Z_N^{1j *}
                          \right)
                       + \left( i \leftrightarrow j \right)\,
                    \Bigr]
                    \pm c.c. \, ,
\\
 v^{(0), Z}_{ij}  ( a^{(0), Z}_{ij})  = \
&
            \frac{1}{4 c_W} \Bigl(
                 \, \widetilde Z_N^{3i} \widetilde Z_N^{3j *} 
                  - \widetilde Z_N^{4i} \widetilde Z_N^{4j *}
                         \mp (i \leftrightarrow j) \
                        \Bigr) \ .
\label{eq:app_couplings_nnX}
\end{align}
The (axial-) vector and (pseudo-) scalar coupling factors in
(\ref{eq:app_couplings_ccX}) and (\ref{eq:app_couplings_nnX}), which are all
related to interactions with neutral SM and Higgs particles, satisfy
\begin{align}
 \nonumber
v^*_{ij} \ &=  \phantom{-} v_{ji} \ , 
&a^*_{ij} \ =  \phantom{-} a_{ji} \ ,
\\
s^*_{ij} \ &=  \phantom{-} s_{ji} \ ,
&p^*_{ij} \ =  - p_{ji} \ .
\end{align}
as a consequence of the hermiticity of the underlying SUSY Lagrangian.

The generic form of the vertex factor for three-point interactions of a
neutralino or chargino with a SM fermion 
and a sfermion is given in 
the right-most diagram in Fig.~\ref{fig:generic_vertices}. In case of
interactions of an incoming SM fermion $f^I$ with a sfermion
$\tilde S_j$ and an outgoing neutralino $\chi^0_i$, 
the specific (pseudo-) scalar
couplings, that have to be replaced for the generic $s_{Iij}$ and $p_{Iij}$
expressions in Fig.~\ref{fig:generic_vertices} read
\begin{align}
\nonumber
 s^{u \tilde U}_{Iij}  ( p^{u \tilde U}_{Iij} ) \ = \
&
        \frac{1}{\sqrt{2}} ~ q_u ~ \tan\theta_W \widetilde Z_N^{1i *} Z_U^{(I+3)j *}
           -\frac{m_u^I}{2\sqrt{2} \sin\beta M_W} 
                     \left(
                       \widetilde Z_N^{4i *} Z_U^{Ij *}
                  \pm  \widetilde Z_N^{4i} Z_U^{(I+3)j *}
                     \right)
\\\nonumber
&
           \mp \frac{1}{\sqrt{2}}
                      \left(
                        T_u \widetilde Z_N^{2i}
                      + (q_u - T_u) \widetilde Z_N^{1i} \tan\theta_W 
                       \right) Z_U^{Ij *}
                         \ ,
\\
\nonumber
 s^{d \tilde D}_{Iij}  ( p^{d \tilde D}_{Iij} ) \ = \
&
         \frac{1}{\sqrt{2}} ~ q_d ~ \tan\theta_W \widetilde Z_N^{1i *} Z_D^{(I+3)j}
           -\frac{m_d^I}{2\sqrt{2} \cos\beta M_W} 
                     \left(
                        \widetilde Z_N^{3i *} Z_D^{Ij}
                    \pm \widetilde Z_N^{3i} Z_D^{(I+3)j}
                     \right)
\\\nonumber
&
           \mp \frac{1}{\sqrt{2}}
                      \left(
                        T_d \widetilde Z_N^{2i}
                     + (q_d - T_d) \widetilde Z_N^{1i} \tan\theta_W 
                       \right) Z_D^{Ij}
                         \ ,
\\
\nonumber
 s^{\nu \tilde \nu}_{Iij}  ( p^{\nu \tilde \nu}_{Iij} ) \ = \
&
           \mp \frac{1}{2\sqrt{2}}
                     \left(
                       \widetilde Z_N^{2i} - \widetilde Z_N^{1i} \tan\theta_W
                     \right)  Z_\nu^{Ij *}
                         \ ,
\\
\nonumber
 s^{l \tilde L}_{Iij}  ( p^{l \tilde L}_{Iij} ) \ = \
&
            \frac{1}{\sqrt{2}} ~ q_l ~ \tan\theta_W \widetilde Z_N^{1i *} Z_L^{(I+3)j}
           -\frac{m_l^I}{2\sqrt{2} \cos\beta M_W} 
                     \left(
                       \widetilde Z_N^{3i *} Z_L^{Ij}
                  \pm  \widetilde Z_N^{3i} Z_L^{(I+3)j}
                     \right)
\\
&
           \mp \frac{1}{\sqrt{2}}
                      \left(
                        T_l \widetilde Z_N^{2i}
                      + (q_l - T_l) \widetilde Z_N^{1i} \tan\theta_W 
                       \right) Z_L^{Ij}
                         \ .
\label{eq:app_chi0fScouplings}
\end{align}
$I = 1,2,3$ denotes the generation index for the fermions, and 
$j = 1,\ldots, 6$ labels the sfermion states ($j=1,2,3$ in case of 
sneutrinos $\tilde\nu_j$). $T_{f}$ and $q_f$ are defined as
\begin{align}
\nonumber
T_{u}\ &= \ - T_d \ = \ - T_l = \ \frac{1}{2} \ ,
\\
 q_u \ &= \ \frac{2}{3} \ , \ \ q_d \ = \ -\frac{1}{3}
 \ , \ \ q_l \ = \ -1 \ .
\end{align}
The superscripts, $f \tilde S$, on the couplings in
(\ref{eq:app_chi0fScouplings}), refer to the fermion~($f$)- and
sfermion~($\tilde S$)-type involved in the underlying interaction.
In case of chargino-fermion-sfermion interactions we find (a sum over repeated
indices is implicit)
\begin{align}
\nonumber
 s^{u \tilde D}_{Iij}  ( p^{u \tilde D}_{Iij} ) \ = \
&
           \frac{m_u^I}{2\sqrt{2} \sin\beta M_W} 
                        K^{IJ *} \widetilde Z_+^{2i *} Z_D^{Jj}
           ~\pm \frac{m_d^J}{2\sqrt{2} \cos\beta M_W}
                        K^{IJ *} \widetilde Z_-^{2i} Z_D^{(J+3)j}
\\\nonumber
&
            \mp \frac{1}{2} K^{IJ *} \widetilde Z_-^{1i} Z_D^{Jj}
                         \ ,
\\
\nonumber
 s^{d \tilde U}_{Iij}  ( p^{d \tilde U}_{Iij} ) \ = \
&
           \frac{m_d^I}{2\sqrt{2} \cos\beta M_W} 
                        K^{JI} \widetilde Z_-^{2i *} Z_U^{Jj *}
           ~\pm \frac{m_u^J}{2\sqrt{2} \sin\beta M_W}
                        K^{JI} \widetilde Z_+^{2i} Z_U^{(J+3)j *}
\\\nonumber
&
            \mp \frac{1}{2} K^{JI} \widetilde Z_+^{1i} Z_U^{Jj *}
                         \ ,
\\\nonumber
 s^{\nu \tilde L}_{Iij}  ( p^{\nu \tilde L}_{Iij} ) \ = \
&
           \pm \frac{m_l^I}{2\sqrt{2} \cos\beta M_W} \widetilde Z_-^{2i} Z_L^{(I+3)j}
           ~\mp \frac{1}{2} \widetilde Z_-^{1i} Z_L^{Ij}
                         \ ,
\\
 s^{l \tilde \nu}_{Iij}  ( p^{l \tilde \nu}_{Iij} ) \ = \
&
           \frac{m_l^I}{2\sqrt{2} \cos\beta M_W} 
                        \widetilde Z_-^{2i *} Z_\nu^{Jj *}
           \mp \frac{1}{2} \widetilde Z_+^{1i} Z_\nu^{Ij *}
                         \ .
\label{eq:chifScouplings}
\end{align}
The coupling factors with $f\tilde S = u \tilde D, \nu \tilde L$ refer to the
interaction of an incoming up-type quark~($u^I$) or neutralino~($\nu^I$) with a
$\tilde D_j$- or $\tilde L_j$-sfermion and an outgoing $\chi^+_i$. In case of
$f\tilde S = d \tilde U, l\tilde\nu$, the coupling factors in
(\ref{eq:chifScouplings}) are related to interactions of an incoming down-type
quark~($d^I$) or lepton~($l^I$) with an $\tilde U_j$- or $\tilde\nu_j$-sfermion
and an outgoing $\chi^{+\,C}_i$ (denoting the charge-conjugate field of $\chi^+_i$, see \cite{Rosiek:1989rs}).

For the specific $c_{ABX_i}$ factors that emerge at the three-point vertex of
the $X_A X_B$ particle pair with the single $s$-channel exchanged particle $X_i$
in Fig.~\ref{fig:genericamplitudes}, we refer to the Feynman rules in
\cite{Rosiek:1989rs}: a specific $c_{ABX_i}$ is obtained as the factor
that multiplies the structure $i g_2 L_{ABX_i}$ in the respective
Feynman rule therein.
The generic forms of the Lorentz structures $L_{ABX_i}$ are
collected in Tab.~\ref{tab:app_Labi_structures}.
\begin{table}
\centering
\begin{tabular}{c | c}
 \hline
 $X_A X_B X_i$
&
 $L_{ABX_i}$
  \\
 \hline\hline
 $V_\alpha V_\beta V_\mu$
&
 $g^{\alpha\beta} (k_A - k_B)^\mu
  + g^{\mu\alpha} (k_i - k_A)^\beta
  + g^{\beta\mu} (k_B - k_i)^\alpha$
  \\
 $V_\alpha V_\beta\, S$
&
 $M_W ~ g^{\alpha\beta}$
  \\
 $V_\alpha\, S\, S$
&
 $(k_B - k_i)^\alpha$
  \\
 $S\, S\, V_\mu$
&
 $(k_B - k_A)^\mu$
  \\
 $S\, S\, S$
&
 $M_W$
  \\
 $\eta\, \overline\eta\, V_\mu$
&
 $k_B^\mu$
 \\
 $\eta\, \overline\eta~ S$
&
 $M_W$
\\
 \hline
\end{tabular}
\caption{The generic form of the Lorentz structures $L_{ABX_i}$,
that are part of the Feynman rule $i g_2 c_{ABX_i} \, L_{ABX_i}$ for the
$X_A X_B X_i$ three-point vertex in Fig~\ref{fig:genericamplitudes}.
We assume all four-momenta, $k_A, k_B, k_i$, to be outgoing at the vertex.
The case of $X_A X_B X_i = V_\alpha S\, V_\mu$ is trivially related to the case
$V_\alpha V_\beta\, S$.}
\label{tab:app_Labi_structures}
\end{table}
Finally, (axial-) vector and (pseudo-) scalar coupling factors $r$ and $q$
of two SM fermions with a gauge- or Higgs-boson
(see Fig.~\ref{fig:genericamplitudes_fermions}) can be directly taken from the
corresponding Feynman rules in \cite{Rosiek:1989rs}.

In order to illustrate how the above rules should be applied, let us consider 
an example. Suppose we wish to know the coupling factors
$c^{(\alpha)}_{n,\, i_1 V}$, $\alpha=1,\dots 4$, of diagrams 
with $s$-channel exchange of a $Z$-boson for the
$\chi^-_{e_1} \chi^+_{e_2} \to W^+ G^- \to \chi^0_{e_4} \chi^0_{e_3}$ processes.
Following the recipe above, we draw all tree-level diagrams for the 
$\chi^-_{e_1} \chi^+_{e_2} \to W^+ G^-$ as well as the 
$\chi^0_{e_4} \chi^0_{e_3} \to W^+ G^-$ process and assign the same
fermion flow as given in the corresponding generic diagrams,
Fig.~\ref{fig:genericamplitudes}.
Referring to that fixed fermion flow, we determine the following vertex factors
in the diagrams $t_1$ and $s$,
associated with tree-level $\chi^-_{e_1} \chi^+_{e_2} \to W^+ G^-$ annihilations:
\begin{align}
\nonumber
 V^{\beta(t_1)}_{e_1 i_1} \ &= \ s^G_{e_1 i_1} + p^G_{e_1 i_1} \gamma_5 \ ,
&
 V^{\alpha(t_1)}_{e_2 i_1} \ &= \ \gamma^\alpha \left(
                              - v^{W *}_{e_2 i_1} + a^{W *}_{e_2 i_1} \gamma_5
                                        \right) \ ,
 \\
 V^{\mu(s)}_{e_1 e_2} \ &= \ \gamma^\mu \left(
                              - v^Z_{e_1 e_2} + a^Z_{e_1 e_2} \gamma_5
                                   \right) \ ,
&
 c_{WGZ} \ &= \ - \frac{s_W^2}{c_W^2} \ .
\label{eq:app_example_couplings1}
\end{align}
The coupling factors are those from (\ref{eq:app_couplings_ncX}) 
and (\ref{eq:app_couplings_ccX}).
Note that there is no $t$-channel exchange diagram $t_2$ for the above process,
as it is forbidden by charge conservation.
Further, note that the sign in front of the vector-coupling factor in
$V^{\alpha(t_1)}_{e_2 i_1}$ and $V^{\mu(s)}_{e_1 e_2}$ follows from rule $2b)$ above.
In case of diagrams contributing to $\chi^0_{e_4} \chi^0_{e_3} \to W^+ G^-$ we
find
\begin{align}
\nonumber
 V^{\beta(t_1)}_{e_4 i_1} \ &= \ s^G_{i_1 e_4} + p^G_{i_1 e_4} \gamma_5 \ ,
&
 V^{\alpha(t_1)}_{e_3 i_1} \ &= \ \gamma^\alpha \left(
                               v^{W *}_{i_1 e_3} + a^{W *}_{i_1 e_3} \gamma_5
                                     \right) \ ,
 \\
\nonumber
 V^{\alpha(t_2)}_{e_4 i_1} \ &= \ \gamma^\alpha \left(
                             - v^{W *}_{i_1 e_4} + a^{W *}_{i_1 e_4} \gamma_5 
                                      \right) \ ,
&
 V^{\beta(t_2)}_{e_3 i_1} \ &= \ s^G_{i_1 e_3} + p^G_{i_1 e_3} \gamma_5 \ ,
 \\
 V^{\mu(s)}_{e_4 e_3} \ &= \ \gamma^\mu \left(
                               v^{(0)Z}_{e_3 e_4} + a^{(0)Z}_{e_3 e_4} \gamma_5 
                                     \right) \ ,
&
 c_{WGZ} \ &= \ - \frac{s_W^2}{c_W^2} \ .
\label{eq:app_example_couplings2}
\end{align}
To obtain the building blocks for the non-vanishing $c^{(\alpha)}_{n,\, i_1 V}$
with $\alpha = 1$, one has to combine the coupling factor expressions in the
first row of (\ref{eq:app_example_couplings1})
(the factors related to diagram $t_1$ in $\chi_{e_1}^-\chi_{e_2}^+\to W^+ G^-$
annihilations) with the coupling factor expressions in the last row of
(\ref{eq:app_example_couplings2}) (expressions originating from diagram $s$ in
$\chi_{e_4}^0\chi_{e_3}^0\to W^+ G^-$), as $\alpha = 1$ refers to the $t_1 s$
product of tree-level diagrams.
Similarly, for $\alpha = 2$ and $4$, one has to build the combinations of
expressions referring to $s t_1$ and $s t_2$. Therefore, the building-blocks for
the non-vanishing $c^{(\alpha)}_{n\, i_1 V}$ related to single $s$-channel $V=Z$
exchange read:
\begin{align}
\label{eq:app_example_couplings_alpha1}
&\alpha = 1 :
 &&\{
 \{s^G_{e_1 i_1}, \ p^G_{e_1 i_1} \} \ , \ \
 \{-v^{W *}_{e_2 i_1}, \ a^{W *}_{e_2 i_1} \} \ , \ \
 \{v^{(0) Z *}_{e_3 e_4}, \ a^{(0) Z *}_{e_3 e_4} \} \ , \ \
 \{ - \frac{s_W^2}{c_W^2} \}
 \} \ \ ,
\\
&\alpha = 2 :
 &&\{
 \{-v^Z_{e_1 e_2}, \ a^Z_{e_1 e_2} \} \ , \ \
 \{v^W_{i_1 e_3}, \ a^W_{i_1 e_3} \} \ , \ \
 \{s^{G *}_{i_1 e_4}, \ p^{G *}_{i_1 e_4} \} \ , \ \
 \{ - \frac{s_W^2}{c_W^2} \}
 \} \ \ ,
\\\label{eq:app_example_couplings_alpha4}
&\alpha = 4 :
 &&\{
 \{-v^Z_{e_1 e_2}, \ a^Z_{e_1 e_2} \} \ , \ \
 \{s^{G *}_{i_1 e_3}, \ p^{G *}_{i_1 e_3} \} \ , \ \
 \{-v^W_{i_1 e_4}, \ a^W_{i_1 e_4} \} \ , \ \
 \{ - \frac{s_W^2}{c_W^2} \}
 \} \ \ .
\end{align}
In selecting one element from each of the above given subsets 
and multiplying the selected elements for fixed $\alpha$ with each 
other, the $c^{(\alpha)}_{n,\, i_1 V}$ expressions in
$\chi^-_{e_1}\chi^+_{e_2}\to W^+ G^- \to \chi^0_{e_4} \chi^0_{e_3}$ reactions
are obtained. Proceeding in that way, we obtain eight 
different coupling factor combinations for fixed $\alpha$, that are labelled
with index $n$. Following our convention for this label, $n$ ranges over
$n = rrr, rrq, rqr, qrr, rqq, qrq, qqr, qqq$. The $c^{(2)}_{qqr,\, i_1 V}$
expression, for example, reads
\begin{align}
c^{(2)}_{qqr, \, i_1 V} \ = \
 -\frac{s_W^2}{c_W^2} ~ a^Z_{e_1 e_2} ~ a^W_{i_1 e_3} ~ s^{G *}_{i_1 e_4} \ .
\end{align}

\subsection{Kinematic factors}
\label{sec:app_kinematicfactors}
The kinematic factor expressions that refer to a specific
$\chi_{e_1} \chi_{e_2} \to X_A X_B \to \chi_{e_4} \chi_{e_3}$ 
scattering reaction depend on the external particles' 
mass scales $m$, $\overline m$
and $M = m + \overline m$. We remind the reader of our convention
(see Sec.~\ref{subsec:massdiffexp})
\begin{align}
\nonumber
 m \ =& \ \frac{m_{e_1} + m_{e_4}}{2} \ ,
&
 \overline m \ = \ \frac{m_{e_2} + m_{e_3}}{2} \ ,
\\
 \delta m \ =& \ \frac{m_{e_4 }- m_{e_1}}{2} \ ,
&
 \delta \overline m \ = \ \frac{m_{e_3} - m_{e_2}}{2} \ .
\end{align}
Further recall that we expand the scattering amplitudes in $\delta m$, 
$\delta \overline m$ and count these quantities as ${\cal O}(v^2)$. 
Hence, for the leading-order $S$-wave results presented below, the 
mass differences $\delta m = \delta \overline m = 0$, such that
there are only two mass scales, $m$ and $\overline m$, left, which
characterize the external chargino or neutralino states. The masses of the 
particles $X_A$ and $X_B$ will be denoted with $m_A$ and $m_B$.
Let us introduce the general notation $\widehat m_a$ for the
rescaling of any mass $m_a$ in units of the mass scale $M$,
\begin{align}
 \rescm m_a \ = \ \frac{m_a}{M} \ .
\end{align}
Define the dimensionless quantities
\begin{align}
\nonumber
 \Delta_{AB} \ =& \ ~\widehat m_A^2 - \widehat m_B^2 \ ,
\\
 \beta \ =& \
  ~\sqrt{ 1 - 2 ~ (\widehat m_A^2 + \widehat m_B^2) + \Delta_{AB}^2 } \ \ ,
\end{align}
where in case that $X_A = X_B$, $\beta$ is the leading-order term in the
expansion of the velocity
of the $X_A X_B$ particle pair in the non-relativistic momenta and mass
differences.
The expansion of single $s$-channel (gauge or Higgs boson $X_i$) exchange
propagators in $\delta m$, $\delta \overline m$ and the non-relativistic
3-momenta leads to the following denominator-structure at leading order:
\begin{align}
 P^s_i \ = \ 1 - \widehat m_i^2 \ .
\end{align}
Similarly, the leading-order expansion of  $t$- and $u$-channel 
gaugino and sfermion propagators in
$\delta m$, $\delta \overline m$ and the non-relativistic 3-momenta gives rise
to the denominator-structures 
\begin{align}
\nonumber
 P_{i\,AB} \ =& \
         \rescm{m} ~ \rescm{\overline m} + \rescm m_i^2
       - \rescm m ~ \rescm m_A^2
       - \rescm{\overline m} ~ \rescm m_B^2 \ ,
\\
 P_{i\,BA} \ =& \
         P_{i\,AB} \ \vert_{ A \leftrightarrow B} \ .
\end{align}
Using the above definitions, the kinematic factors for the 
leading order $S$-wave 
Wilson coefficients related to the selfenergy-topology in
Fig.~\ref{fig:genericselfenergy} are conveniently written as
\begin{align}
B^{\, X_A  X_B}_{n, \, i_1 i_2}(^{2s+1}S_J) \ = \
  \frac{ \beta}{ P^s_{i_1} \, P^s_{i_2}} \ 
  \tilde B^{\, X_A X_B}_{n, \, i_1 i_2}(^{2s+1}S_J) \ ,
\label{eq:app_B_generic}
\end{align}
where the labels $i_1$ and $i_2$ refer to the particle species that are 
exchanged in the left and right $s$-channel propagator. As generically either
gauge-boson~($V$) or Higgs~($S$) $s$-channel exchange occurs 
in the processes under consideration, the combination $i_1 i_2$ is given by
$i_1 i_2 = VV, VS, SV, SS$. Kinematic factors arising from 
the triangle-topologies shown in Fig.~\ref{fig:generictriangles} 
have the following generic form
\begin{align}
\nonumber
C^{(\alpha)\, X_A  X_B}_{n, \, i_1 X}(^{2s+1}S_J) \ =& \
  \frac{ \beta}{ P_{i_1 AB} \, P^s_{X}} \ 
  \tilde C^{(\alpha)\, X_A X_B}_{n, \, i_1 X}(^{2s+1}S_J)  
\qquad \alpha=1,2 \quad ,
\\
C^{(\alpha)\, X_A  X_B}_{n, \, i_1 X}(^{2s+1}S_J) \ =& \
  \frac{ \beta}{ P_{i_1 BA} \, P^s_{X}} \ 
  \tilde C^{(\alpha)\, X_A X_B}_{n, \, i_1 X}(^{2s+1}S_J) 
\qquad \alpha=3,4 \quad .
\label{eq:app_C_generic}
\end{align}
The index $i_1$ in the above expressions is related to the $t$- or $u$-channel
exchanged particle species, whereas the subscript-index $X$ indicates the type
of the single $s$-channel exchanged particle-species, $X = V,S$.
Finally, kinematic factors associated with the box-topologies 
of Fig.~\ref{fig:genericboxes} and
Fig.~\ref{fig:genericboxes_fermions} generically read
\begin{align}
\nonumber
D^{(1)\, X_A  X_B}_{n, \, i_1 i_2}(^{2s+1}S_J) \ = \
  \frac{ \beta}{ P_{i_1 AB} \, P_{i_2 BA}} \ 
  \tilde D^{(1)\, X_A X_B}_{n, \, i_1 i_2}(^{2s+1}S_J) \ ,
 \\\nonumber
D^{(2)\, X_A X_B}_{n, \, i_1 i_2}(^{2s+1}S_J) \ = \
  \frac{ \beta}{ P_{i_1 AB} \, P_{i_2 AB}} \ 
  \tilde D^{(2)\, X_A X_B}_{n, \, i_1 i_2}(^{2s+1}S_J) \ ,
 \\\nonumber
D^{(3)\, X_A X_B}_{n, \, i_1 i_2}(^{2s+1}S_J) \ = \
  \frac{ \beta}{ P_{i_1 BA} \, P_{i_2 AB}} \ 
  \tilde D^{(3)\, X_A X_B}_{n, \, i_1 i_2}(^{2s+1}S_J) \ ,
 \\
D^{(4)\, X_A X_B}_{n, \, i_1 i_2}(^{2s+1}S_J) \ = \
  \frac{ \beta}{ P_{i_1 BA} \, P_{i_2 BA}} \ 
  \tilde D^{(4)\, X_A X_B}_{n, \, i_1 i_2}(^{2s+1}S_J) \ .
\label{eq:app_D_generic}
\end{align}
Indices $i_1$ and $i_2$ in (\ref{eq:app_D_generic}) refer to the exchanged 
particle species in the left and
right $t$- and $u$-channels of the 1-loop box-amplitudes, respectively.

Throughout this appendix, the labels $A$ and $B$ are related to the
particles $X_A$ and $X_B$. Recall that these are the actual final-state 
particles in a $\chi_i \chi_j \to X_A X_B$ (tree-level) annihilation
reaction.
The overall prefactors in (\ref{eq:app_B_generic}--\ref{eq:app_D_generic})
arise from the phase-space integration $(\beta)$ and from the leading-order
expansion of $s$- or $t$- and $u$-channel propagators in the 
non-relativistic limit.

Finally recall, that each individual index $n$ in 
(\ref{eq:app_B_generic}--\ref{eq:app_D_generic}) is given by a character 
string, whose elements indicate the type ($r$ or $q$) of the 
corresponding generic coupling structures 
at the vertices of the respective underlying 1-loop amplitude in
Figs.~\ref{fig:genericselfenergy}--\ref{fig:genericboxes_fermions}. 
In the results that we quote next we only write explicitly the kinematic 
factors for those $n$ which are non-vanishing.

\subsubsection{Kinematic factors for $X_A X_B = V V$}
The kinematic factors $\tilde B^{VV}_{n,\, i_1 i_2}$
in case of $^1S_0$ partial wave reactions are given by
\begin{align}
\label{eq:B_VV_first}
\tilde B^{VV}_{qq,\, VV }(^1S_0) \ =& \ 
 - \frac{\beta^2}{2}
 + 3~\Delta_{AB}^2 \ ,
\\
\tilde B^{VV}_{qq,\, VS }(^1S_0) \ =& \ 
 \tilde B^{VV}_{qq,\, SV }(^1S_0) \ = \ 
 3~\widehat{m}_W \Delta_{AB} \ ,
\\
\tilde B^{VV}_{qq,\, SS }(^1S_0) \ =& \ 
 4~\widehat{m}_W^2 \ .
\end{align}
In case of $^3S_1$ partial-wave reactions we have
\begin{align}
\tilde B^{VV}_{rr,\, VV }(^3S_1) \ =& \ 
 - \frac{9}{2}
 + \frac{4}{3}~\beta^2
 - \frac{1}{2}~\Delta_{AB}^2 \ .
\label{eq:B_VV_last}
\end{align}
Only the kinematic factors $\tilde B^{VV}_{n,\, i_1 i_2}$ given explicitly 
in (\ref{eq:B_VV_first}--\ref{eq:B_VV_last}) with $n = rr, qq$ are
non-vanishing. In case of $X_A X_B = VV$, the kinematic factors for the 
triangle- and box-diagram topologies $\alpha = 3,4$ are related 
to the corresponding expressions for diagram-topologies $\alpha = 1,2$
(see Figs.~\ref{fig:generictriangles}--\ref{fig:genericboxes}).
The relations read
\begin{align}
\nonumber
\tilde C^{(3)\, VV}_{n, \, i_1 V}(^{2s+1}S_J) \ =& \ 
 - \tilde C^{(1)\, VV}_{n, \, i_1 V}(^{2s+1}S_J) \
                             \vert_{A \leftrightarrow B} \ ,
\\\nonumber
\tilde C^{(4)\, VV}_{n, \, i_1 V}(^{2s+1}S_J) \ =& \ 
 - \tilde C^{(2)\, VV}_{n, \, i_1 V}(^{2s+1}S_J) \
                             \vert_{A \leftrightarrow B} \ ,
\\\nonumber
\tilde C^{(3)\, VV}_{n, \, i_1 S}(^{2s+1}S_J) \ =& \ 
 \tilde C^{(1)\, VV}_{n, \, i_1 S}(^{2s+1}S_J) \
                             \vert_{A \leftrightarrow B} \ ,
\\\nonumber
\tilde C^{(4)\, VV}_{n, \, i_1 S}(^{2s+1}S_J) \ =& \ 
 \tilde C^{(2)\, VV}_{n, \, i_1 S}(^{2s+1}S_J) \
                             \vert_{A \leftrightarrow B} \ ,
\\\nonumber
\tilde D^{(3)\, VV}_{n, \, i_1 i_2}(^{2s+1}S_J) \ =& \ \tilde D^{(1)\, VV}_{n, \, i_1 i_2}(^{2s+1}S_J) \
                             \vert_{A \leftrightarrow B} \ ,
\\
\tilde D^{(4)\, VV}_{n, \, i_1 i_2}(^{2s+1}S_J) \ =& \ \tilde D^{(2)\, VV}_{n, \, i_1 i_2}(^{2s+1}S_J) \
                             \vert_{A \leftrightarrow B} \ .
\label{eq:app_VV_34_relations}
\end{align}
The minus sign in the relation for the triangle coefficients
$\tilde C^{(\alpha)\, VV}_{n,\, i_1 V}$ in (\ref{eq:app_VV_34_relations}) arises from
interchanging the two gauge-bosons $X_A$ and $X_B$ at the internal
three-gauge boson vertex.
The expressions $\tilde C^{(\alpha)\,VV}_{n,\, i_1 V}$ for diagram-topologies
$\alpha = 1,2$, that refer to leading-order $^1S_0$ partial waves read
\begin{align}
\label{eq:app_C_VV_V_1}
\tilde C^{(1)\,VV}_{rqq,\, i_1 V}(^1S_0) \ =& \ 
   \frac{\beta^2}{2}
 - \frac{3}{2}~(\widehat{m}- \widehat{\overline{m}}) \Delta_{AB}
 - \frac{3}{2}~\Delta_{AB}^2 
 + 3~\widehat{m}_{i_1} \Delta_{AB} \ ,
\\
 \tilde C^{(2)\,VV}_{qqr,\, i_1 V}(^1S_0) \ =& \ 
 \tilde C^{(1)\,VV}_{rqq,\, i_1 V}(^1S_0) \ .
\end{align}
In case of $C^{(\alpha)VV}_{n,\, i_1 V}$ expressions related to $^3S_1$ partial waves
and diagram-topologies $\alpha = 1,2$ we find
\begin{align}
\label{eq:app_C_VV_V_1_rrr}
\tilde C^{(\alpha)\,VV}_{rrr,\, i_1 V}(^3S_1) \ =& \ 
 -\frac{5}{6}~\beta^2
 + (\widehat{m}- \widehat{\overline{m}}) \frac{\Delta_{AB}}{2}
 + \frac{~\Delta_{AB}^2}{2}
 + 3~\widehat{m}_{i_1} \ .
\end{align}
We deduce the following expressions for $\tilde C^{(\alpha)\,VV}_{n,\, i_1 S}$
coefficients and diagram topologies $\alpha = 1,2$:
\begin{align}
\label{eq:app_C_VV_S_1_rqq}
\tilde C^{(1)\,VV}_{rqq,\, i_1 S}(^1S_0) \ =& \ 
 \tilde C^{(2)\,VV}_{qqr,\, i_1 S}(^1S_0)
 \ = \
 - \widehat{m}_W (\widehat{m}- \widehat{\overline{m}} + \Delta_{AB})
 + 4~ \widehat{m}_W\widehat{m}_{i_1} \ .
\end{align}
There are additional non-vanishing $C^{(\alpha)VV}_{n,\, i_1 X}$ expressions
for $X = V,S$ in both the case of $^1S_0$ and $^3S_1$ partial wave reactions,
which are related to the expressions in
(\ref{eq:app_C_VV_V_1}--\ref{eq:app_C_VV_S_1_rqq})
in the following way:
\begin{align}
\nonumber
\tilde C^{(1)\, VV}_{qqr,\, i_1 X}(^{2s+1}S_J) \ =& \ \tilde C^{(2)\, VV}_{rqq,\, i_1 X}(^{2s+1}S_J) \ = \
  \tilde C^{(1)\, VV}_{rrr,\, i_1 X}(^{2s+1}S_J)
  \vert_{m_{i_1} \to - m_{i_1}} \ ,
\\
\tilde C^{(1)\, VV}_{qrq,\, i_1 X}(^{2s+1}S_J) \ =& \ \tilde C^{(2)\, VV}_{qrq,\, i_1 X}(^{2s+1}S_J) \ = \
  \tilde C^{(1)\, VV}_{rqq,\, i_1 X}(^{2s+1}S_J)
  \vert_{m_{i_1} \to - m_{i_1}} \ .
\label{eq:app_C_(V)VV_relations}
\end{align}
Turning to terms related to box-diagrams, the non-vanishing expressions
$\tilde D^{(\alpha)\,VV}_{n, \, i_1 i_2}$ for $\alpha = 1,2$ are given by
\begin{align}
\tilde D^{(\alpha)\, VV}_{rrrr, \, i_1 i_2}(^1S_0) \ =& \ \frac{\beta^2}{2} \ , 
\\\nonumber
\tilde D^{(1)\, VV}_{rqqr, \, i_1 i_2}(^1S_0) \ =& \ 
  \phantom{-} ~ \frac{\beta^2}{2}
  + ( \widehat m - \widehat{\overline m} )^2
  - \Delta_{AB}^2
  + 4~  \widehat m_{i_1} \widehat m_{i_2}
  \\ & \
  -  \widehat m_{i_1} ( \widehat m - \widehat{\overline m} - \Delta_{AB})
  -  \widehat m_{i_2} ( \widehat m - \widehat{\overline m} + \Delta_{AB})
 \ ,
\\\nonumber
\tilde D^{(2),\,VV}_{rqqr, \, i_1 i_2}(^1S_0) \ =& \ 
  - \frac{\beta^2}{2}
  + \left(
         \widehat m- \widehat{\overline m} + \Delta_{AB}
    \right)^2
  +  4~\widehat m_{i_1} \widehat m_{i_2}
  \\ & \
  - ( \widehat m_{i_1} + \widehat m_{i_2}) ~
    ( \widehat m - \widehat{\overline m} + \Delta_{AB})
 \ ,
\end{align}
and
\begin{align}
\tilde D^{(1)\, VV}_{rrrr, \, i_1 i_2}(^3S_1) \ =& \
  - \frac{2}{3} ~ \beta^2
  - \frac{1}{2} ~ ( \widehat m - \widehat{\overline m})^2
  + \frac{1}{2} ~ \Delta_{AB}^2
  +   2~ \widehat m_{i_1} \widehat m_{i_2}
 \ ,
\\
\tilde D^{(2)\, VV}_{rrrr, \, i_1 i_2}(^3S_1) \ =& \
  \phantom{-}~\frac{2}{3}~\beta^2
  -\frac{1}{2} ~\left( \widehat m- \widehat{\overline m}
                      + ~\Delta_{AB}\right)^2 
  - 2~ \widehat m_{i_1} \widehat m_{i_2}
\ ,
\\\nonumber
\tilde D^{(1)\, VV}_{rqqr, \, i_1 i_2}(^3S_1) \ =& \
  -\frac{1}{3} ~ \beta^2
  - \frac{1}{2}~( \widehat m - \widehat{\overline m})^2
  + \frac{1}{2}~\Delta_{AB}^2
  -2 ~ \widehat m_{i_1} \widehat m_{i_2}
 \\ & \
  + \widehat m_{i_1} ( \widehat m- \widehat{\overline m} - \Delta_{AB})
  + \widehat m_{i_2} ( \widehat m- \widehat{\overline m} + \Delta_{AB})
 \ ,
\\\nonumber
\tilde D^{(2)\, VV}_{rqqr, \, i_1 i_2}(^3S_1) \ =& \
  -\frac{1}{3}~{\beta}^2
  + \frac{1}{2} \left( m- \overline{m} + \hat\Delta_{AB}\right)^2
  + 2 ~ \widehat m_{i_1} \widehat m_{i_2}
 \\ & \
  - \widehat m_{i_1} ( \widehat m - \widehat{\overline m} + \Delta_{AB})
  - \widehat m_{i_2} ( \widehat m - \widehat{\overline m} + \Delta_{AB})
  \ .
\end{align}
The remaining non-vanishing kinematic factors 
$\tilde D^{VV}_{n,\, i_1 i_2}$ for both spin-0 and spin-1 $\chi\chi$ 
states are related to the expressions given above by
\begin{align}
\nonumber
\tilde D^{(\alpha)\, VV}_{qqqq,\, i_1 i_2}(^{2s+1}S_J) \ =& \ 
  \tilde D^{(\alpha)\, VV}_{rrrr,\, i_1 i_2}(^{2s+1}S_J) \ ,
\\\nonumber
\tilde D^{(\alpha)\, VV}_{rrqq,\, i_1 i_2}(^{2s+1}S_J) \ =& \ \tilde D^{(\alpha)\, VV}_{qqrr,\, i_1 i_2}(^{2s+1}S_J) \ = \
  \tilde D^{(\alpha)\, VV}_{rrrr,\, i_1 i_2}(^{2s+1}S_J)
  \vert_{\ \hat m_{\,i_1} \hat m_{\,i_2}  \rightarrow \ - \hat m_{\,i_1} \hat m_{\,i_2}} \ ,
\\\nonumber
\tilde D^{(\alpha)\, VV}_{qrrq,\, i_1 i_2}(^{2s+1}S_J) \ =& \ 
  \tilde D^{(\alpha)\, VV}_{rqqr,\, i_1 i_2}(^{2s+1}S_J)
  \vert_{\ \hat m_{\,i_{1,2}}  \rightarrow \ - \hat m_{\,i_{1,2}}} \ ,
\\\nonumber
\tilde D^{(\alpha)\, VV}_{rqrq,\, i_1 i_2}(^{2s+1}S_J) \ =& \ 
  \tilde D^{(\alpha)\, VV}_{rqqr,\, i_1 i_2}(^{2s+1}S_J)
  \vert_{\ \hat m_{\,i_{2}}  \rightarrow \ - \hat m_{\,i_{2}}} \ ,
\\
\tilde D^{(\alpha)\, VV}_{qrqr,\, i_1 i_2}(^{2s+1}S_J) \ =& \ 
  \tilde D^{(\alpha)\, VV}_{rqqr,\, i_1 i_2}(^{2s+1}S_J)
  \vert_{\ \hat m_{\,i_{1}}  \rightarrow \ - \hat m_{\,i_{1}}} \ .
\label{eq:app_D_VV_relations}
\end{align}
The notation in the second line of (\ref{eq:app_D_VV_relations})
means that the product $\widehat m_{i_1} \widehat m_{i_2}$ is replaced, but all
other appearances of either $\widehat m_{i_1}$ or $\widehat m_{i_2}$ are
untouched.

\subsubsection{Kinematic factors for $X_A X_B = V S$}
We find the following expressions for $\tilde B^{VS}_{n,\, i_1 i_2}$ terms in
$^1S_0$ partial-wave reactions with $i_1 i_2 = VV, VS, SV, SS$:
\begin{align}
\tilde B^{VS}_{qq,\, VV }(^1S_0) \ =& \ 
 -\widehat{m}_W^2 \ ,
\\
\tilde B^{VS}_{qq,\, VS }(^1S_0) \ =& \ 
 \tilde B^{VS}_{qq,\, SV }(^1S_0) \ = \ 
 \frac{\widehat{m}_W}{2} ~ (-3 + \Delta_{AB}) \ ,
\\
\tilde B^{VS}_{qq,\, SS }(^1S_0) \ =& \ 
   \frac{\beta^2}{4}
 - \frac{9}{4}
 + \frac{3}{2}~\Delta_{AB}
 - \frac{\Delta_{AB}^2}{4} \ .
\end{align}
In case of $^3S_1$ partial-wave processes the corresponding
 $\tilde B^{VS}_{n,\, i_1 i_2}$ coefficients read
\begin{align}
\tilde B^{VS}_{rr,\, VV }(^3S_1) \ =& \ \widehat{m}_W^2 \ .
\end{align}
Kinematic factors $\tilde C^{(\alpha)\, VS}_{n,\, i_1 V}$, that are related to the four
generic triangle-topologies $\alpha$ with gauge-boson~($V$) exchange in the
single $s$-channel (see Fig.~\ref{fig:generictriangles}) are given by
\begin{align}
\label{eq:app_C_VS_V}
\tilde C^{(1)\,VS}_{rqq,\, i_1 V}(^1S_0) \ =& \ 
 \tilde C^{(2)\,VS}_{qqr,\, i_1 V}(^1S_0) \ = \ 
 - \frac{\widehat{m}_W}{2} (\widehat{m}- \widehat{\overline{m}} + \Delta_{AB})
 - \widehat{m}_W \widehat{m}_{i_1} \ ,
\\
\tilde C^{(3)\,VS}_{rqq,\, i_1 V}(^1S_0) \ =& \ 
 \tilde C^{(4)\,VS}_{qqr,\, i_1 V}(^1S_0) \ = \
 \phantom{-} \frac{\widehat{m}_W}{2}
             (\widehat{m}- \widehat{\overline{m}} - \Delta_{AB})
 + \widehat{m}_W \widehat{m}_{i_1} \ ,
\end{align}
as well as
\begin{align}
\tilde C^{(1)\,VS}_{rrr,\, i_1 V}(^3S_1) \ =& \ 
 \tilde C^{(2)\,VS}_{rrr,\, i_1 V}(^3S_1) \ = \ 
 -~\tilde C^{(1)\,VS}_{rqq,\, i_1 V}(^1S_0) \ ,
\\
\tilde C^{(3)\,VS}_{rrr,\, i_1 V}(^3S_1) \ =& \ 
 \tilde C^{(4)\,VS}_{rrr,\, i_1 V}(^3S_1) \ = \ 
 -~\tilde C^{(3)\,VS}_{rqq,\, i_1 V}(^1S_0)
   \vert_{\widehat m_{i_1} \to -\widehat m_{i_1}} \ .
\end{align}
In case of $\tilde C^{(\alpha)\,VS}_{n,\, i_1 S}$ expressions we find
\begin{align}
\nonumber
\tilde C^{(1)\,VS}_{rqq,\, i_1 S}(^1S_0) \ =& \ 
 -  \frac{\beta^2}{4}
 - \frac{3}{4}~( \widehat m - \widehat{\overline m})
 + ( \widehat m - \widehat{\overline m} - 3)~\frac{\Delta_{AB}}{4}
 \\ & \
 + \frac{\Delta_{AB}^2}{4}
 - \frac{\widehat{m}_{i_1}}{2}~(3 - \Delta_{AB}) \ ,
\\
\tilde C^{(2)\,VS}_{qqr,\, i_1 S}(^1S_0) \ =& \ 
 \tilde C^{(1)\,VS}_{rqq,\, i_1 S}(^1S_0) \ ,
\\\nonumber
\tilde C^{(3)\,VS}_{rqq,\, i_1 S}(^1S_0) \ =& \
 -  \frac{\beta^2}{4}
 + \frac{3}{4}~(\widehat m - \widehat{\overline m})
 - (\widehat m - \widehat{\overline m} + 3)~\frac{\Delta_{AB}}{4}
 \\ & \
 + \frac{\Delta_{AB}^2}{4}
 + \frac{\widehat{m}_{i_1}}{2}~(3 - \Delta_{AB}) \ ,
\\
\tilde C^{(4)\,VS}_{qqr,\, i_1 S}(^1S_0) \ =& \ 
 \tilde C^{(3)\,VS}_{rqq,\, i_1 S}(^1S_0) \ .
\label{eq:app_CVS_Slast}
\end{align}
There are additional non-vanishing kinematic factors for
$\tilde C^{(\alpha)\,VS}_{n,\, i_1 X}$ with $X = V$ or $S$, 
related to the corresponding expressions in
(\ref{eq:app_C_VS_V}--\ref{eq:app_CVS_Slast})
in the following way:
\begin{align}
\nonumber
\tilde C^{(1)\,VS}_{qqr,\, i_1 X}(^{2s+1}S_J) \ =& \ \tilde C^{(2)\,VS}_{rqq,\, i_1 X}(^{2s+1}S_J) \ = \
 -~\tilde C^{(1)\,VS}_{rrr,\, i_1 X}(^{2s+1}S_J)
 \vert_{\widehat m_{i_1} \to - \widehat m_{i_1}} \ ,
\\\nonumber
\tilde C^{(3)\,VS}_{qqr,\, i_1 X}(^{2s+1}S_J) \ =& \ \tilde C^{(4)\,VS}_{rqq,\, i_1 X}(^{2s+1}S_J)  \ = \
 \tilde C^{(3)\,VS}_{rrr,\, i_1 X}(^{2s+1}S_J)
 \vert_{\widehat m_{i_1} \to - \widehat m_{i_1}} \ ,
\\\nonumber
\tilde C^{(1)\,VS}_{qrq,\, i_1 X}(^{2s+1}S_J) \ =& \ \tilde C^{(2)\,VS}_{qrq,\, i_1 X}(^{2s+1}S_J) \ = \
 -~\tilde C^{(1)\,VS}_{rqq,\, i_1 X}(^{2s+1}S_J)
 \vert_{\widehat m_{i_1} \to - \widehat m_{i_1}} \ ,
\\
\tilde C^{(3)\,VS}_{qrq,\, i_1 X}(^{2s+1}S_J) \ =& \ \tilde C^{(4)\,VS}_{qrq,\, i_1 X}(^{2s+1}S_J) \ = \
 \tilde C^{(3)\,VS}_{rqq,\, i_1 X}(^{2s+1}S_J)
 \vert_{\widehat m_{i_1} \to - \widehat m_{i_1}} \ .
\end{align}
The non-vanishing kinematic factors for $X_A X_B = VS$ and the four
box-topologies $\alpha$ are given by
\begin{align}
\nonumber
\tilde D^{(1)\,VS}_{rqqr,\, i_1 i_2}(^1S_0) \ =& \
   \frac{1}{4} ~ \beta^2
 + \frac{1}{4}(\widehat{m} - \widehat{\overline{m}})^2
 - \frac{1}{4} ~ \Delta_{AB}^2
 + \widehat{m}_{i_1} \widehat{m}_{i_2}
 \\ & \
 + \frac{1}{2} ~ \widehat{m}_{i_1}
               (\widehat{m}- \widehat{\overline{m}} - \Delta_{AB})
 + \frac{1}{2} ~ \widehat{m}_{i_2}
               (\widehat{m}- \widehat{\overline{m}} + \Delta_{AB}) \ ,
\\\nonumber
\tilde D^{(2)\,VS}_{rqqr,\, i_1 i_2}(^1S_0) \ =& \
   \frac{1}{4} ~ \beta^2
 - \frac{1}{4} ~ \left(\widehat{m} -\widehat{\overline{m}} + \Delta_{AB}\right)^2
 - \widehat{m}_{i_1} \widehat{m}_{i_2}
 \\ & \
 - \frac{1}{2} ~ ( \widehat{m}_{i_1} + \widehat{m}_{i_2} )
               (\widehat{m} - \widehat{\overline{m}} + \Delta_{AB}) \ ,
\\
\tilde D^{(3)\,VS}_{rqqr,\, i_1 i_2}(^1S_0) \ =& \
 \tilde D^{(1)\,VS}_{rqqr,\, i_1 i_2}(^1S_0)
 \vert_{A \leftrightarrow B} \ ,
\\
\tilde D^{(4)\,VS}_{rqqr,\, i_1 i_2}(^1S_0) \ =& \
 \tilde D^{(2)\,VS}_{rqqr,\, i_1 i_2}(^1S_0)
 \vert_{A \leftrightarrow B} \ .
\end{align}
In case of $^3S_1$ partial waves we have
\begin{align}
\nonumber
\tilde D^{(1)\,VS}_{rrrr,\, i_1 i_2}(^3S_1) \ =& \
 - \frac{1}{12}~\beta^2
 - \frac{1}{4} ~ (\widehat m - \widehat{\overline m})^2
 + \frac{1}{4}~\Delta_{AB}^2
 + \widehat{m}_{i_1} \widehat{m}_{i_2}
 \\ & \
 - \frac{1}{2}~\widehat{m}_{i_1}
              (\widehat{m}- \widehat{\overline{m}} - \Delta_{AB})
 + \frac{1}{2}~\widehat{m}_{i_2}
              (\widehat{m}- \widehat{\overline{m}} + \Delta_{AB}) \ ,
\\\nonumber
\tilde D^{(2)\,VS}_{rrrr,\, i_1 i_2}(^3S_1) \ =& \
 - \frac{1}{12}~\beta^2
 + \frac{1}{4} \left(
                \widehat m - \widehat{\overline m} + \Delta_{AB}
                \right)^2
 + \widehat{m}_{i_1} \widehat{m}_{i_2}
 \\ & \
 + \frac{1}{2}~(\widehat{m}_{i_1} + \widehat{m}_{i_1} )
               (\widehat{m}- \widehat{\overline{m}} + \Delta_{AB}) \ ,
\\
\tilde D^{(3)\,VS}_{rrrr,\, i_1 i_2}(^3S_1) \ =& \
 \tilde D^{(1)\,VS}_{rrrr,\, i_1 i_2}(^3S_1)
 \vert_{\widehat m \leftrightarrow \widehat{\overline m}} \ ,
\\
\tilde D^{(4)\,VS}_{rrrr,\, i_1 i_2}(^3S_1) \ =& \
 \tilde D^{(2)\,VS}_{rrrr,\, i_1 i_2}(^3S_1)
 \vert_{\widehat m \leftrightarrow \widehat{\overline m}} \ ,
\\
\tilde D^{(\alpha)\,VS}_{rqqr,\, i_1 i_2}(^3S_1) \ =& \
 \frac{(-1)^{\,\alpha}}{6}~\beta^2 \ .
\end{align}
Relations for the remaining kinematic factors for both $^1S_0$ and $^3S_1$
partial wave reactions read in case of diagram-topologies $\alpha = 1, 2$:
\begin{align}
\nonumber
\tilde D^{(\alpha)\,VS}_{qqqq,\, i_1 i_2}(^{2s+1}S_J) \ =& \ 
 (-1)^{\,\alpha} ~ \tilde D^{(\alpha)\,VS}_{rrrr,\, i_1 i_2}(^{2s+1}S_J)
 \vert_{\ \widehat m_{\,i_{1,2}}  \rightarrow \ - \widehat m_{\,i_{1,2}}} \ ,
\\\nonumber
\tilde D^{(\alpha)\,VS}_{rrqq,\, i_1 i_2}(^{2s+1}S_J) \ =& \ 
 (-1)^{\,\alpha+1} ~ \tilde D^{(\alpha)\,VS}_{rrrr,\, i_1 i_2}(^{2s+1}S_J)
 \vert_{\ \hat m_{\,i_{2}}  \rightarrow \ - \hat m_{\,i_{2}}} \ ,
\\\nonumber
\tilde D^{(\alpha)\,VS}_{qqrr,\, i_1 i_2}(^{2s+1}S_J) \ =& \ 
 - \tilde D^{(\alpha)\,VS}_{rrrr,\, i_1 i_2}(^{2s+1}S_J)
 \vert_{\ \hat m_{\,i_{1}}  \rightarrow \ - \hat m_{\,i_{1}}} \ ,
\\\nonumber
\tilde D^{(\alpha)\,VS}_{qrrq,\, i_1 i_2}(^{2s+1}S_J) \ =& \ 
 (-1)^{\,\alpha} ~ \tilde D^{(\alpha)\,VS}_{rqqr,\, i_1 i_2}(^{2s+1}S_J)
  \vert_{\ \hat m_{\,i_{1,2}}  \rightarrow \ - \hat m_{\,i_{1,2}}} \ ,
\\\nonumber
\tilde D^{(\alpha)\,VS}_{rqrq,\, i_1 i_2}(^{2s+1}S_J) \ =& \ 
 (-1)^{\,\alpha+1} ~ \tilde D^{(\alpha)\,VS}_{rqqr,\, i_1 i_2}(^{2s+1}S_J)
  \vert_{\ \hat m_{\,i_{2}}  \rightarrow \ - \hat m_{\,i_{2}}} \ ,
\\
\tilde D^{(\alpha)\,VS}_{qrqr,\, i_1 i_2}(^{2s+1}S_J) \ =& \ 
 - \tilde D^{(\alpha)\,VS}_{rqqr,\, i_1 i_2}(^{2s+1}S_J)
  \vert_{\ \hat m_{\,i_{1}}  \rightarrow \ - \hat m_{\,i_{1}}} \ .
\end{align}
The corresponding relations for diagram-topologies $\alpha = 3,4$
are given by
\begin{align}
\nonumber
\tilde D^{(\alpha)\,VS}_{qqqq,\, i_1 i_2}(^{2s+1}S_J) \ =& \ 
 (-1)^{\,\alpha} ~ \tilde D^{(\alpha)\,VS}_{rrrr,\, i_1 i_2}(^{2s+1}S_J)
 \vert_{\ \widehat m_{\,i_{1,2}}  \rightarrow \ - \widehat m_{\,i_{1,2}}} \ ,
\\\nonumber
\tilde D^{(\alpha)\,VS}_{rrqq,\, i_1 i_2}(^{2s+1}S_J) \ =& \ 
 (-1)^{\,\alpha} ~ \tilde D^{(\alpha)\,VS}_{rrrr,\, i_1 i_2}(^{2s+1}S_J)
 \vert_{\ \hat m_{\,i_{2}}  \rightarrow \ - \hat m_{\,i_{2}}} \ ,
\\\nonumber
\tilde D^{(\alpha)\,VS}_{qqrr,\, i_1 i_2}(^{2s+1}S_J) \ =& \ 
 \tilde D^{(\alpha)\,VS}_{rrrr,\, i_1 i_2}(^{2s+1}S_J)
 \vert_{\ \hat m_{\,i_{1}}  \rightarrow \ - \hat m_{\,i_{1}}} \ ,
\\\nonumber
\tilde D^{(\alpha)\,VS}_{qrrq,\, i_1 i_2}(^{2s+1}S_J) \ =& \ 
 (-1)^{\,\alpha} ~ \tilde D^{(\alpha)\,VS}_{rqqr,\, i_1 i_2}(^{2s+1}S_J)
  \vert_{\ \hat m_{\,i_{1,2}}  \rightarrow \ - \hat m_{\,i_{1,2}}} \ ,
\\\nonumber
\tilde D^{(\alpha)\,VS}_{rqrq,\, i_1 i_2}(^{2s+1}S_J) \ =& \ 
 (-1)^{\,\alpha} ~ \tilde D^{(\alpha)\,VS}_{rqqr,\, i_1 i_2}(^{2s+1}S_J)
  \vert_{\ \hat m_{\,i_{2}}  \rightarrow \ - \hat m_{\,i_{2}}} \ ,
\\
\tilde D^{(\alpha)\,VS}_{qrqr,\, i_1 i_2}(^{2s+1}S_J) \ =& \ 
 \tilde D^{(\alpha)\,VS}_{rqqr,\, i_1 i_2}(^{2s+1}S_J)
  \vert_{\ \hat m_{\,i_{1}}  \rightarrow \ - \hat m_{\,i_{1}}} \ .
\end{align}

\subsubsection{Kinematic factors for $X_A X_B = SS$}
The non-vanishing $\tilde B^{SS}_{n,\, i_1 i_2}$ terms with
$i_1 i_2 = VV, VS, SV, VV$ read
\begin{align}
\tilde B^{SS}_{qq,\, VV }(^1S_0) \ =& \ 
 \Delta_{AB}^2 \ ,
\\
\tilde B^{SS}_{qq,\, VS }(^1S_0) \ =& \ 
 \tilde B^{SS}_{qq,\, SV }(^1S_0) \ = \
 -~\widehat{m}_W \Delta_{AB} \ ,
\\
\tilde B^{SS}_{qq,\, SS }(^1S_0) \ =& \ 
 \widehat{m}_W^2 \ ,
\end{align}
and in case of $^3S_1$ reactions
\begin{align}
\tilde B^{SS}_{rr,\, VV }(^3S_1) \ =& \ 
 \frac{\beta^2}{3} \ .
\end{align}
As in the case of $X_A X_B = VV$, the kinematic factors for $X_A X_B = SS$ and 
diagram topologies $\alpha = 3,4$ are related to the corresponding expressions
that arise from diagram-topologies $\alpha = 1,2$. This applies to both 
triangle- and box-topologies (see Fig.~\ref{fig:generictriangles} and
Fig.~\ref{fig:genericboxes}):
\begin{align}
\nonumber
\tilde C^{(3)\, SS}_{n, \, i_1 V}(^{2s+1}S_J) \ =& \ 
 -~\tilde C^{(1)\, SS}_{n, \, i_1 V}(^{2s+1}S_J) \
                             \vert_{A \leftrightarrow B} \ ,
\\\nonumber
\tilde C^{(4)\, SS}_{n, \, i_1 V}(^{2s+1}S_J) \ =& \ 
 -~ \tilde C^{(2)\, SS}_{n, \, i_1 V}(^{2s+1}S_J) \
                             \vert_{A \leftrightarrow B} \ ,
\\\nonumber
\tilde C^{(3)\, SS}_{n, \, i_1 S}(^{2s+1}S_J) \ =& \ 
 \tilde C^{(1)\, SS}_{n, \, i_1 S}(^{2s+1}S_J) \
                             \vert_{A \leftrightarrow B} \ ,
\\\nonumber
\tilde C^{(4)\, SS}_{n, \, i_1 S}(^{2s+1}S_J) \ =& \ 
 \tilde C^{(2)\, SS}_{n, \, i_1 S}(^{2s+1}S_J) \
                             \vert_{A \leftrightarrow B} \ ,
\\\nonumber
\tilde D^{(3)\, SS}_{n, \, i_1 i_2}(^{2s+1}S_J) \ =& \ \tilde D^{(1)\, SS}_{n, \, i_1 i_2}(^{2s+1}S_J) \
                             \vert_{A \leftrightarrow B} \ ,
\\
\tilde D^{(4)\, SS}_{n, \, i_1 i_2}(^{2s+1}S_J) \ =& \ \tilde D^{(2)\, SS}_{n, \, i_1 i_2}(^{2s+1}S_J) \
                             \vert_{A \leftrightarrow B} \ .
\end{align}
In case of expressions $\tilde C^{(\alpha)\, SS}_{n,\, i_1 V}$ 
for diagram-topologies $\alpha = 1,2$ we find
\begin{align}
\label{eq:app_C_SS_V_1}
\tilde C^{(1)\,SS}_{rqq,\, i_1 V}(^1S_0) \ =& \
  \tilde C^{(2)\,SS}_{qqr,\, i_1 V}(^1S_0) \ = \ 
   \frac{\Delta_{AB}}{2} (\widehat{m}- \widehat{\overline{m}} + \Delta_{AB})
 + \widehat{m}_{i_1} \Delta_{AB} \ ,
\\
\tilde C^{(1)\,SS}_{rrr,\, i_1 V}(^3S_1) \ =& \
  \tilde C^{(2)\,SS}_{rrr,\, i_1 V}(^3S_1) \ = \
  -~\frac{\beta^2}{6} \ .
\end{align}
The $\tilde C^{(\alpha)\, SS}_{n,\, i_1 S}$ expressions with 
$\alpha = 1,2$ are given by
\begin{align}
\tilde C^{(1)\,SS}_{rqq,\, i_1 S}(^1S_0) \ =& \ 
 \tilde C^{(2)\,SS}_{qqr,\, i_1 S}(^1S_0) \ = \ 
 - \frac{\widehat{m}_W}{2}~(\widehat{m}- \widehat{\overline{m}} + \Delta_{AB})
 -  \widehat{m}_W \widehat{m}_{i_1} \ .
\label{eq:app_C_SS_S}
\end{align}
All other non-vanishing expressions for $\tilde C^{(\alpha)\, SS}_{n,\, i_1 X}$
with $X = V,S$ and $\alpha = 1,2$ can be related to the terms in
(\ref{eq:app_C_SS_V_1}--\ref{eq:app_C_SS_S}) in the following way:
\begin{align}
\nonumber
\tilde C^{(1)\,SS}_{qqr,\, i_1 X}(^{2s+1}S_J) \ =& \ 
\tilde C^{(2)\,SS}_{rqq,\, i_1 X}(^{2s+1}S_J) \ = \
 -~\tilde C^{(1)\,SS}_{rrr,\, i_1 X}(^{2s+1}S_J)
 \vert_{\widehat m_{i_1} \to - \widehat m_{i_1}} \ ,
\\
\tilde C^{(1)\,SS}_{qrq,\, i_1 X}(^{2s+1}S_J) \ =& \ 
\tilde C^{(2)\,SS}_{qrq,\, i_1 X}(^{2s+1}S_J)  \ = \
 -~\tilde C^{(1)\,SS}_{rqq,\, i_1 X}(^{2s+1}S_J)
 \vert_{\widehat m_{i_1} \to - \widehat m_{i_1}} \ .
\end{align}
The expressions $\tilde D^{(\alpha)\,SS}_{n, \, i_1 i_2}$ for 
diagram-topologies $\alpha = 1,2$ and $^1S_0$ partial waves read
\begin{align}
\nonumber
\tilde D^{(1)\,SS}_{rqqr,\, i_1 i_2}(^1S_0) \ =& \ 
  \frac{1}{4}~(\widehat{m}- \widehat{\overline{m}})^2
 - \frac{\Delta_{AB}^2}{4}
 + \widehat{m}_{i_1} \widehat{m}_{i_2}
 \\ & \
 + \frac{\widehat{m}_{i_1}}{2}~(\widehat{m}- \widehat{\overline{m}} - \Delta_{AB})
 + \frac{\widehat{m}_{i_2}}{2}~(\widehat{m}- \widehat{\overline{m}}
 + \Delta_{AB}) \ ,
\\\nonumber
\tilde D^{(2)\,SS}_{rqqr,\, i_1 i_2}(^1S_0) \ =& \
 \frac{1}{4} \left(\widehat{m}- \widehat{\overline{m}} + \Delta_{AB}\right)^2
 + \widehat{m}_{i_1} \widehat{m}_{i_2}
 \\ & \
 + \frac{1}{2} ~ (\widehat{m}_{i_1} + \widehat{m}_{i_2})
                 (\widehat{m}- \widehat{\overline{m}} + \Delta_{AB}) \ .
\end{align}
In case of a $^3S_1$ partial wave configuration we find
\begin{align}
\tilde D^{(\alpha)\,SS}_{rrrr,\, i_1 i_2}(^3S_1) \ =& \
 (-1)^{\,\alpha}~ \frac{\beta^2}{12} \ .
\end{align}
The remaining non-vanishing kinematic factors related to both
$^1S_0$ and $^3S_1$ partial-wave reactions read
\begin{align}
\nonumber
\tilde D^{(\alpha)\,SS}_{qqqq,\, i_1 i_2}(^{2s+1}S_J) \ =& \ 
  \tilde D^{(\alpha)\,SS}_{rrrr,\, i_1 i_2}(^{2s+1}S_J) \ ,
\\\nonumber
\tilde D^{(\alpha)\,SS}_{rrqq,\, i_1 i_2}(^{2s+1}S_J) \ =& \ \tilde D^{(\alpha)\,SS}_{qqrr,\, i_1 i_2}(^{2s+1}S_J) \ = \
 - \tilde D^{(\alpha)\,SS}_{rrrr,\, i_1 i_2}(^{2s+1}S_J) \ ,
\\\nonumber
\tilde D^{(\alpha)\,SS}_{qrrq,\, i_1 i_2}(^{2s+1}S_J) \ =& \ 
 \tilde D^{(\alpha)\,SS}_{rqqr,\, i_1 i_2}(^{2s+1}S_J)
 \vert_{\ \hat m_{\,i_{1,2}}  \rightarrow \ - \hat m_{\,i_{1,2}}} \ ,
\\\nonumber
\tilde D^{(\alpha)\,SS}_{rqrq,\, i_1 i_2}(^{2s+1}S_J) \ =& \ 
 - \tilde D^{(\alpha)\,SS}_{rqqr,\, i_1 i_2}(^{2s+1}S_J)
 \vert_{\ \hat m_{\,i_{2}}  \rightarrow \ - \hat m_{\,i_{2}}} \ ,
\\
\tilde D^{(\alpha)\,SS}_{qrqr,\, i_1 i_2}(^{2s+1}S_J) \ =& \ 
 - \tilde D^{(\alpha)\,SS}_{rqqr,\, i_1 i_2}(^{2s+1}S_J)
 \vert_{\ \hat m_{\,i_{1}}  \rightarrow \ - \hat m_{\,i_{1}}} \ .
\end{align}

\subsubsection{Kinematic factors for $X_A X_B = ff$}
The non-vanishing $\tilde B^{ff}_{n,\, i_1 i_2}$ terms with
$i_1i_2 = VV, VS, SV, SS$ are given by
\begin{align}
\label{eq:app_B_ff_first}
\tilde B^{ff}_{qqqq,\, VV }(^1S_0) \ =& \ 
 1 - \beta^2
 + 4~\widehat{m}_{A} \widehat{m}_{B}
 - \Delta_{AB}^2 \ ,
\\
\tilde B^{ff}_{qqqq,\, VS }(^1S_0) \ =& \ 
 \tilde B^{ff}_{qqqq,\, SV }(^1S_0) \ = \
 2 ~ \Bigl(
       \widehat{m}_{A} + \widehat{m}_{B}
     - (\widehat{m}_{A} - \widehat{m}_{B}) \Delta_{AB}
   \Bigr) \ ,
\\
\tilde B^{ff}_{qqqq,\, SS }(^1S_0) \ =& \ 
 1 + \beta^2
 + 4~\widehat{m}_{A} \widehat{m}_{B}
 - \Delta_{AB}^2 \ ,
\end{align}
and in case of $^3S_1$ partial-wave reactions
\begin{align}
\tilde B^{ff}_{rrrr,\, VV }(^3S_1) \ =& \ 
 1 + \frac{\beta^2}{3}
 + 4~\widehat{m}_{A} \widehat{m}_{B}
 - \Delta_{AB}^2 \ .
\label{eq:app_B_ff_last}
\end{align}
There are additional non-vanishing terms $\tilde B^{ff}_{n,\, i_1 i_2}$ 
related to the expressions in
(\ref{eq:app_B_ff_first}--\ref{eq:app_B_ff_last}).
In case of $i_1 i_2 = VV, SS$, the corresponding relations read
\begin{align}
\tilde B^{ff}_{rqqr,\, i_1 i_2 }(^{2s+1}S_J) \ =& \
 \tilde B^{ff}_{rrrr,\, i_1 i_2 }(^{2s+1}S_J)
 \vert_{\widehat m_A \widehat m_B \to - \widehat m_A \widehat m_B} \ ,
\\
\tilde B^{ff}_{qrrq,\, i_1 i_2 }(^{2s+1}S_J) \ =& \
 \tilde B^{ff}_{qqqq,\, i_1 i_2 }(^{2s+1}S_J)
 \vert_{\widehat m_A \widehat m_B \to - \widehat m_A \widehat m_B} \ ,
\end{align}
and our notation implies, that the product $\widehat m_A \widehat m_B$ has to
be replaced, but all other occurrences of $\widehat m_A$ or $\widehat m_B$ are
untouched. Similarly, in case of $i_1 i_2 = VS, SV$, the additional
non-vanishing $\tilde B^{ff}_{n,\, i_1 i_2}$ terms are given by
\begin{align}
\tilde B^{ff}_{rqqr,\, i_1 i_2 }(^{2s+1}S_J) \ =& \
 -\tilde B^{ff}_{rrrr,\, i_1 i_2 }(^{2s+1}S_J)
 \vert_{\widehat m_A \to - \widehat m_A} \ ,
\\
\tilde B^{ff}_{qrrq,\, i_1 i_2 }(^{2s+1}S_J) \ =& \
 -\tilde B^{ff}_{qqqq,\, i_1 i_2 }(^{2s+1}S_J)
 \vert_{\widehat m_A \to - \widehat m_A} \ .
\end{align}
The relations among kinematic factors for diagram topologies $\alpha = 3,4$ and
diagram-topologies $\alpha = 1,2$ in both the cases of box- and
triangle-topologies are given by ($X = V,S$)
\begin{align}
C^{(3)\, ff}_{n, \, i_1 X}(^{2s+1}S_J) \ =& \ 
  C^{(1)\, ff}_{n, \, i_1 X}(^{2s+1}S_J) \
  \vert_{A \leftrightarrow B} \ ,
\\
C^{(4)\, ff}_{n, \, i_1 X}(^{2s+1}S_J) \ =& \ 
  C^{(2)\, ff}_{n, \, i_1 X}(^{2s+1}S_J) \
  \vert_{A \leftrightarrow B} \ ,
\\
D^{(3)\, ff}_{n, \, i_1 i_2}(^{2s+1}S_J) \ =& \ D^{(1)\, ff}_{n, \, i_1 i_2}(^{2s+1}S_J) \
                             \vert_{A \leftrightarrow B} \ ,
\\
D^{(4)\, ff}_{n, \, i_1 i_2}(^{2s+1}S_J) \ =& \ D^{(2)\, ff}_{n, \, i_1 i_2}(^{2s+1}S_J) \
                             \vert_{A \leftrightarrow B} \ ,
\end{align}
compare to the generic diagrams in
Fig.~\ref{fig:generictriangles} and Fig.~\ref{fig:genericboxes_fermions}.
The structures $\tilde C^{(\alpha)\,ff}_{n,\, i_1 V}$ for topologies
$\alpha = 1,2$ are given by
\begin{align}
\label{eq:app_C_ff_V1}
\nonumber
\tilde C^{(\alpha)\,ff}_{qqqq,\, i_1 V}(^1S_0) \ =& \ 
 \frac{\beta^2}{4}
 - \frac{1}{4}~(1 - 2~\widehat m_A) (1 - 2~\widehat m_B)
 \\ &
 - (\widehat{m}_{A} - \widehat{m}_{B})~\frac{\Delta_{AB}}{2}
 - \frac{\Delta_{AB}^2}{4} \ ,
\end{align}
and in case of $^3S_1$ partial wave reactions the respective expressions read
\begin{align}
\nonumber
\tilde C^{(\alpha)\,ff}_{rrrr,\, i_1 V}(^3S_1) \ =& \ 
 -\frac{\beta^2}{12}
 - \frac{1}{4}~(1 + 2~\widehat{m}_{A}) (1 + 2~\widehat{m}_{B})
 \\ &
 + (\widehat{m}_{A} - \widehat{m}_{B})~\frac{\Delta_{AB}}{2}
 + \frac{~\Delta_{AB}^2}{4} \ .
\label{eq:app_C_ff_Vlast}
\end{align}
The relations of the additional non-vanishing
$\tilde C^{(\alpha)\,ff}_{n,\, i_1 V}$ expressions to the respective terms in
(\ref{eq:app_C_ff_V1}--\ref{eq:app_C_ff_Vlast}) read
\begin{align}
\nonumber
\tilde C^{(1)\,ff}_{qqrr,\, i_1 V}(^{2s+1}S_J) \ =& \
 \tilde C^{(2)\,ff}_{rrqq,\, i_1 V}(^{2s+1}S_J) \ = \
 -~\tilde C^{(1)\,ff}_{rrrr,\, i_1 V}(^{2s+1}S_J)
 \vert_{m_{A,B} \to - m_{A,B}} \ ,
\\\nonumber
\tilde C^{(\alpha)\,ff}_{rqqr,\, i_1 V}(^{2s+1}S_J) \ =& \
 \tilde C^{(\alpha)\,ff}_{rrrr,\, i_1 V}(^{2s+1}S_J)
 \vert_{m_A \to - m_A} \ ,
\\\nonumber
\tilde C^{(1)\,ff}_{qrqr,\, i_1 V}(^{2s+1}S_J) \ =& \
 \tilde C^{(2)\,ff}_{rqrq,\, i_1 V}(^{2s+1}S_J) \ = \
 -~\tilde C^{(1)\,ff}_{rrrr,\, i_1 V}(^{2s+1}S_J)
 \vert_{m_B \to - m_B} \ ,
\\\nonumber
\tilde C^{(1)\,ff}_{rqrq,\, i_1 V}(^{2s+1}S_J) \ =& \
 \tilde C^{(2)\,ff}_{qrqr,\, i_1 V}(^{2s+1}S_J) \ = \
 -~\tilde C^{(\alpha)\,ff}_{qqqq,\, i_1 V}(^{2s+1}S_J)
 \vert_{m_B \to - m_B} \ ,
\\\nonumber
\tilde C^{(\alpha)\,ff}_{qrrq,\, i_1 V}(^{2s+1}S_J) \ =& \
 \tilde C^{(\alpha)\,ff}_{qqqq,\, i_1 V}(^{2s+1}S_J)
 \vert_{m_A \to - m_A} \ ,
\\
\tilde C^{(1)\,ff}_{rrqq,\, i_1 V}(^{2s+1}S_J) \ =& \
 \tilde C^{(2)\,ff}_{qqrr,\, i_1 V}(^{2s+1}S_J) \ = \
 -~\tilde C^{(\alpha)\,ff}_{qqqq,\, i_1 V}(^{2s+1}S_J)
 \vert_{m_{A,B} \to - m_{A,B}} \ .
\end{align}
The terms $\tilde C^{(\alpha)\,ff}_{n,\, i_1 S}$ for $\alpha = 1,2$ read
\begin{align}
\nonumber
\tilde C^{(\alpha)\,ff}_{qqqq,\, i_1 S}(^1S_0) \ =& \
  \frac{\beta^2}{4}
 + \frac{1}{4}~(1 - 2~\widehat m_A) (1 - 2~\widehat m_B)
 \\&
 + (\widehat{m}_{A} - \widehat{m}_{B})~\frac{\Delta_{AB}}{2}
 - \frac{\Delta_{AB}^2}{4} \ ,
\label{eq:app_C_ff_S_last}
\end{align}
and all remaining non-vanishing $C^{(\alpha)\,ff}_{n,\, i_1 S}$ terms are obtained
from (\ref{eq:app_C_ff_S_last}) in the following
way:
\begin{align}
\nonumber
\tilde C^{(1)\,ff}_{rqrq,\, i_1 S}(^{2s+1}S_J) \ =& \
 \tilde C^{(2)\,ff}_{qrqr,\, i_1 S}(^{2s+1}S_J) \ = \
 \tilde C^{(\alpha)\,ff}_{qqqq,\, i_1 S}(^{2s+1}S_J)
 \vert_{m_B \to - m_B} \ ,
\\\nonumber
\tilde C^{(\alpha)\,ff}_{qrrq,\, i_1 S}(^{2s+1}S_J) \ =& \
 \tilde C^{(\alpha)\,ff}_{qqqq,\, i_1 S}(^{2s+1}S_J)
 \vert_{m_A \to - m_A} \ ,
\\
\tilde C^{(1)\,ff}_{rrqq,\, i_1 S}(^{2s+1}S_J) \ =& \
 \tilde C^{(2)\,ff}_{qqrr,\, i_1 S}(^{2s+1}S_J) \ = \
 \tilde C^{(\alpha)\,ff}_{qqqq,\, i_1 S}(^{2s+1}S_J)
 \vert_{m_{A,B} \to - m_{A,B}} \ .
\end{align}
In case of box-diagram topologies $\alpha = 1,2$, we find the following
$\tilde D^{(\alpha)\,ff}_{n,\, i_1 i_2}$ structures for the $^1S_0$
partial waves:
\begin{align}
\tilde D^{(\alpha)\,ff}_{rrrr,\, i_1 i_2}(^1S_0) \ =& \ 
 \frac{1}{8} ~ (1 + 2~\widehat{m}_{B} - \Delta_{AB}) 
               (1 + 2~\widehat{m}_{A} + \Delta_{AB}) \ ,
&\\
\tilde D^{(\alpha)\,ff}_{rrqq,\, i_1 i_2}(^1S_0) \ =& \ 
 \frac{\beta^2}{8} - \frac{1}{2}~\widehat{m}_{A} \widehat{m}_{B} \ . &
\end{align}
For $^3S_1$ partial-wave configurations we have
\begin{align}
\tilde D^{(\alpha)\,ff}_{rrrr\, i_1 i_2}(^3S_1) \ =& \ 
  (-1)^{\,\alpha} \, \tilde D^{(\alpha)\,ff}_{rrrr,\, i_1 i_2}(^1S_0) \ , &
\\
\tilde D^{(\alpha)\,ff}_{rrqq,\, i_1 i_2}(^3S_1) \ =& \ 
 (-1)^{\,\alpha+1} \left(
               \frac{\beta^2}{24}
             + \frac{1}{2}~\widehat{m}_{A} \widehat{m}_{B}
               \right) \ . &
\end{align}
Relations for the remaining non-vanishing kinematic factors related to both
$^1S_0$ and $^3S_1$ partial-wave processes read in case of diagram topology
$\alpha = 1$
\begin{align}
\nonumber
\tilde D^{(1)\,ff}_{qqqq,\, i_1 i_2}(^{2s+1}S_J) \ =& \ 
  \tilde D^{(1)\,ff}_{rrrr,\, i_1 i_2}(^{2s+1}S_J)
 \vert_{\ \hat m_{A,B}  \rightarrow \ - \hat m_{A,B}} \ ,
\\\nonumber
\tilde D^{(1)\,ff}_{qqrr,\, i_1 i_2}(^{2s+1}S_J) \ =& \ 
  \tilde D^{(1)\,ff}_{rrqq,\, i_1 i_2}(^{2s+1}S_J) \ ,
\\\nonumber
\tilde D^{(1)\,ff}_{rqqr,\, i_1 i_2}(^{2s+1}S_J) \ =& \ \tilde D^{(1)\,ff}_{qrrq,\, i_1 i_2}(^{2s+1}S_J) \ = \
  \tilde D^{(1)\,ff}_{rrqq,\, i_1 i_2}(^{2s+1}S_J)
 \vert_{\ \hat m_A \hat m_B  \rightarrow \ - \hat m_A \hat m_B} \ ,
\\\nonumber
\tilde D^{(1)\,ff}_{rqrq,\, i_1 i_2}(^{2s+1}S_J) \ =& \ 
  \tilde D^{(1)\,ff}_{rrrr,\, i_1 i_2}(^{2s+1}S_J)
 \vert_{\ \hat m_A \rightarrow \ - \hat m_A} \ ,
\\
\tilde D^{(1)\,ff}_{qrqr,\, i_1 i_2}(^{2s+1}S_J) \ =& \ 
  \tilde D^{(1)\,ff}_{rrrr,\, i_1 i_2}(^{2s+1}S_J)
 \vert_{\ \hat m_B \rightarrow \ - \hat m_B} \ .
\end{align}
In case of diagram topology $\alpha = 2$, the corresponding relations are given
by
\begin{align}
\nonumber
\tilde D^{(2)\,ff}_{qqqq,\, i_1 i_2}(^{2s+1}S_J) \ =& \ 
  \tilde D^{(2)\,ff}_{rrrr,\, i_1 i_2}(^{2s+1}S_J)
 \vert_{\ \hat m_{A,B}  \rightarrow \ - \hat m_{A,B}} \ ,
\\\nonumber
\tilde D^{(2)\,ff}_{qqrr,\, i_1 i_2}(^{2s+1}S_J) \ =& \ 
  \tilde D^{(2)\,ff}_{rrqq,\, i_1 i_2}(^{2s+1}S_J) \ ,
\\\nonumber
\tilde D^{(2)\,ff}_{rqqr,\, i_1 i_2}(^{2s+1}S_J) \ =& \ 
  \tilde D^{(2)\,ff}_{rrrr,\, i_1 i_2}(^{2s+1}S_J)
 \vert_{\ \hat m_A \rightarrow \ - \hat m_A} \ ,
\\\nonumber
\tilde D^{(2)\,ff}_{qrrq,\, i_1 i_2}(^{2s+1}S_J) \ =& \ 
  \tilde D^{(2)\,ff}_{rrrr,\, i_1 i_2}(^{2s+1}S_J)
 \vert_{\ \hat m_B \rightarrow \ - \hat m_B} \ ,
\\
\tilde D^{(2)\,ff}_{rqrq,\, i_1 i_2}(^{2s+1}S_J) \ =& \ \tilde D^{(2)\,ff}_{qrqr,\, i_1 i_2}(^{2s+1}S_J) \ = \
  \tilde D^{(2)\,ff}_{rrqq,\, i_1 i_2}(^{2s+1}S_J)
 \vert_{\ \hat m_A \hat m_B \rightarrow \ - \hat m_A \hat m_B} \ .
\end{align}

\subsubsection{Kinematic factors for $X_A X_B = \eta \overline{\eta}$}
In case of $X_A X_B = \eta\overline\eta$ one cannot directly construct the
coupling factors $b_{n\,\, i_1 i_2}$ using the recipe given in
Sec.~\ref{sec:app_couplingfactors}, which is based on considering the
$\chi_{e_1}\chi_{e_2} \to X_A X_B$ and $\chi_{e_4}\chi_{e_3} \to X_A X_B$
tree-level annihilation amplitudes.
In order to obtain the coupling factor expressions $b_{n\,\, i_1 i_2}$, that
correspond to the kinematic factors presented below, one should
proceed as follows:
First extract the (axial-) vector and (pseudo-) scalar coupling factors 
associated with the interaction of the $\chi_{e_1}\chi_{e_2}$ or
$\chi_{e_4}\chi_{e_3}$ pair and the $s$-channel exchanged particle species.
This is done following the steps 1. and 2. in the recipe given in
Sec.~\ref{sec:app_couplingfactors}. Next, complex-conjugate the 
couplings related to the $\chi_{e_4}\chi_{e_3}$ particle pair.
In order to determine the couplings to the ghosts, consider the 1-loop
amplitude
$\chi_{e_1}\chi_{e_2} \to \eta \overline{\eta} \to \chi_{e_4}\chi_{e_3}$, 
similar to the selfenergy-amplitude in Fig.~\ref{fig:genericselfenergy}.
Assign a ghost flow to the lower line of the 1-loop amplitude (labelled with
$X_A$ in Fig.~\ref{fig:genericselfenergy}), that flows from left to right.
Consequently there is a ghost flow from right to left on the upper line,
which is labelled with $X_B$.
Assume that the coupling factors at each of the two ghost vertices
are generically of the form $i g_2 c_{ABX_i} L_{ABX_i}$, where the
Lorentz structures $L_{ABX_i}$ are defined in
Tab.~\ref{tab:app_Labi_structures}.
Determine the expressions that replace the generic $c_{ABX_i}$ factors for
the specific process under consideration.
Now build all possible combinations of two-coupling factor products
from the set of
the neutralino/chargino couplings to the $s$-channel exchanged particles  
(including factors of $-1$ in front of vector couplings) in
the $\chi_{e_1}\chi_{e_2} \to \eta 
\overline{\eta} \to \chi_{e_4}\chi_{e_3}$
reaction, and multiply them by the $c_{ABX_{i_1}}$ and  $c_{ABX_{i_2}}$ factors.
The convention for the naming of the resulting coupling factor expressions
$b_{n,\, i_1 i_2}$ with subscripts $n=rr,qq$ is the same as in the cases
$X_A X_B = VV, VS, SS$, see Sec.~\ref{sec:app_couplingfactors}.
The coupling factors $b_{n,\, i_1 i_2}$ derived in this way correspond to
the kinematic factors given below. Note that the mass parameter $m_A$
in the expressions below refers to the mass of the ghost flowing in the lower line, and $m_B$ to the 
mass of the ghost in the upper line. 

The non-vanishing $\tilde B^{\eta \overline{\eta}}_{n,\, i_1 i_2}$ terms 
with $i_1 i_2 = VV, VS, SV, SS$ read
\begin{align}
\tilde B^{\eta \overline\eta}_{qq,\, VV }(^1S_0) \ =& \ 
 \frac{1}{4}~(1 - \Delta_{AB}^2) \ ,
\\
\tilde B^{\eta \overline\eta}_{qq,\, VS }(^1S_0) \ =& \ 
 -\frac{\widehat{m}_W}{2} (1 + \Delta_{AB}) \ ,
\\
\tilde B^{\eta \overline\eta}_{qq,\, SV }(^1S_0) \ =& \ 
 \frac{\widehat{m}_W}{2} (1 - \Delta_{AB}) \ ,
\\
\tilde B^{\eta \overline\eta}_{qq,\, SS }(^1S_0) \ =& \ 
 -\widehat{m}_W^2 \ .
\end{align}
Similarly,
\begin{align}
\tilde B^{\eta \overline\eta}_{rr,\, VV }(^3S_1) \ =& \ -\frac{\beta^2}{12} \ .
\end{align}


\providecommand{\href}[2]{#2}\begingroup\raggedright\endgroup

\end{document}